\documentclass[useAMS,usenatbib,usegraphics]{mn2e}

\usepackage{natbib}
\usepackage{graphicx}
\usepackage{float}
\usepackage{latexsym,amssymb}
\usepackage{amsmath}
\usepackage{gensymb}
\usepackage[draft]{hyperref}
\usepackage{hyperref}
\usepackage{capt-of}
\usepackage{textpos}
\usepackage{journalnames}
\usepackage{url}
\usepackage{multirow}
\usepackage{color}
\usepackage[T1]{fontenc}
\usepackage{float}
\usepackage{rotating}
\usepackage{tikz}
\usepackage{tablefootnote}
\usepackage{longtable}
\usepackage{footnote}
\usepackage{framed}
\usepackage{pgfgantt,rotating}
\usepackage{subfloat}
\usepackage{multirow}
\usepackage{blindtext}
\usepackage{booktabs}
\usepackage{tabu}
\usepackage{threeparttablex}
\usepackage{pdflscape}

\hypersetup{
    bookmarks=true,         % show bookmarks bar?
    unicode=false,          % non-Latin characters in Acrobat�s bookmarks
    pdftoolbar=true,        % show Acrobat�s toolbar?
    pdfmenubar=true,        % show Acrobat�s menu?
    pdffitwindow=false,     % window fit to page when opened
    pdfstartview={FitH},    % fits the width of the page to the window
    pdftitle={My title},    % title
    pdfauthor={Author},     % author
    pdfsubject={Subject},   % subject of the document
    pdfcreator={Creator},   % creator of the document
    pdfproducer={Producer}, % producer of the document
    pdfkeywords={keyword1} {key2} {key3}, % list of keywords
    pdfnewwindow=true,      % links in new window
    colorlinks=false,       % false: boxed links; true: coloured links
    linkcolor=black,          % colour of internal links (change box colour with linkbordercolour)
    citecolor=blue,        % colour of links to bibliography
    filecolor=black,      % colour of file links
    urlcolor=black,           % colour of external links
    linkbordercolor={1 0 0},
    citebordercolor={0 1 0}
}

\makeatletter
\Hy@AtBeginDocument{%
  \def\@pdfborder{0 0 1}% Overrides border definition set with colourlinks=true
  \def\@pdfborderstyle{/S/U/W 0}% Overrides border style set with colourlinks=true
}
\makeatother

\title[Spectral library of age-benchmark low-mass stars and brown dwarfs]{Spectral library of age-benchmark low-mass stars and brown dwarfs \thanks{Based on observations of VLT/XSHOOTER, under the program ID 098.C-0277.}}
\author[E. Manjavacas et al.]
{\parbox{\textwidth}{E. Manjavacas$^{1,2}$	\thanks{E-mail:  emanjavacas@keck.hawaii.edu},
N. Lodieu$^{3, 4}$, 
V. J. S. B\'ejar$^{3, 4}$,
M. R. Zapatero-Osorio$^{5}$,
S. Boudreault$^{6}$,
M. Bonnefoy$^{7}$
}\vspace{0.4cm}\\
\parbox{\textwidth}{
$^{1}$W. M. Keck Observatory, 65-1120 Mamalahoa Highway, Kamuela, HI 96743, USA.\\
$^{2}$Department of Astronomy/Steward Observatory, The University of Arizona, 933 N. Cherry Avenue, Tucson, AZ, 85721, USA.
$^{3}$Instituto de Astrof\'isica de Canarias, C/ V\'ia L\'actea, s/n, E38205, La Laguna (Tenerife), Spain.\\
$^{4}$Dpt. de Astrof\'isica, Univ. de La Laguna, Avda. Astrof\'isico Francisco S\'anchez s/n, 38206, La Laguna (Tenerife), Spain.\\
$^{5}$Centro de Astrobiolog\'ia (CSIC-INTA), Crta. Ajalvir km 4, E-28850 Torrej\'on de Ardoz, Madrid, Spain.\\
$^{6}$Max-Planck-Institut f\"ur Sonnensystemforschung, Justus-von-Liebig-Weg 3, 37077, G\"ottingen, Germany.\\
$^{7}$Universit\'e Grenoble Alpes, CNRS, IPAG, 38000 Grenoble, France.}}

\begin{document}

%\date{In prep. for MNRAS}
%\date{Accepted 1988 December 15. Received 1988 December 14; in original form 1988 October 11}

\pagerange{\pageref{firstpage}--\pageref{lastpage}} \pubyear{\today{}}

\def\LaTeX{L\kern-.36em\raise.3ex\hbox{a}\kern-.15em
    T\kern-.1667em\lower.7ex\hbox{E}\kern-.125emX}

\maketitle

\label{firstpage}

\begin{abstract}

 In the past years, some extremely red brown dwarfs were found. They were believed to have low surface gravity, but many of their spectral characteristics were similar to those of high surface gravity brown dwarfs, showing that youth spectral characteristics are poorly understood. We aim to test surface gravity indicators in late-M and early-L brown dwarf spectra using data obtained with the X-shooter spectrograph at the Very Large Telescope. We selected a benchmark sample of brown dwarfs members of Chamaeleon\,I ($\sim$2 Myr), Upper Scorpius (5$-$10~Myr), Pleiades (132$\pm$27 Myr), and Praesepe (590$-$790 Myr) with well-constrained ages, and similar metallicities. 
We provided a consistent spectral classification of the sample in the optical and in the near-infrared. We measured the equivalent widths of their alkali lines, finding that they have a moderate correlation with age, especially for objects with spectral types M8 and later. We used spectral indices defined in the literature to estimate surface gravity, finding that their gravity assignment is accurate for 75\% of our sample. We investigated the correlation between red colours and age, finding that after $\sim$10~Myr, the colour does not change significantly for our sample {with spectral types M6.0-L3.0}. In this case, red colours might be associated with circumstellar disks, {ring structures, extinction, or viewing angle}. Finally, we calculated the {bolometric luminosity}, {and $J$ and $K$ bolometric corrections} for our sample. We found that six objects are overluminous compared to other members of the same association. Those objects are also flagged as binary candidates by the $Gaia$ survey.

\end{abstract}


%============================= section 1 =============================

\section{Introduction}\label{introduction}

Brown dwarfs are substellar objects that are unable to sustain hydrogen fusion. Since they are born, brown dwarfs  cool down with time and contract. During this process, brown dwarfs change spectral types, through the M, L, T and  Y, changing their luminosity and chemistry \citep{Kirkpatrick2012}. Due to this evolution, {spectral types do not constrain substellar masses, and we additionally need ages to estimate masses  using evolutionary models for substellar objects \citep{1997ApJ...491..856B,Baraffe2015}}. 

{Brown dwarfs contract as they cool down over their lifetime, therefore, younger brown dwarfs (<100~Myr) have larger radii and lower gravity ($g=GM/R^{2}$) than their older counterpart in the field (>500~Myr)}.  
Young brown dwarfs and exoplanet atmospheres are expected to {share  similar colours, temperatures and surface gravities}  \citep{2004A&A...425L..29C, 2008Sci...322.1348M, Faherty2013}. Nevertheless, {isolated} young brown dwarfs, unlike exoplanets, are isolated and not close to their parent star, which make them easier to observe, {especially} for spectroscopic studies.    {The study of young free-floating brown dwarf are excellent proxies to improve our view of the atmospheres of imaged young giant exoplanets. }

Gravity affects brown dwarf {atmospheres}, changing their spectral characteristics.  Many atomic lines and molecular bands are weaker in spectra of young brown dwarfs than for their higher gravity counterparts. The neutral alkali lines are weaker over the whole optical and near-infrared spectrum: Rb\,I at 794.8~nm, Na\,I at 818.3~nm, Na\,I at 819.5~nm, Cs\,I at 852.0~nm, K\,I at 1169~nm, K\,I at 1177~nm, K\,I at 1243~nm and K\,I at 1253~nm   ({\citealt{Martin1996, Gorlova2003, McGovern,Martin2017}, and references therein. See \citealt{Schlieder2012}, for a theoretical explanation}). In addition, FeH absorptions in the $J$ and $H$-bands are also less prominent for low gravity dwarfs {\citep[among others]{Allers2013, Lodieu2018}}. In contrast, {the VO bands at {0.74, 0.96}, 1.06~$\mu$m and 1.18~$\mu$m, and the TiO bands at 0.71,
0.76, 0.82, 0.84, and 1.25 $\mu$m  are stronger for low-surface gravity objects {\citep[and references therein]{Martin1996, Zapatero_Osorio1997, Allers2013}, particularly in L-dwarfs \citep{Allers2013}}, due to {inefficient} condensation  \citep{Lodders2002}}.
In addition, very red colours were also believed to be an indication of low surface gravity, as it was expected to prevent the dust settlement of the upper atmosphere.
In the past years, several tens of very red L brown dwarfs \citep[among others]{Kirkpatrick2006, Stephens2009, 2012AJ....144...94G, 2013ApJ...777L..20L, 2014MNRAS.439..372M,Filippazzo2015,Faherty2016,Liu2016,Schneider2014, Schneider2017, Best2017} have been found in large surveys, like 2MASS (The Two Micron All Sky Survey, \citealt{2MASS}), SDSS (Sloan Digital Sky Survey, \citealt{sdss}), and PanStarrs (Panoramic Survey Telescope and Rapid Response System, \citealt{panstarrs}). {These very red brown dwarfs were believed to be young, nonetheless, some  objects did not exhibit {clear spectroscopic signposts of low gravity atmospheres in their spectra}, like for example, weak alkali lines. This is the case of ULAS J222711-004547 \citep{2014MNRAS.439..372M}, 2MASS~J035523.37$+$113343.7  \citep{Faherty2013}, and WISEP J004701.06$+$680352.1  \citep{2015ApJ...799..203G}, among others. These examples suggest that red colours \textit{alone} cannot be used as a youth signature.}

%the kinematics for some of these ultra-red objects indicated that some  may belong to moving groups with ages much older than expected for such an object, or even to the field, this is the case of 2MASS~J035523.37$+$113343.7 \citep{Faherty2013}, PSO J318.5338-22.8603 \citep{Liu2013}, ULAS J222711-004547 \citep{2014MNRAS.439..372M}, WISEP J004701.06$+$680352.1 \citep{2015ApJ...799..203G}, VHS J1256-1257b \citep{Gauza2015} among others. These examples suggest that colours alone cannot constrain the age precisely \citep{Marocco2013,Liu2016,Bowler2017,Zapatero-Osorio2017}, and  that {the youth characteristics in brown dwarf spectra are poorly understood}. Therefore, a systematic study in a wide wavelength range of a {well characterized sample with known ages} is needed to trace the effect of gravity in brown dwarf spectra.

In this paper, we aim to test the most common methods to estimate surface gravity (i.e. age) in brown dwarfs. To achieve this aim, we targeted spectra of members of open clusters and associations with well-determined ages.  We selected  20  members to the Chamaeleon\,I (Cha\,I) region ($\sim$2~Myr, \citealt{Luhman2007}), the Upper Scorpius (UppSco) association (5--10~Myr, \citealt{Slesnick2008, Song2012, Feiden2016, Pecaut2016,Rizzuto2016, Fang2017}), the Pleiades open cluster (132$\pm$27~Myr, \citealt{Stauffer1998, Barrado2004, Dahm2015, Lodieu2019c}), and the Praesepe open cluster (590$-$790~Myr, \citealt{Fossati2008, Delorme2011, Brandt2015, Gossage2018}). The ages of these associations or open clusters correspond respectively to the gravities that defined the $\gamma$, $\beta$ and  $\alpha$  for low, intermediate and field gravity classification of M and L dwarfs \citep{2009AJ....137.3345C}, or the VL-G (Very Low Gravity), INT-G (Intermediate Gravity) and FLD-G (Field Gravity) classes defined by \cite{Allers2013}.

% We will focus on very low-mass stars and brown dwarfs with spectral types between M6 and L3.5 and $J=$14.9--18.5. 

%Our objective is to test the spectral features that are believed to change with age from 600 to 2500 nm, with intermediate spectral resolution and good S/N. In particular, we are going to test the spectral indices developed by Allers \& Liu (2013) for young brown dwarfs, but  contrary to their sample, where only 12\% of the objects have accurate age estimates, {we are going to use a sample of known members of open clusters with well-determined ages, multiplying by 4 the sample of objects} (21 objects requested and 14 from the ESO Archive). 

%We need a minimum resolution of R$\sim$1000 to resolve  the alkali lines in the optical (Rb\,I at 794.8 nm, Na\,I doublet at 818.3 nm and 819.5 nm, Cs\,I at 852.0 nm) and in the near-infrared (K\,I 1169 nm, K\,I 1177 nm, K\,I 1243 nm  and K\,I 1253 nm), with a S/N $\sim$10--30, depending on the magnitude of the object. 

%The relationship between red colours, age and spectral characteristics for young brown dwarfs will help to understand/reconstruct the evolution of their atmospheres in the early stages of their lives and to shed light into the spectral characteristics of giant exoplanets atmospheres.

In Section \ref{sample}, we describe the selected sample of objects presented in this paper. In Section \ref{obs_reduction}, we provide details on the X-shooter  observations and data reduction. In Section \ref{analysis_results} we explain the analysis carried out and our results. In Section \ref{physical_param} we derived physical parameters for  our sample using evolutionary models for substellar objects. Finally, in Section \ref{conclusions_remarks} we summarise our {results}.

\begin{landscape}
\begin{table}
	\footnotesize
	\caption{List of observed targets with their {full names, coordinates}, magnitudes, spectral types, cluster or association memberships, and flags for disk bearing sources and binary candidates.}  
	\label{literature}
	\centering
	\renewcommand{\footnoterule}{}  % to avoid a line before footnotes
	\begin{center}
		\begin{tabular}{lllllllllcl}
			\hline 
			
			Num. &Name & RA  &  DEC  & $J$ [mag]  & Opt SpT & NIR SpT &  Open Cluster/Association & Disk &Binary candidate?   & Ref.\\		
			
			\hline              		
			            1& UGCS J083748.00$+$201448.5 & 08 37 48.01 & $+$20 14 48.6 & 16.50$\pm$0.01 & 	&  & Praesepe (590$-$790~Myr) & $\mathrm{NR^{a}}$ & $\mathrm{NR^{b}}$  & 1\\
			            2& 2MASS J08370215$+$1952074 & 08 37 02.13 & $+$19 52 07.4 & 15.69$\pm$0.01 & M7 [29] 	&  & Praesepe (590$-$790~Myr)& $\mathrm{NR^{a}}$ & Yes  & 2, 31 \\
			            3&	UGCS J083654.60$+$195415.7 & 08 36 54.60 & $+$19 54 15.7 &17.11$\pm$0.02 & 	&  & Praesepe (590$-$790~Myr)& $\mathrm{NR^{a}}$ & $\mathrm{NR^{b}}$  & 2\\
			            4& 2MASS J08410852$+$1954018& 08 41 08.54 & $+$19 54 01.0 & 16.45$\pm$0.01 & 	&  & Praesepe (590$-$790~Myr)& $\mathrm{NR^{a}}$ & Yes  & 2, 31 \\
			            5& 2MASS J08370450$+$2016033 & 08 37 04.49 & $+$20 16 03.2 & 16.82$\pm$0.02 & 	&  & Praesepe (590$-$790~Myr)& $\mathrm{NR^{a}}$ & Yes  & 3, 31\\
			            6 & UGCS J084510.65$+$214817.0 & 08 45 10.66 & $+$21 48 17.1 & 17.42$\pm$0.03 & L0.5 [28] &  & Praesepe (590$-$790~Myr)&$\mathrm{NR^{a}}$ & Yes  &  3, 31 \\

			           7& 2MASS J03484469$+$2437236 & 03 48 44.69 & $+$24 37 23.6 & 14.90$\pm$0.01 &	&  M5.5 [25] & Pleiades (132$\pm$27~Myr)  &$\mathrm{NR^{a}}$ & Yes  &  4, 31 \\
			            8& 2MASS J03491512$+$2436225 & 03 49 15.12 & $+$24 36 22.5 & 15.17$\pm$0.01 &	M6.5 [27]&   M6.5 [25] & Pleiades (132$\pm$27~Myr) &$\mathrm{NR^{a}}$ & $\mathrm{NR^{b}}$  & 5 \\
			           9 & 2MASS J03512557$+$2345214 & 03 51 25.57 & $+$23 45 21.4 &16.13$\pm$0.03 &	M8.0 [26]&   M8.0 [25] & Pleiades (132$\pm$27~Myr) &$\mathrm{NR^{a}}$ &  $\mathrm{NR^{b}}$  &  6 \\
			           10 &2MASS J03443516$+$2513429 & 03 44 35.16 & $+$25 13 42.9 & 15.66$\pm$0.01 &	&  M9.3 [25] & Pleiades (132$\pm$27~Myr) & $\mathrm{NR^{a}}$ & Yes   &  7, 31\\
			          11 & 2MASS J03463425$+$2350036 & 03 46 34.25 & $+$23 50 03.6 & 17.46$\pm$0.03 &	&   & Pleiades (132$\pm$27~Myr) & $\mathrm{NR^{a}}$ & $\mathrm{NR^{b}}$   & 8 \\
			          12 & 2MASS J03461406$+$2321565 & 03 46 14.06 & $+$23 21 56.5 & 15.67$\pm$0.02 &	&   & Pleiades (132$\pm$27~Myr) & $\mathrm{NR^{a}}$ & $\mathrm{NR^{b}}$   &  9 \\
			          13 & 2MASS J03541027$+$2341402 & 03 54 10.27 & $+$23 41 40.2 & 18.14$\pm$0.05 & 	&  L3.0 [25]& Pleiades (132$\pm$27~Myr) &$\mathrm{NR^{a}}$ & $\mathrm{NR^{b}}$  & 10 \\
			            
			         14 & 2MASS  J15591135--2338002 & 15 59 11.35  & --23 38 00.2 & 14.40$\pm$0.04 & M7.0 [11]	&   & UppSco (5--10~Myr)& Yes & $\mathrm{NR^{b}}$   & 11, 17\\
			          15 & 2MASS J16060391--2056443 & 16 06 03.91 & --20 56 44.3  &13.53$\pm$0.03  & M7.0 [24]	&  & UppSco (5--10~Myr) & Yes & $\mathrm{NR^{b}}$   &  12, 18\\ 
			         16 & 2MASS  J16060629--2335133 & 16 06 06.29 & --23 35 13.3  & 16.23$\pm$0.01  & M9 [23]	&  L0.0 [13] & UppSco (5--10~Myr) & $\mathrm{NR^{a}}$ & $\mathrm{NR^{b}}$  & 13\\

%			        17 & 2MASS   J16073799--2242468 & 16 07 37.99  & --22 42 46.8 & 16.79$\pm$0.01 & L0.0 & Upper Scorpius (5--10~Myr) &12 \\
                        
%                     18 & 2MASS  J16082847--2315103 & 16 08 28.47 & --23 15 10.3 & 15.48$\pm$0.06 & L1.0 & Upper Scorpius (5--10~Myr) &12 \\
                        
			        17 &	2MASS J11085497--7632410  & 11 08 54.97 & --76 32 41.0 &13.06$\pm$0.03  & M5.5 [22]	&    & Cha\,I ($\sim$2~Myr)  & Yes & $\mathrm{NR^{b}}$   & 14, 19 \\    
					18 &	2MASS J11123099--7653342 & 11 12 30.99 & --76 53 34.2 & 14.07$\pm$0.03 & M7.0 [21]	& M7.0 [21] & Cha\,I ($\sim$2~Myr) & $\mathrm{NR^{a}}$ & $\mathrm{NR^{b}}$   & 15 \\
					19 &	2MASS J11074656--7615174   & {11 07 46.56}  & {--76 15 17.4} & 13.94$\pm$0.03& M6.5 [30] 	&   & Cha\,I ($\sim$2~Myr)  & Yes &  $\mathrm{NR^{b}}$  &  30\\
					20 &	2MASS J11062554--7633418  & {11 06 25.54} & {--76 33 41.8} & 13.00$\pm$0.03 & M5.5 [30]	&    & Cha\,I ($\sim$2~Myr)  & Yes & $\mathrm{NR^{b}}$   &  30 \\
	
			\hline

		\end{tabular}
	\end{center}
	
	\begin{tablenotes}
		\small
		\item References: [1] - \cite{Boudreault2010}, [2] - \cite{Hodgkin1999}, [3] - \cite{Boudreault2012}, [4] - \cite{Bouvier1998}, [5] - \cite{Stauffer1998}, [6] - \cite{Bouvier1998}, [7] - \cite{Bouvier1998}, [8] - \cite{Nagashima2003}, [9] - \cite{Lodieu2012},   [10] - \cite{Bihain2006}, [11] - \cite{Ardila2000}, [12] - \cite{Martin2004}, [13] - \cite{Lodieu2007}, [14] - \cite{Persi2000}, [15] - \cite{Lopez2004}. \\
        References of disks detections: [16] - \cite{Luhman2008}, [17] - \cite{Vanderplas2016}, [18] - \cite{Herczeg2009}, [19] - \cite{Long2017}.\\
        a: NR: The existence of a disk has not been reported for this object. \\
        References for spectral type: [20] - \cite{AlvesdeOliveira2012}, [21] - \cite{Luhman2007}, [22] - \cite{Luhman2004}, [23] - \cite{Lodieu2018}, [24] - \cite{Slesnick2008}, [25] - \cite{Bihain2010}, [26] - \cite{Martin1996}, [27] - \cite{Stauffer1998}, [28] - \cite{Boudreault2013}, [29] - \cite{West2011}, [30] - \cite{Manara2017} \\
		References binary candidates: [31] - \cite{Gaia_Collaboration2018}\\
		b: NR: Not reported as a binary candidate in the literature.
		
	\end{tablenotes}
\end{table}
\end{landscape}

\section{Sample Selection}\label{sample}

We have selected a sample of 20 benchmark objects {with confirmed membership to} open clusters or associations, with known ages, metallicities and distances. We have chosen a sample  with photometry and optical spectral types confirming their membership.  In the case of the Pleiades and Praesepe, members are not only confirmed by photometry and optical spectroscopy, but also by proper motion \citep{2012MNRAS.422.1495L,Boudreault2012,  2015A&A...577A.148B}. We selected objects to each of the associations or open clusters with magnitude ranging from {$J$=14.9 to $J$=18.1 mag, which allow us to obtain spectroscopy of a  signal-to-noise of between 4 and  53 in the near-infrared, using exposure times between 40 min and 1.5~h}. Our sample have spectral types between M5.5 and L3.5, having at least one target per open cluster or association in each spectral type between M5.5 and L0. To complete our sample in age range, we added three members to the UppSco association, and four from the Cha\,I  region from proposals:  093.C-0109(A) (P.I. Van der Plas),   093.C-0769(A) (P.I. Bonnefoy), and 095.C-0378(A) (P.I. Testi).
We list the objects in our sample with their main characteristics in Table~\ref{literature}.

{We selected four  members of the Cha\,I region \citep[and references therein]{Prusti1991,Boulanger1998, Mamajek2000, Winston2012}. Cha\,I is a molecular cloud with ongoing star formation due to its young age ($\sim$2~Myr \citealt{Luhman2007})}, but old enough to have relatively low extinction (A$_{V}$\,$\leq$\,5 mag).  \cite{Roccatagliata2018} found that Cha\,I has a double population, one north and one south, with slightly difference distances. The north is at 192.7$\pm$0.4~pc, while the southern one is at 186.5$\pm$0.7~pc, with {nearly solar} metallicity ([Fe/H] = --0.08$\pm$0.04 dex; \citealt{Spina2014}). {All our Cha\,I targets lie in the northern cloud, with exception of 2MASS J11123099$-$765334.}

We have three {members} in our sample that belong to the UppSco association. UppSco \citep[and references therein]{deZeeuw1999, Kraus2008, Pecaut2012, Lodieu2013} is part of the Scorpius Centaurus association, with  an estimated age of 5--10~Myr \citep{Slesnick2008, Song2012, Feiden2016, Pecaut2016,Rizzuto2016, Fang2017}, and {nearly solar metallicity} ([Fe/H] = 0.02 dex, \citealt{Carpenter2014}). Its recent {average} updated distance from $Gaia$ is {d=144$\pm$2~pc, with a standard deviation of $\pm$17.2~pc} \citep{Fang2017}.

Seven objects of our sample are members of the Pleiades open cluster. The Pleiades \citep[and references therein]{Zapatero_Osorio1997, Bouvier1998, Moraux2003, Deacon2004, Lodieu2012}  is an intermediate age open cluster  (132$\pm$27~Myr, according to the most recent estimate from \citealt{Lodieu2019c}), with solar metallicity \citep{Soderblom2009} at a distance of {139.41$\pm0.08$~pc \citep{Gaia_Collaboration2018}, although more recently, \cite{Lodieu2019c} provide a distance of 135.15$\pm$0.43~pc}. The cluster has low foreground extinction, E(B-V) = 0.04~mag, as determined by \cite{Breger1986}.

{Finally, six of the members of {our} sample belong to the Praesepe open cluster} \citep[and references therein]{Jones1991, Pinfield2003, Boudreault2012, Brandt2015}. The Praesepe is the oldest cluster of our sample, with an age between 590$-$790~Myr, \citep{Fossati2008, Delorme2011, Brandt2015, Gossage2018, Martin2018}, and a metallicity of [Fe/H] = 0.12--0.16 dex \citep{Netopil2016}. Its most updated distance provided by $Gaia$ \citep{Gaia_Collaboration2018, Lodieu2019c} is 186.18$\pm$0.11~pc. The  Praesepe has a low reddening on the line of sight of E(B-V) = 0.027$\pm$0.04 \citep{Taylor2006}.

\section{Observations and Data Reduction}\label{obs_reduction}

Our targets were observed between October
2016 and January 2017, under the proposal number ID: 098.C-0277(A) (P.I.  Manjavacas), using the X-shooter spectrograph \citep{Vernet2011}, mounted at the Kueyen (UT2,VLT) telescope at the Paranal Observatory.   X-shooter  
is composed of three arms: UVB (300-550 nm), optical (550-1000 nm) and
near-infrared (1000-2500 nm). It was operated in echelle slit nod mode, using
the 1.3" slit width for the UVB arm, and the 1.2" slit width for the optical and the
near-infrared arms. This setup provides resolutions of $\sim$2030 in the UVB, $\sim$3360 in the VIS,
and $\sim$3900 in the near-infrared. The average signal to noise of each spectra in the optical and near-infrared is shown in Table \ref{log0} of the Appendix. Observations were performed at the parallactic angle to mitigate the effect of differential chromatic refraction. We moved the object along the slit between
two positions following an ABBA pattern with a size of 6 arcsec. The flux expected in the UVB arm is
extremely low, therefore {we chose  not to use} spectra taken in this range. Telluric
standards were observed before or after every target at similar airmass. 
Bias, darks and flats were taken  every night. Arc frames
were taken  every second day. The  observing log  including telluric standard stars and {the raw seeing during the observations} is shown in  Table \ref{log0} {of the Appendix}.

The spectra were reduced using the ESO X-shooter pipeline version 1.3.7. In
the reduction cascade, the pipeline deletes the non-linear pixels and subtracts
bias in the optical or dark frames in the near-infrared. It generates a
guess order from a format-check frame, a reference list of arc line and a
reference spectral format table. It refines the  guess order table by illuminating the instrument pinhole with a continuum lamp. The master flat frame and the order
tables tracing the flat edges are created. 
Finally, the pipeline determines the instrumental response. 

In the case of the near
infrared, we extracted the 2D spectrum provided by the pipeline with the  {\tt{apall}} routine in IRAF (Image Reduction and Analysis Facility, \citealt{Tody1986,Tody1993})\footnote{IRAF is distributed by the National Optical Astronomy Observatory, which is operated by the Association of Universities for Research in Astronomy (AURA) under a cooperative agreement with the National Science Foundation}. We used the spectrum of the telluric calibration star of the corresponding 
science target observed in the same night to {correct from instrumental response and remove telluric lines}.  
First, we removed artifacts and cosmic rays from the
calibration stars. {We also removed}  the H and He absorption lines {from} their spectra using a Legendre
polynomial fit of the pseudo-continuum around the line. We then derived a response function by dividing the non-flux calibrated clean spectrum of the calibration standard by a black body synthetic spectrum with the same temperature as the
telluric star \citep{Theodossiour_Danezis1991}.  Finally, we divided our spectra by the corresponding response function calculated with the spectrum of respective calibration star, eliminating the {telluric absorption bands} and correcting for the instrumental response simultaneously.

We {flux calibrated} our near-infrared spectra using the $J$-band {magnitudes} given by
2MASS (Two Micron All Sky Survey; \citealt{2MASS}) or UKIDSS (UKIRT Infrared Deep Sky Survey; \citealt{Lawrence2007}) for our targets. We {convolved} our near-infrared spectra with the
$J$ filter transmission curves of 2MASS or UKIDSS, depending on the object. 
{The convolved spectra were integrated over the $J$-band wavelength range and the results were taken to the observed photometric fluxes. This procedure is affected by the error bars of the $J$-band magnitudes and the uncertainties introduced during the response correction.}
 
To {flux} calibrate  the optical X-shooter spectra, we calculated a scaling factor in the overlapping wavelengths of the optical and near-infrared spectra (995$-$1020 nm), to match the optical and the near-infrared {data}. 

%The reduced spectra are shown in Fig. \ref{all_spectra}\footnote{These spectra will be available in the ESO Phase 3 data release}.  Wavelengths  affected by telluric absorption are removed from the figure....

In addition, we carried out low-resolution optical spectroscopy with the 
Optical System for Imaging and low Resolution Integrated Spectroscopy 
\citep[OSIRIS;][]{cepa00} mounted on 
the 10.4~m Gran Telescopio Canarias (GTC) in the Roque de Los 
Muchachos Observatory in La Palma (Canary Islands) under program GTC66-12B 
(PI Boudreault). We used the 
R300R grism and a 1.0 arcsec slit with a 2$\times$2 binning {of the detector}, yielding 
a spectral resolution of $\sim$300 at 680~\AA{}. 
This configuration shows 
contamination from the second-order light redwards of 9000~\AA{}. 
Therefore, we restrain the range of analysis of our spectra to the 
5500--9000~\AA{} wavelength range. We observed six members of Praesepe and one member of the Pleiades 
(see Table \ref{log1} in the Appendix for a log of observations) under grey time, spectroscopic 
conditions, and seeing better than 1.2 arcsec. We offset the object along 
the slit in case of several exposures. We reduced the optical spectra 
using standard IRAF packages \citep{Tody1986,Tody1993}. To summarize, we 
first subtracted the overscan and removed the flat-field contribution 
before trimming the images with the {\tt{ccdred}} package. Then we {optimally} extracted 
the spectrum that we calibrated in wavelength using a combination of 
HgAr, Ne and Xe lamps. 
%We calibrate our targets in flux with the spectrophotometric standard stars G191-B2 \citep{Oke1990} and Ross\,640 \citep{Oke1974}. 
To {flux} calibrate GTC/OSIRIS optical spectra, we followed a similar procedure  for the near-infrared {data}, but we used the Pan-STARRS (Panoramic Survey Telescope and Rapid Response System; \citealt{panstarrs}) $i$-filter.
The final {Xshooter and OSIRIS} spectra are presented in Figures \ref{all_spectra_pra}, \ref{all_spectra_ple}, \ref{all_spectra_uppsco} and \ref{all_spectra_cha} of the Appendix.

%We measured the pseudo-equivalent widths of the sodium doublet around 
%$\sim$8200~\AA which is unresolved at our resolution for all OSIRIS spectra. 
%We report the values with their associated errors (in \AA{}) {in Table X}.

\section{Analysis and Results}\label{analysis_results}

\subsection{Spectral Types}\label{SpT}

We aimed at performing a consistent spectral classification of the objects in our sample in the optical and in the near-infrared independently. Within both wavelength ranges, we performed two different classifications: one comparing our  spectra to field {late-M and early L-type dwarfs} published in different spectral libraries, and a second one determined by comparing our sample to young brown dwarf spectra {of the same types}. 
For this purpose, we compared our optical  spectra to optical young spectral libraries \citep{Luhman2010, Luhman2018, Lodieu2018}, and field brown dwarf spectral libraries \citep{Kirkpatrick1999}, determining the best match to both young and field optical spectra. For the comparison, we degraded the resolution of our X-shooter spectra to the resolution of the corresponding comparison spectra. To determine the best match for each object and wavelength range we used the following expression as in  \cite{Cushing2008}:

\begin{equation}\label{eq_cushing}
    G = \sum_{\lambda}\left[\displaystyle\frac{C(\lambda)-\alpha T(\lambda)}{\sigma_{c}(\lambda)}\right]^{2},
    \end{equation}
    where $C(\lambda)$ is the spectrum of our object, $T(\lambda)$ is the {comparison} spectrum,  $\alpha$ is a scaling factor that {minimises} $G$, and $\sigma_{c}(\lambda)$ are the uncertainties of the {target} spectrum.

    %{To perform a spectral classification in the near-infrared, we followed the same procedure than for the optical, comparing to near infrared libraries of young brown dwarfs \citep{Allers2007, Allers&Liu}, and field brown dwarfs (SpeX Prism Libraries\footnote{\url{http://www.browndwarfs.org/spexprism/}}) independently.
    %We  additionally checked all the best spectral matches by visual inspection. }
    
    {In Table \ref{spts} we summarise the optical and near-infrared spectral types obtained from the best matches. We show the best matches  for Praesepe, the Pleiades, USco, and Cha in Figures 1$+$2, 3$+$4, 5$+$6, and 7$+$8, respectively.}
    
    % Table \ref{spts}, we summarise the spectral types found when comparing to field and young objects in the optical and in the near infrared. In Figures \ref{field_prae} and \ref{young_prae}, we show the best matches of members of the Praesepe open cluster to optical and near infrared field and young brown dwarfs. In Figures \ref{field_plei}, and \ref{young_plei}, we show a similar plot for the Pleiades open cluster of our sample. In Figures \ref{field_upps} and \ref{young_upps}, we show a similar plot for members of the Upper Scorpius open cluster. Finally, in Figures \ref{field_cha}, and \ref{young_cha}, we show the best matches to field and young brown dwarfs in the optical and near infrared to Chamaelon\,I confirmed members of our sample.

\begin{table*}
	\footnotesize
	\caption{List of  derived optical and near-infrared spectral types using field and young brown dwarfs.}  
	\label{spts}
	\centering
	\renewcommand{\footnoterule}{}  % to avoid a line before footnotes
	\begin{center}
		\begin{tabular}{llcccc}
			\hline 
			
						Num. &Name & Opt SpT field & Opt SpT young & NIR SpT Field & NIR SpT Young\\			
			\hline              		
			            1& UGCS J083748.00$+$201448.5 & M7.0 & M5.5 & M6.0 & M8.0 \\
			            2& 2MASS J08370215$+$1952074  & M6.0 & M7.0  & M7.5 & M8.0  \\
			            3& UGCS J083654.60$+$195415.7 & M8.0 & M7.25 & M7.0 & M8.0 \\
			            4& 2MASS J08410852$+$1954018  & M8.0 & M8.0  & M9.0 &  M8.0\\
			            5& 2MASS J08370450$+$2016033  & L0.0 & L1.0 & L0.0 & L0.0  \\
			            6 & UGCS J084510.65$+$214817.0 & L1.5 & L1.0 & L1.5 & L2.0\\
			            
			           7 & 2MASS J03484469$+$2437236 & M6.0 & M5.5 & M6.0 & M6.0  \\
			            8& 2MASS J03491512$+$2436225 & M7.0 & M6.5  & M6.0 & M6.0  \\
			           9 & 2MASS J03512557$+$2345214 & M6.5 & M6.5 & M7.0 & M7.25 \\
			           10 &2MASS J03443516$+$2513429 & M7.5 & M6.5 & M7.0 & M7.0\\
			          11 & 2MASS J03463425$+$2350036 & L0.0 & L1.0 & L1.0 & L2.0  \\
			          12 & 2MASS J03461406$+$2321565 & M7.0 & M6.5 & M7.0 &  M6.0 \\
			          13 & 2MASS J03541027$+$2341402 & NA & NA  & L3.0 & L3.0 \\
			            
			         14 & 2MASS  J15591135--2338002 & M7.5 & M7.0 &  M8.5  & M7.25 \\
			          15 & 2MASS J16060391--2056443 & M8.0 & M7.5 & M8.5 & M7.0 \\ 
			         16 & 2MASS  J16060629--2335133 & L0.0 & L1.0 & M9.5 & M9.0 \\
                        
			        17 &	2MASS J11085497--7632410 & M7.0 & M6.0 & M8.0 & M6.0\\    
					18 &	2MASS J11123099--7653342 & M7.0 & M6.5 & M7.0 &  M7.0 \\
					19 &	2MASS J11074656-7615174   & M7.0 & M6.5 & M8.0 & M8.0 \\
					20 &	2MASS J11062554-7633418  & M6.0 & M5.75 & M8.0 & M6.0  \\
	
			\hline

		\end{tabular}
        
    \begin{tablenotes}
		\small
		\item NA: Spectrum not available.

	\end{tablenotes}
	\end{center}

\end{table*}

We find a maximum dispersion of $\pm$2.5 spectral types between the different spectral classifications, for the  members of the Praesepe, and for the members of the Chamaelon\,I association. Differences in spectral classification using different wavelengths and type of objects are expected, as the spectral characteristics of brown dwarfs and low-mass stars evolve with age (i.e. surface gravity). This {likely} explains why fitting field brown dwarf spectra to younger brown dwarf spectra and vice-versa might provide slightly different spectral types.

{In addition, we need to highlight that the three Cha\,I members (2MASS J11062554--76334, 2MASS J11074656--76151, and 2MASS J11085497--763241), and two UppSco members (2MASS J15591135--233800 and 2MASS J16060391--20564) harbours disks (see Table \ref{literature} for details). As a consequence, their Spectral Energy Distribution (SED) usually show near or mid-infrared excesses, which might lead to later spectral type estimates.}  

For the subsequent analysis in this paper, we will adopt as final spectral classification the one provided by the comparison to near-infrared field objects (third column in Table \ref{spts}). The reason to choose this spectral classification is that most of the flux of the objects in our sample is in this wavelength range.  The reason we choose field objects for the spectral classification is that they provide a reasonable fit for most of the object's near-infrared spectra. In addition, the spectral classification sequence of field low-mass stars and brown dwarfs have been well-defined in the literature so far \citep[and references therein]{Kirkpatrick1999, Cushing, Burgasser2006}, in contrast with the spectral classification for young low-mass stars or brown dwarfs.

\subsection{Spectral Characteristics} \label{Spt_char}

\subsubsection{Pseudo Equivalent Widths of Alkali Lines}\label{pEWs}

Surface gravity has been found to be correlated with the pseudo equivalent widths (pEWs) of the alkali lines \citep[and references therein]{Steele_Jameson1995, Martin1996,Gorlova2003, Cushing,Allers2007,Allers2013,Bonnefoy2014b}. 

To estimate {relative} surface gravities of our sample, we measured the pEWs of the following alkali lines in the optical: Rb\,I (794.8~nm), Na\,I (818.3 nm), Na\,I (819.5~nm), and Cs\,I (852.0~nm). We also measured the Li\,I line (670.8~nm) and the H$\alpha$ line (656.3~nm). In the near-infrared, we measure the 1169.2~nm, 1177.8~nm, 1243.7 and 1252.9~nm K\,I lines. These lines  are blended with FeH, Fe\,I, and $\mathrm{H_{2}O}$ features in the $J$-band for objects with spectral types similar to those in our sample, which {might constrain} the reliability of these features as age indicators.
We measured the pEWs using the same pseudo continuum windows as  defined in \cite{Allers2013} for the alkali lines in the near-infrared. For alkali lines in the optical, we use the windows shown in Table~\ref{conti}.

\begin{table}
	\caption{List of measured lines in the optical spectra  with central wavelength of the line and continuum edges.}  
	\label{conti}
	\centering
    \small
	\renewcommand{\footnoterule}{}  % to avoid a line before footnotes
	\begin{center}
		\begin{tabular}{lccc}
			\hline 
			
			Line  & $\lambda_{line}$ (nm)& Continuum 1 (nm)  &  Continuum 2 (nm) \\		
			
			\hline              		
			        Li\,I & 670.8 & 669.0 & 672.0 \\
			        H$\alpha$& 656.3 & 655.0& 657.0 \\
			        Rb\,I & 794.8 & 792.7 & 797.5\\
			        Na\,I & 818.3 & 817.7 & 818.8 \\
			        Na\,I & 819.5 & 819.0 & 819.5 \\			                           Cs\,I & 852.0 & 851.0 & 853.4 \\

			\hline		
			
		\end{tabular}
	\end{center}
	
\end{table}

%------
% OPT
%-----

	\begin{table*}
		\scriptsize
		\caption{Equivalent widths in nm for alkali lines, and H-$\alpha$ emission measured in the optical. Negative values indicate lines in emission.}  
		\label{ew_all_lines_op}
		\centering
		\begin{center}
			\begin{tabular}{lllllllll}
				\hline
				\hline 
 
				Name & NIR SpT &H$\alpha$ [1] & He\,I [2]  & Li\,I [2]& Rb\,I [4] & Na\,I [5] & Na\,I [6] & Cs\,I [7]\\		
				\hline 
				 UGCS J083748.00$+$201448.5 & M7.0 &	$<$0.02			&		NA     &	NA	& <0.01				&		NA			&		NA				&			NA			\\
                2MASS J08370215$+$1952074 & M8.0 &	-1.79$\pm$0.20	&	$<$0.14	&	<0.15	                    &0.05$\pm$0.02	&0.14$\pm$0.01	& 0.15$\pm$0.01		& 0.06$\pm$0.01		\\
                UGCS J083654.60$+$195415.7& M8.0	&		$<$0.02		&		NA	   &	NA		&<0.02			&		NA			&			NA			&		NA				\\
                2MASS J08410852$+$1954018 & M9.0  &		$<$0.02			&		NA	   &	NA		&NA			&		NA			&		NA				&			NA			\\
                2MASS J08370450$+$2016033& L0.0		&		NA			&		NA	   &	NA		&NA			&		NA			&		NA				&			NA			\\
                UGCS J084510.65$+$214817.0& L1.5	&		NA			&		NA	   &	NA		&NA			&		NA			&		NA				&		NA				\\
                
                2MASS J03484469$+$2437236 & M6.0 & -0.75$\pm$0.05	   & NA	 &	<0.04    	&0.06$\pm$0.10	& 0.15$\pm$0.01	&	0.14$\pm$0.01	&	0.03$\pm$0.01	\\
                2MASS J03491512$+$2436225& M6.0	& -0.97$\pm$0.06     & NA&	<0.06			&0.07$\pm$0.01	&	0.14$\pm$0.01	&	0.14$\pm$0.01 &	0.07$\pm$0.01	\\
                2MASS J03512557$+$2345214 & M7.0 &-1.14$\pm$0.32		   & $<$0.14   &  <0.58		&0.09$\pm$0.03	&	0.15$\pm$0.01	&	0.16$\pm$0.02 &	0.07$\pm$0.02	\\
                2MASS J03443516$+$2513429& M8.5	 & -0.87$\pm$0.41	   & -0.37$\pm$0.27  &	<0.38	    	&0.08$\pm$0.04		& 0.13$\pm$0.01		&	0.12$\pm$0.02	&	0.10$\pm$0.02	\\
                2MASS J03463425$+$2350036 & L1.0	&			NA		&	NA			&	NA	&NA				&		NA			&	NA					&	NA					\\
                2MASS J03461406$+$2321565 & M7.0	&-0.89$\pm$0.19		& -0.09$\pm$0.15 &	<0.14   		&0.09$\pm$0.02	& 0.13$\pm$0.01	&	0.17$\pm$0.01	&	0.07$\pm$0.01		\\
                2MASS J03541027$+$2341402 & L3.0		&	NA			&		NA		&		NA	&NA			&		NA			&		NA				&		NA				\\
                
                2MASS J15591135--2338002  &	M8.5  & -4.93$\pm$0.03	    & -0.17$\pm$0.03 &0.04$\pm$0.01		&0.06$\pm$0.01	&	0.09$\pm$0.01	&	0.08$\pm$0.01	&	0.06$\pm$0.01	\\
                2MASS J16060391--2056443  & M8.5	&-12.15$\pm$0.07	& -0.09$\pm$0.01 &0.02$\pm$0.01		&0.06$\pm$0.01	&	0.07$\pm$0.01	&	0.07$\pm$0.01	&	0.03$\pm$0.01	\\              
                2MASS  J16060629--2335133 &	M9.5  & -2.16$\pm$0.30	     & $<$0.02       &0.35$\pm$0.10 	&0.13$\pm$0.06	&	0.12$\pm$0.01	&	0.13$\pm$0.02	&	0.09$\pm$0.04	\\
                2MASS J11085497--7632410 &	M6.0	&-4.16$\pm$0.01		& -0.024$\pm$0.10 & $<$0.01			& <0.02			&	0.06$\pm$0.01				&		0.07$\pm$0.01			&		0.04$\pm$0.01		\\
                2MASS J11123099--7653342&	M7.0	&-1.32$\pm$0.08		&-0.15$\pm$0.06	&<0.07			&0.04$\pm$0.01				&	0.09$\pm$0.01			&	0.05$\pm$0.02				&	0.06$\pm$0.01				\\
                2MASS J11074656--7615174 &	M8.0	&	-2.26$\pm$0.03	&	$<$0.02&0.05$\pm$0.02  	&0.04$\pm$0.01			&	0.08$\pm$0.01		&		0.06$\pm$0.01			&	0.03$\pm$0.01		\\
                2MASS J11062554--7633418 &	M8.0	&-1.96$\pm$0.01		& NA &0.04$\pm$0.01		&0.03$\pm$0.01				&	0.07$\pm$0.01			&	0.03$\pm$0.01				&	0.04$\pm$0.01		\\

				\hline				
				
			\end{tabular}
		\end{center}
        	\begin{tablenotes}
		\small
		\item Lines wavelengths: [1] H$\alpha$: 656.3 nm - [2] He\,I: 667.7~nm - [3] Li\,I: 670.8 nm  - [4] Rb\,I: 794.8~nm - [5] Na\,I: 818.3~nm - [6] Na\,I: 819.5~nm - [7] Cs\,I: 852.0~nm\\
		
	\end{tablenotes}
	\end{table*}

%---------
% NIR
%--------

	\begin{table*}
		\small
		\caption{Equivalent widths in nm for alkali lines measured  in the near-infrared.}  
		\label{ew_all_lines_nir}
		\centering
		\begin{center}
			\begin{tabular}{llllllll}
				\hline
				\hline 
				
				Name & NIR SpT                      & K\,I~(1169~nm)  & K\,I~(1177~nm) & K\,I~(1243~nm) & K\,I~(1253~nm) & GS $\mathrm{^{a, b}}$& OC/A$\mathrm{^{c}}$\\		
				\hline              
				UGCS J083748.00$$+$$201448.5 & M6.0 & {0.35$\pm$0.02} & 0.42$\pm$0.04   & 0.19$\pm$0.02   & 0.25$\pm$0.01    &  00--0   / FLD-G  &  Praesepe  \\
                2MASS J08370215$+$1952074 & M7.5    & 0.21$\pm$0.02   & 0.48$\pm$0.03   &0.31$\pm$0.01    & 0.30$\pm$0.01    &    11--1    / INT-G  &  Praesepe   \\
                UGCS J083654.60$+$195415.7  & M7.0  & 0.33$\pm$0.01   & 0.52$\pm$0.05   & 0.79$\pm$0.04   & {0.40$\pm$0.02}  &    11--1    / INT-G  &  Praesepe    \\
                2MASS J08410852$+$1954018& M9.0     & {0.72$\pm$0.04} & {0.69$\pm$0.05} & {0.63$\pm$0.01} &	0.39$\pm$0.02    &  01--1 / FLD-G  &  Praesepe      \\
                2MASS J08370450$+$2016033& L0.0     & 0.54$\pm$0.01   & 0.63$\pm$0.04   & 0.56$\pm$0.04   & 0.53$\pm$0.02    &   11--1     / INT-G  &  Praesepe   \\
                UGCS J084510.65$+$214817.0  & L1.5  & 0.91$\pm$0.01   &	0.75$\pm$0.05   & 0.47$\pm$0.04   & 0.62$\pm$0.03    &   01--1    / FLD-G  &  Praesepe    \\
                
                2MASS J03484469$+$2437236 & M6.0	& 0.18$\pm$0.01   & 0.17$\pm$0.03   & 0.07$\pm$0.02   & 0.18$\pm$0.01    &   02--1  / INT-G  &  Pleiades   \\
                2MASS J03491512$+$2436225& M6.0	    & 0.18$\pm$0.01   & 0.33$\pm$0.02   & 0.11$\pm$0.02   & 0.20$\pm$0.01    &  00--1   / FLD-G  &  Pleiades   \\
                2MASS J03512557$+$2345214 & M7.0    & 0.30$\pm$0.01   & 0.50$\pm$0.03   & 0.11$\pm$0.04   & 0.39$\pm$0.02    &   00--0  / FLD-G  &  Pleiades  \\
                2MASS J03443516$+$2513429& M7.0     & 0.39$\pm$0.01   & 0.54$\pm$0.04   & 0.50$\pm$0.02   & 0.39$\pm$0.01    &  00--0   / FLD-G  &  Pleiades   \\
                2MASS J03463425$+$2350036 & L1.0    & {0.63$\pm$0.01} & 0.69$\pm$0.04   & {0.58$\pm$0.04} & {0.52$\pm$0.03}  &   11--1  / INT-G &  Pleiades    \\
                2MASS J03461406$+$2321565 & M7.0    & 0.26$\pm$0.01   & 0.30$\pm$0.02   & 0.13$\pm$0.02   & 0.24$\pm$0.01    &   12--1  / INT-G   &  Pleiades   \\
                2MASS J03541027$+$2341402 & L3.0    & 0.80$\pm$0.08   & 0.75$\pm$0.02   & 0.56$\pm$0.06   & {0.30$\pm$0.04}  &    11--2  / INT-G  &  Pleiades   \\
                
                2MASS  J15591135--2338002& M8.5     & 0.15$\pm$0.02   & 0.23$\pm$0.02   & $<$0.02         & 0.10$\pm$0.01    &   22--2  / VL-G  &  UpSco    \\
				2MASS J16060391--2056443 & M8.5	    &0.12$\pm$0.01	  &	0.22$\pm$0.01   & NA              & 0.06$\pm$0.01    &   22-- --/ VL-G  &  UpSco    \\  
                2MASS J16060629--2335133 &	M9.5    & 0.24$\pm$0.01	  & 0.27$\pm$0.04   & 0.10$\pm$0.03   & 0.14$\pm$0.02    &  22--2   / VL-G   &  UpSco   \\

				2MASS J11085497--7632410 & M8.0     &	0.03$\pm$0.01 & 0.17$\pm$0.01	& NA	          & $<$0.03	         &     22-- --   / VL-G  & Cha\,I  	\\
                2MASS J11123099--7653342&	M7.0	&	0.09$\pm$0.01 &	0.02$\pm$0.04   &<0.03	          &	$<$0.03          &    22--2 / VL-G   & Cha\,I   	\\
                2MASS J11074656-7615174  &	M8.0    &	0.04$\pm$0.01 &	0.18$\pm$0.01	&		NA	      &		NA	     	 & 22-- -- / VL-G   & Cha\,I  	      \\
                2MASS J11062554-7633418 &	M8.0    & 0.05$\pm$0.01	  & 0.17$\pm$0.02	&0.04$\pm$0.01    & <35.74	         &   222 --   / VL-G &  Cha\,I     \\
				
				\hline
				
			\end{tabular}
		\end{center}
				\begin{tablenotes}
		\small
		\item a: GS: Gravity scores calculated as in \cite{Allers2013}. b: Gravity scores are ordered according to the alkali line that they correspond to. The dash symbol indicates that none gravity score can be determined with that particular line. c: OC/A: Open Cluster/Association to which the target belongs to.\\
		
	\end{tablenotes}		
	\end{table*}

\begin{figure}
\centering
\includegraphics[width=0.5\textwidth]{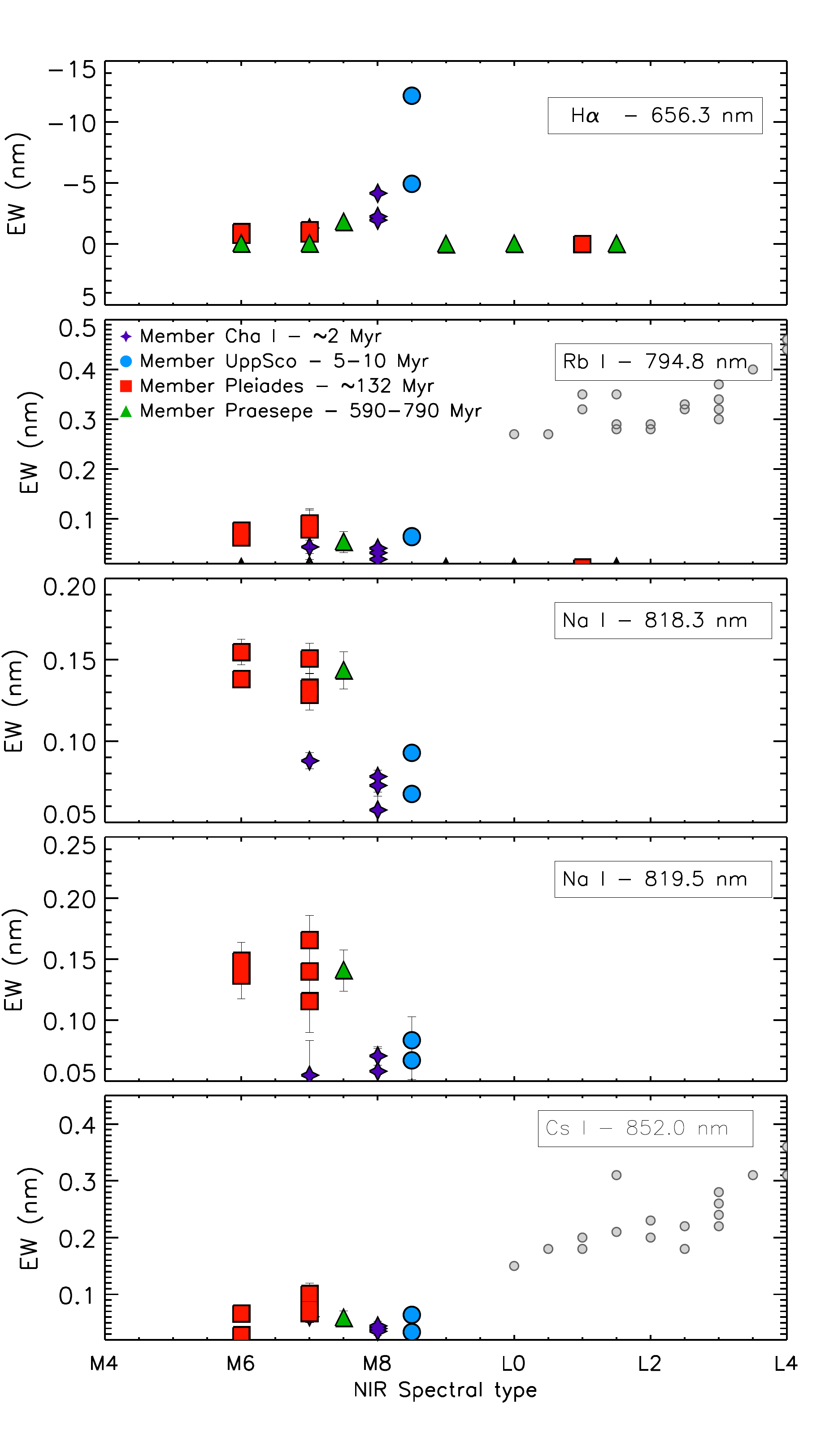}
\caption{\label{pEWs_opt}  Equivalent widths of the detected alkali lines, and the H-$\alpha$ line in the optical for our benchmark objects with different ages: purple stars, blue circles, red squares, and green triangles. Grey circles represent other L-dwarfs from \citet{Chiu2006}, \citet{Golimowski_2004}, and \citet{Knapp_2004}.}
\end{figure}

\begin{figure*}
\centering
\includegraphics[width=0.9\textwidth]{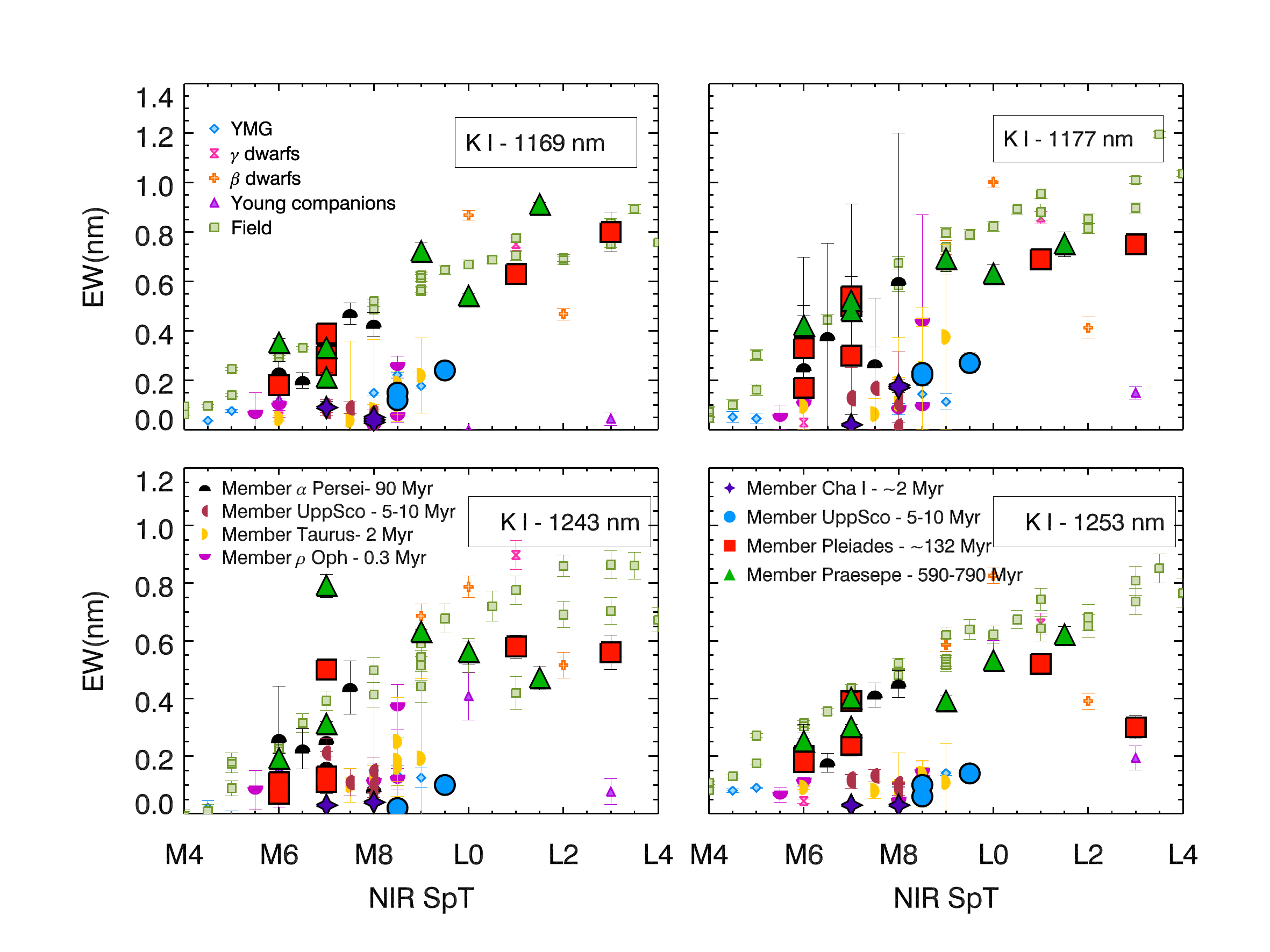}
\caption{\label{pEWs_nir0} Equivalent widths of the detected alkali lines in the  near-infrared for our benchmark objects with different ages: purple stars, blue circles, red squares, and green triangles. We overplot field objects \citep{McLean, Cushing}, objects that belong to TW Hydrae Association (TWA), young companions \citep{Allers2013,Bonnefoy2014a}, young $\beta$-dwarfs and $\gamma$-dwarfs as a comparison \citep{Allers2013}. We include the pEWs published by \citet{Martin2017} of confirmed members of $\alpha$ Persei (90 Myr, black half upper circle), of UppSco (5--10 Myr, red half left circles), of Taurus (2 Myr, yellow half right circles), and from $\rho$ Ophiuchi (0.3 Myr, pink half down circles).}
\end{figure*}

In Figures \ref{pEWs_opt} and \ref{pEWs_nir0} we present the pEWs of the alkali lines in the optical and in the near-infrared for objects in our sample with ages from 2~Myr to 590--790~Myr expanding through low-gravity, intermediate gravity and field gravity classes defined in \cite{Allers2013}. In Tables \ref{ew_all_lines_op}, and \ref{ew_all_lines_nir}, we present the pEWs measured values for the alkali lines. In the near-infrared, we overplotted field objects \citep{McLean, Cushing}, objects that belong to TW Hydrae Association (TWA), young companions \citep{Allers2013,Bonnefoy2014a}, young $\beta$-dwarfs and $\gamma$-dwarfs as a comparison \citep{Allers2013}. In addition, we include high probability members to open clusters from \cite{Martin2017}, those involve five objects from $\alpha$ Persei (90$\pm$10~Myr; \citealt{Stauffer1999, Barrado2004}), seven objects from Rho Ophiuchi ($\mathrm{0.3^{+2.7}_{-0.2}}$ Myr; \citealt{Wilking2005}), two of TWA (10$\pm$3~Myr; \citealt{Mamajek2005}), 13 objects from Taurus (1.5$\pm$0.5~Myr; \citealt{Briceno2002}), and 12 from the UppSco association (5$-$10~Myr).

{In the optical we found H$\alpha$ emission at 656.3~nm for at least 12 of the 20 objects, most of them with ages up to 132$\pm$27~Myr, indicating chromospheric activity \citep{Stauffer1986}. In addition, {we detect lithium in absorption at 670.8~nm  in the four Cha\,I members and the UppSco objects}, indicating that no fusion processes have taken place in the interiors at the age of the association. We do not detect Li in any of the Pleiades members,  probably due to the low resolution of the OSIRIS/GTC spectra.}

{As expected} we found that the alkali lines present in the optical spectra (Rb\,I at 794.8~nm, Na\,I at 818.3~nm, Na\,I at 819.5~nm and Cs\,I at 852.0~nm, see Figure~\ref{pEWs_opt}) are weaker for younger objects (see blue points and violet stars belonging to targets from UppSco and Cha I, respectively). {Members of the Praesepe open cluster and the Pleiades have nearly similar pEWs for the alkali lines, even though the Pleiades (132$\pm$27~Myr) is much younger than the Praesepe (590$-$790~Myr)}. That would suggest that gravity does not change much after the age of the Pleiades, and therefore, we can consider that {low mass stars and more massive brown dwarfs} {have significantly approached} their final radii after their first $\sim$100~{Myr}. {This tendency agrees with predictions of evolutionary models \citep{Baraffe2015} that estimate a decrease in radii for a dwarf of 60~$\mathrm{M_{Jup}}$ by a factor of 4.4 between 1 and 100~Myr. Between 100~Myr and 3~Gyr, the predicted decrease radii is a factor of 1.8, which translates to a change in surface gravity from 3.8 dex for the very young objects to 4.9 dex for objects of around 100~Myr and 5.3 dex for objects older than a few hundred Myr.}

In the near-infrared (see Figure~\ref{pEWs_nir0}) we observe that the alkali lines are  weaker for younger objects in general. We observe  that objects belonging to $\alpha$~Persei ($\sim$90~Myr), Pleiades (132$\pm$27~Myr) and Praesepe (590$-$790~Myr) have similar pEWs for all for K\,I alkali lines, confirming the tendency found in the optical that suggests that gravity does not increase significantly after $\sim$100~Myr. The M and L dwarfs  belonging to the Pleiades, have slightly weaker pEWs in the near-infrared than field objects. 

In addition, we observe that the increase of the pEWs is steeper for objects with spectral type later than M8, where the increases in pEWs quadruple from the youngest to the oldest objects (see Section \ref{age_EW}). Therefore, only for objects later than M8, the pEWs of the alkali lines on the $J$-band might be useful age indicators. An extra challenge when we aim to use alkali lines in the $J$-band as age tracers is the existence of remaining {features} from the telluric correction, and the blending of these lines with some other spectroscopic features, like the FeH, Fe\,I and $\mathrm{H_{2}O}$ that might introduce extra noise in the measurements of those lines. We find that the best line as gravity/age tracer in the near-infrared is the K\,I line at 1253~nm, in agreement with \cite{Martin2017} and \cite{Lodieu2018}, {for which the pEWs have uncertainties between 5--10\%}.

Finally, we obtained the gravity scores corresponding to each object as described by \cite{Allers2013}. For each  near-infrared spectral type, they defined intervals of values of pEWs of the K\,I lines in the $J$-band that correspond to very low-gravity (VL-G), intermediate gravity (INT-G) and field gravity (FLD-G) objects (see Table 10 of \citealt{Allers2013}). A gravity score of "0" is the index of an object classified as an old field dwarf. A score of "1" indicates intermediate surface gravity. Finally, a score index of "2" means that the pEW or index indicates low surface gravity. The values of the pEW or indices that define each category can be found in Tables 9 and 10 from \cite{Allers2013}. The final gravity score is calculated as a median of all the scores. The field gravity objects have a final score of $\leqslant$0.5. The intermediate gravity objects have a median gravity score of 1. Finally, objects with a median gravity of $\geqslant$1.5 belong to the very low gravity object category. In Table \ref{ew_all_lines_nir}, we show the gravity scores obtained for our sample.

The gravity scores classified correctly  very low gravity objects (belonging to UppSco and Cha\,I, respectively), but  Praesepe and Pleiades objects have mixed field and intermediate surface gravity classifications (see Table \ref{ew_all_lines_nir}). {Thus, these gravity scores  do not always predict the surface gravity expected for the members to each of the clusters/associations that we consider in this study}.

\subsubsection{Correlation between ages and alkali lines in the near-infrared}\label{age_EW}

\begin{figure}
\centering
\includegraphics[width=0.49\textwidth]{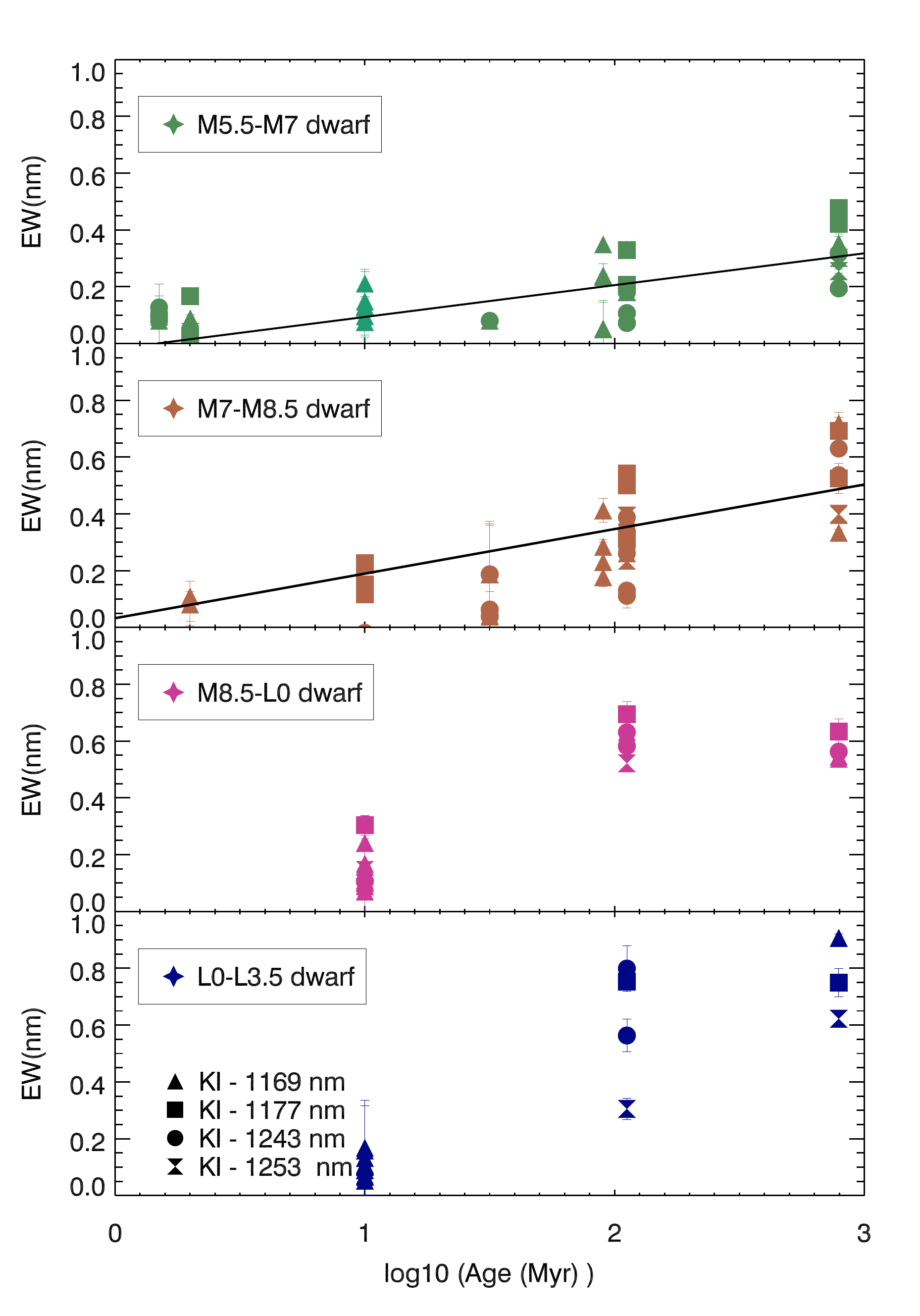}
\caption{\label{age_pew_fig}  Age versus pEWs of the K\,I alkali lines in the $J$-band (line at 1169 nm as triangles, line at 1177 nm as squares, line at 1243 nm  as circles, and line at 1253 nm as hourglass) of objects in our sample and members of $\alpha$~Persei, of UppSco, of Taurus, and from $\rho$~Ophiuchi from \citealt{Martin2017}.}
\end{figure}

In Figure \ref{age_pew_fig}, we plot the age of objects in our sample in logarithmic scale vs the pEWs of the K\,I lines in the $J$-band. We add objects with spectral types from M5.5 and L3.5 with well-determined ages from \cite{Martin2017}. {As suggested by visual inspection in Figures \ref{pEWs_opt} and \ref{pEWs_nir0}, the pEWs of the alkali lines increase from earlier to later spectral types.} Within each spectral type, their pEWs increase from lower to higher surface gravity. We calculate a Kendall correlation index to probe correlation between the pEWs for the K\,I lines in the $J$-band and the age. We obtained Kendall correlations  around  0.63 with significances very close to 0, indicating that there is a moderate correlation between the pEWs of the alkali lines in the $J$-band and the age.

We fitted a first order polynomial  that tentatively relates log(Age) with pEWs of {all} the K\,I lines in the $J$-band at 1169 nm, 1177 nm, 1243 nm and 1253 nm. We do not investigate fits with higher order polynomials as the dispersion of the pEWs is remarkable and we just intend to derive a tentative relation between age and pEWs of the K\,I lines in the $J$-band.   We use the IDL routine {\tt{poly\_fit.pro}}. This routine performs a least-square fit with optional weighting and returns a vector with the coefficients and its $\chi^2$. We added the pEWs vs age plot with the pEWs measured by \cite{Martin2017}, thus, we have M5.5 to L3.5 dwarfs with ages {from $\mathrm{0.3^{+2.5}_{-0.2}}$~Myr ($\rho$ Oph) to 590$-$790~Myr (Praesepe)}, covering the whole VL-G, INT-G and FLD-G classification. In Table \ref{best_fit_pol} we show the best matching first order polynomial for M5.5-M7.0 and M7.0-M8.5 spectral types. We do not provide polynomials for the rest of the spectral types due to the lack of objects within those ranges of spectral types.

%{As mentioned in Section \ref{pEWs} we observe that the slope of the polynomial fit is steeper for  L0-L3.5 (average pEWs at 10~Myr of 100 nm to 800 nm at 790~Myr) dwarfs than for the M5.5-M7 dwarfs (average of 100 nm at 10~Myr to average of 400 nm at 790~Myr).}

\begin{table*}
\begin{center}
	\caption{Tentative linear relation between pEW of {all the alkali lines} and  log10(Age) (Myr) plots as shown in Figure \ref{age_pew_fig}.}  
	\label{best_fit_pol}
	\renewcommand{\footnoterule}{}  % to avoid a line before footnotes
		\begin{tabular}{llllll}
			\hline 
			
			SpT range   & \multicolumn{2}{l}{Polynomial}  & $\chi^{2}_{red}$  & Kendall-$\tau$ coff. & significance\\		
				        & $c_{0}$ & $c_{1}$    \\
			\hline              		
			        M5.5--M7.0 & 0.1383$\pm$0.0177 & 8.9477$\pm$0.1321 & 339.4 & 0.62 & 1.94e-5\\
			        M7.0--M8.5 &  -0.1809$\pm$0.0046 & 8.7026$\pm$0.0462 & 987.5 & 0.63 & 0.0 \\
			        %M8.5--L0.0 &  0.13575$\pm$0.00205 & 4.31179$\pm$0.02269 & 944.2 &  0.63 & 0.0 \\
                   % L0.0--L3.5 & 4.75731$\pm$0.14439 & -1.62653$\pm$0.02269 & 98.3 &  -0.60 & 4.04$e^{-3}$\\  
			\hline		
			
		\end{tabular}
	\end{center}
		\begin{tablenotes}
		\small
		\item The polynomials are defined as: log10(Age) = $c_{0}$ + $c_{1}$ $\times$ pEW\\
		
	\end{tablenotes}
\end{table*}

\subsubsection{Spectral Indices and Gravity Scores}\label{spectral_indices}

We calculated the $\mathrm{FeH_{J}}$, $\mathrm{KI_{J}}$, $\mathrm{FeH_{z}}$, $\mathrm{VO_{z}}$ and $H$-cont spectral indices to estimate surface gravity as defined in \cite{Allers2013}. The $\mathrm{FeH_{J}}$ index at 1200~nm has been found to be correlated with surface gravity \citep{McGovern}. The $\mathrm{KI_{J}}$  measures the depth of the K\,I doublet at 1250~nm. {The VO band dissapears as surface gravity increments with age due to condensation effects \citep{Lodders2002}}. The shape of the continuum of the $H$-band has been found to be affected by surface gravity \citep{Borysow1997, Bowler2012}. For young objects, the shape of the $H$-band is usually triangular-shaped, thus the $H$-cont index measures how much the blue-end of the $H$-band deviates from a straight line. The $H$-cont is small for high gravity objects, and close to 1 for young objects.

{Using the method presented in \cite{Allers2013} we determined the spectral indices to determine the gravity scores from the $\mathrm{FeH_{J}}$, $\mathrm{KI_{J}}$, $\mathrm{FeH_{z}}$, $\mathrm{VO_{z}}$ and $H$-cont indices}. 
Nonetheless, it is important to note that \cite{Lodieu2017} identified   the $H$-cont index  as the most sensitive to gravity (or age) after testing several indices in a sample of field L dwarfs, young $\beta$ and $\gamma$ L dwarfs and L dwarfs members to the UppSco association. We show the gravity scores obtained for the spectral indices for our objects  in Table~\ref{gscores}.

%We calculated the gravity scores for the spectral indices as presented in \cite{Allers2013} and as explained in Section \ref{pEWs}. We obtained that the spectral indices  were able to find very low gravity or young objects (members of Chamaeleon\,I or UppSco), but sometimes failed to find higher gravity objects, specially of intermediate age (Pleiades members), for which the gravity classification varies from VLG-G to FLD-G for different members. 

In Table \ref{final_grav} we {summarise} the final gravity scores given by the alkali lines and the gravity  indices from \cite{Allers2013} for each object. Combining all gravity indicators, we obtained in general a consistent gravity classification with the age of the cluster or association. {Nonetheless, we note that that three of the six Praesepe members would be classified as INT-G instead of FLD-G with the spectral indices and gravity scores. Similarly, one member of the Pleiades obtained a VL-G classification, and two of them obtained a FLD-G classification, which would not correspond to the expected gravity classification for the Pleiades.}

% Gravity scores

	\begin{table*}
		\small
		\caption{Gravity scores for our sample derived from spectral indices defined in the literature \citep{Allers&Liu}.}  
		\label{gscores}
		\centering
		\begin{center}
			\begin{tabular}{lllllllll}
				\hline
				\hline 
				
				Name &  SpT          &               $\mathrm{FeH_{J}}$ & $\mathrm{KI_{J}}$ & $H$-cont          &  $\mathrm{FeH_{z}}$  &  $\mathrm{VO_{z}}$ & G$\mathrm{S^{a}}$   & Member\\		
				\hline              
				UGCS J083748.00$$+$$201448.5 & M6.0 & 1.042$\pm$0.001	&   1.051$\pm$0.001 &	0.956$\pm$0.002 &	0.873$\pm$0.002 & 1.038$\pm$0.002       & 1002--/ FLD-G &  Praesepe \\
                2MASS J08370215$+$1952074    & M7.5 & 1.038$\pm$0.001	&   1.067$\pm$0.001 &	0.908$\pm$0.005 &	1.031$\pm$0.001 & 1.034$\pm$0.002		& 2002-- / INT-G &  Praesepe  \\
                UGCS J083654.60$+$195415.7   & M7.0 & 1.042$\pm$0.001   &   1.075$\pm$0.002 &	0.943$\pm$0.005 &	0.942$\pm$0.003 & 1.061$\pm$0.003		& 2002-- / INT-G  &  Praesepe \\
                2MASS J08410852$+$1954018    & M9.0 & 1.056$\pm$0.001	&   1.092$\pm$0.002 &	1.055$\pm$0.006 &	0.922$\pm$0.003 & 1.039$\pm$0.003		& 2002-- / INT-G &  Praesepe \\
                2MASS J08370450$+$2016033    & L0.0 & 1.091$\pm$0.002   &	1.111$\pm$0.004	&   0.891$\pm$0.001 &	1.102$\pm$0.004 & 1.152$\pm$0.004		& 20200 / FLD-G &  Praesepe \\
                UGCS J084510.65$+$214817.0   & L1.5 & 1.105$\pm$0.002   &	1.118$\pm$0.003 &	0.892$\pm$0.003 &	0.956$\pm$0.003 & 1.165$\pm$0.004		& 20000 / FLD-G &  Praesepe \\
               										 
                2MASS J03484469$+$2437236    & M6.0	& 1.031$\pm$0.001   &   1.021$\pm$0.001 &   0.993$\pm$0.002 &  1.003$\pm$0.001 & 1.023$\pm$0.002		& 2202-- / VL-G  &  Pleiades  \\
                2MASS J03491512$+$2436225    & M6.0 & 1.043$\pm$0.001   &	1.037$\pm$0.001	&   0.951$\pm$0.001 & 1.042$\pm$0.001 & 1.020$\pm$0.002			& 1101-- / INT-G &  Plei  \\
                2MASS J03512557$+$2345214    & M7.0 & 1.067$\pm$0.002   &	1.071$\pm$0.002	&   0.954$\pm$0.001 &	1.081$\pm$0.001 & 1.101$\pm$0.002		& 1001-- / FLD-G &  Pleiades  \\
                2MASS J03443516$+$2513429    & M7.0 & 1.071$\pm$0.002   &	1.062$\pm$0.002 &	0.935$\pm$0.001 &	1.057$\pm$0.001 & 1.108$\pm$0.003		& 1001-- / FLD-G &  Pleiades  \\
                2MASS J03463425$+$2350036    & L1.0 & 1.093$\pm$0.002   &	1.098$\pm$0.004 &	0.937$\pm$0.003 &	0.899$\pm$0.005 & 1.287$\pm$0.006	    & 21022 / INT-G &  Pleiades  \\
                2MASS J03461406$+$2321565    & M7.0 & 1.058$\pm$0.001   &	1.034$\pm$0.002 &	0.954$\pm$0.001 &   1.029$\pm$0.001 & 1.062$\pm$0.002		& 1202-- / INT-G &  Pleiades  \\
                2MASS J03541027$+$2341402    & L3.0 & 1.067$\pm$0.002   &   1.069$\pm$0.002 &	0.853$\pm$0.004 & 	0.819$\pm$0.002 & 1.341$\pm$0.002		& 22022 / VL-G &  Pleiades  \\
                
                2MASS  J15591135--2338002    & M8.5 & 1.016$\pm$0.001   &	1.037$\pm$0.001	&   0.978$\pm$0.003 &  1.040$\pm$0.002 & 1.067$\pm$0.002		& 2212-- / VL-G &  UpSco   \\
                2MASS J16060391--2056443     & M8.5	& 1.013$\pm$0.001   &	1.022$\pm$0.001 &   0.996$\pm$0.002 &  1.041$\pm$0.001 & 1.081$\pm$0.002		& 2222-- / VL-G &  UpSco  \\
                2MASS  J16060629--2335133    & M9.5  & 1.059$\pm$0.002   &	1.057$\pm$0.002 &   0.955$\pm$0.002 &	1.00$\pm$0.001 & 1.180$\pm$0.003	    & 2112-- / VL-G &  UpSco  \\
%                J16082847-23151032	&					&				&					&					\\
												
				2MASS J11085497--7632410    &   M8.0 &	1.008$\pm$0.001 &	1.006$\pm$0.001 &   0.958$\pm$0.002  &	0.986$\pm$0.001 & 1.031$\pm$0.001	    & 2202-- / VL-G & Cha\,I \\
                2MASS J11123099--7653342    &	M7.0 &	1.017$\pm$0.001	&   1.022$\pm$0.001 &	0.974$\pm$0.002  &	1.005$\pm$0.001 & 1.084$\pm$0.001		& 2202-- / VL-G  & Cha\,I \\
                2MASS J11074656-7615174     &	M8.0 & 1.030$\pm$0.001	&   1.005$\pm$0.001 &	0.977$\pm$0.002  &	1.006$\pm$0.001 & 1.021$\pm$0.001		& 2222-- / VL-G & Cha\,I \\
                2MASS J11062554-7633418     &	M8.0 & 	1.023$\pm$0.001 &	1.008$\pm$0.001 &	0.956$\pm$0.001  &	1.001$\pm$0.001 & 1.010$\pm$0.001	& 2202-- / VL-G & Cha\,I \\
                \hline
			\end{tabular}
		\end{center}		
		\begin{tablenotes}
		\small
		\item a: GS: Gravity scores calculated as in \cite{Allers2013}\\		
	\end{tablenotes}		
	\end{table*}

    	\begin{table*}
		\small
		\caption{Summary of gravity scores provided by the alkali lines and gravity indices.}  
		\label{final_grav}
		\centering
		\begin{center}
			\begin{tabular}{llll}
				\hline
				\hline 
				
				Name & NIR SpT                      &  GS $\mathrm{^{a}}$& OC/A$\mathrm{^{c}}$\\		
				\hline              
				UGCS J083748.00$$+$$201448.5 & M6.0 &  00--01002--  / FLD-G  &  Praesepe  \\
                2MASS J08370215$+$1952074 & M7.5    &    11--12002--   / INT-G  &  Praesepe   \\
                UGCS J083654.60$+$195415.7  & M7.0  &    11--12002--     / INT-G  &  Praesepe    \\
                2MASS J08410852$+$1954018& M9.0     &  010012002-- / FLD-G  &  Praesepe      \\
                2MASS J08370450$+$2016033& L0.0      &   11--120200   / INT-G  &  Praesepe   \\
                UGCS J084510.65$+$214817.0  & L1.5   &   01--120000   / FLD-G  &  Praesepe    \\
                
                2MASS J03484469$+$2437236 & M6.0	  &   02--12202-- / INT-G  &  Pleiades   \\
                2MASS J03491512$+$2436225& M6.0	    &  00--11101--  / INT-G  &  Pleiades   \\
                2MASS J03512557$+$2345214 & M7.0  &   00--001001-- / FLD-G  &  Pleiades  \\
                2MASS J03443516$+$2513429& M7.0     &  00--011001--   / FLD-G  &  Pleiades   \\
                2MASS J03463425$+$2350036 & L1.0    &   11--121022 / INT-G &  Pleiades    \\
                2MASS J03461406$+$2321565 & M7.0     &   12--11202-- / INT-G   &  Pleiades   \\
                2MASS J03541027$+$2341402 & L3.0     &    11--222022  / VL-G  &  Pleiades   \\
                
                2MASS  J15591135--2338002& M8.5    &   22--22212--  / VL-G  &  UpSco    \\
				2MASS J16060391--2056443 & M8.5	     &   22-- --2222-- / VL-G  &  UpSco    \\  
                2MASS J16060629--2335133 &	M9.5     &  22--222112--  / VL-G   &  UpSco   \\

				2MASS J11085497--7632410 & M8.0       &     22-- --2202-- / VL-G  & Cha\,I  	\\
                2MASS J11123099--7653342&	M7.0	&    22--22202-- / VL-G   & Cha\,I   	\\
                2MASS J11074656-7615174  &	M8.0    & 22-- --2222-- / VL-G   & Cha\,I  	      \\
                2MASS J11062554-7633418 &	M8.0    &   222--2202--  / VL-G &  Cha\,I     \\
				
				\hline
				
			\end{tabular}
		\end{center}
				\begin{tablenotes}
		\small
		\item a: GS: Gravity scores calculated as in \cite{Allers2013}. c: OC/A: Open Cluster/Association to which the target belongs to.\\
		
	\end{tablenotes}		
	\end{table*}

% EW NIR    

\begin{figure}
\centering
\includegraphics[width=0.5\textwidth]{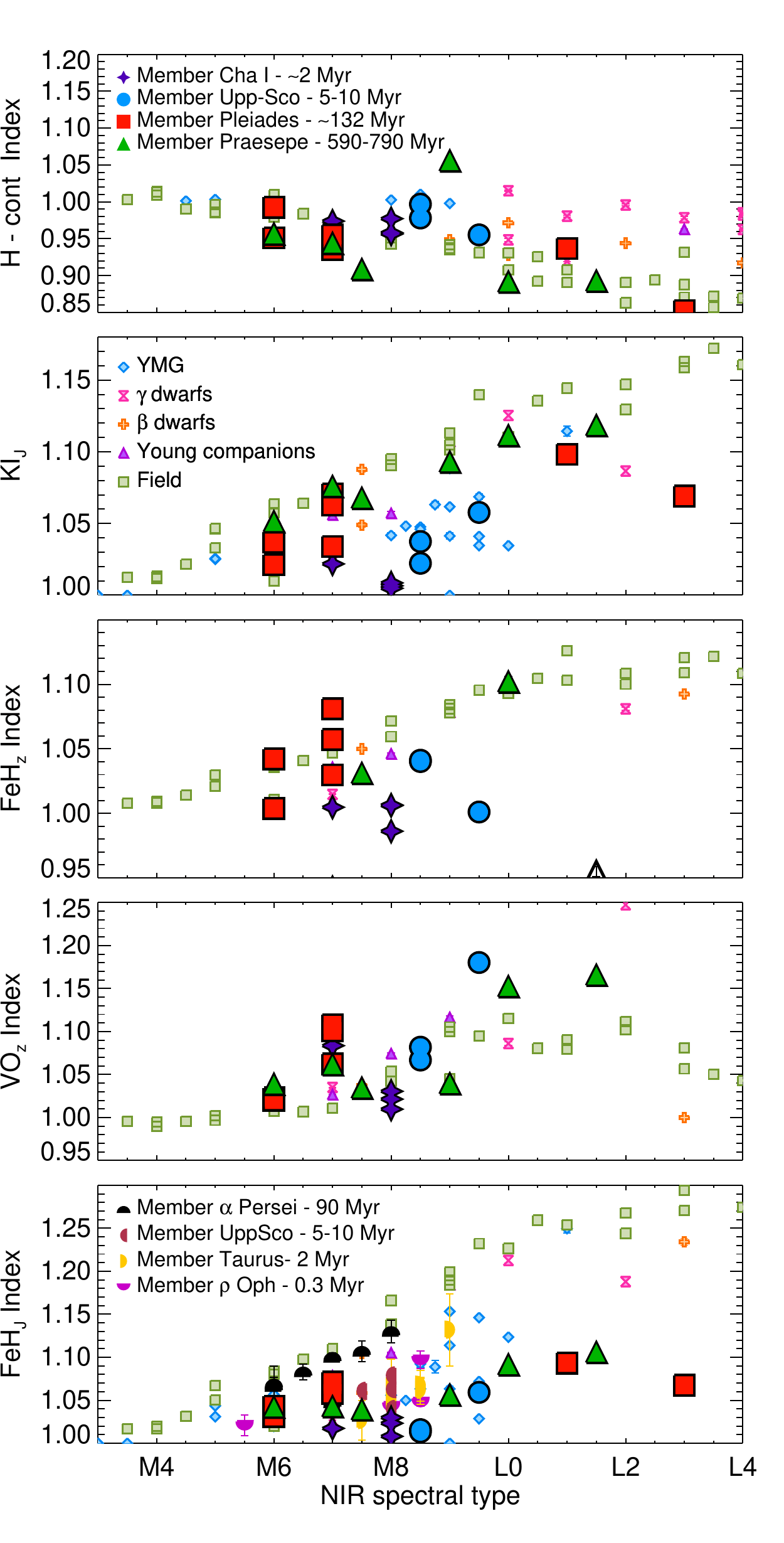}
\caption{\label{indices} $\mathrm{FeH_{J}}$,  $\mathrm{KI_{J}}$ and $H$-continuum indices from \citet{Allers2013} for objects in our sample. We overplot field objects \citep{McLean, Cushing}, members to TW Hydrae Association (TWA), young companions \citep{Allers2007,Bonnefoy2014a}, young $\beta$-dwarfs and $\gamma$-dwarfs as a comparison \citep{Allers2007}. We include the value of the FeH index for  members of Alpha Persei, of UppSco, of Taurus, and from Rho Ophiuchi \citep{Martin2017} }
\end{figure}

% Spectral indices

\subsection{Age sequences}

\begin{figure*}
\centering
\includegraphics[width=0.45\textwidth]{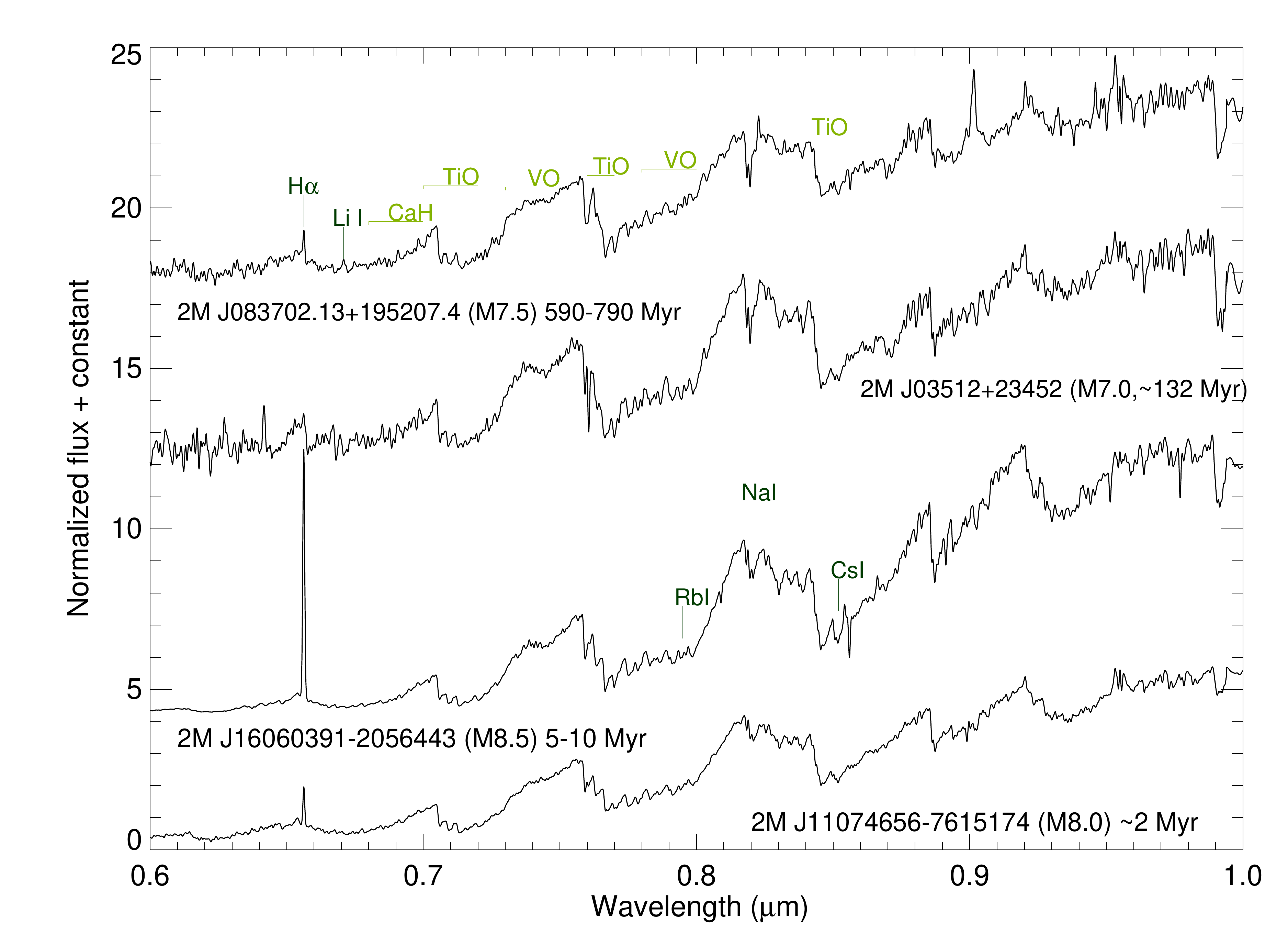}
\includegraphics[width=0.45\textwidth]{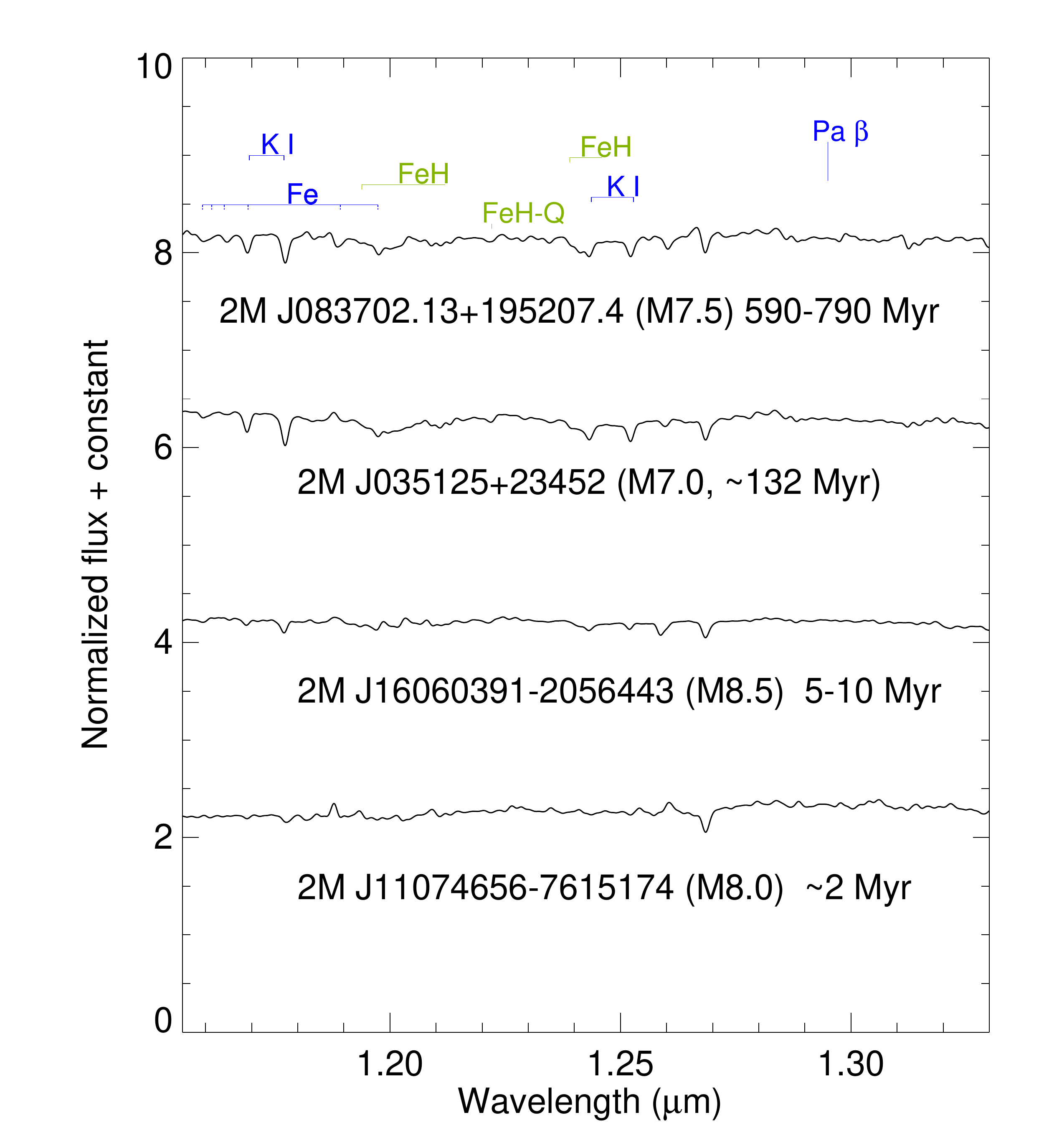}
\includegraphics[width=0.45\textwidth]{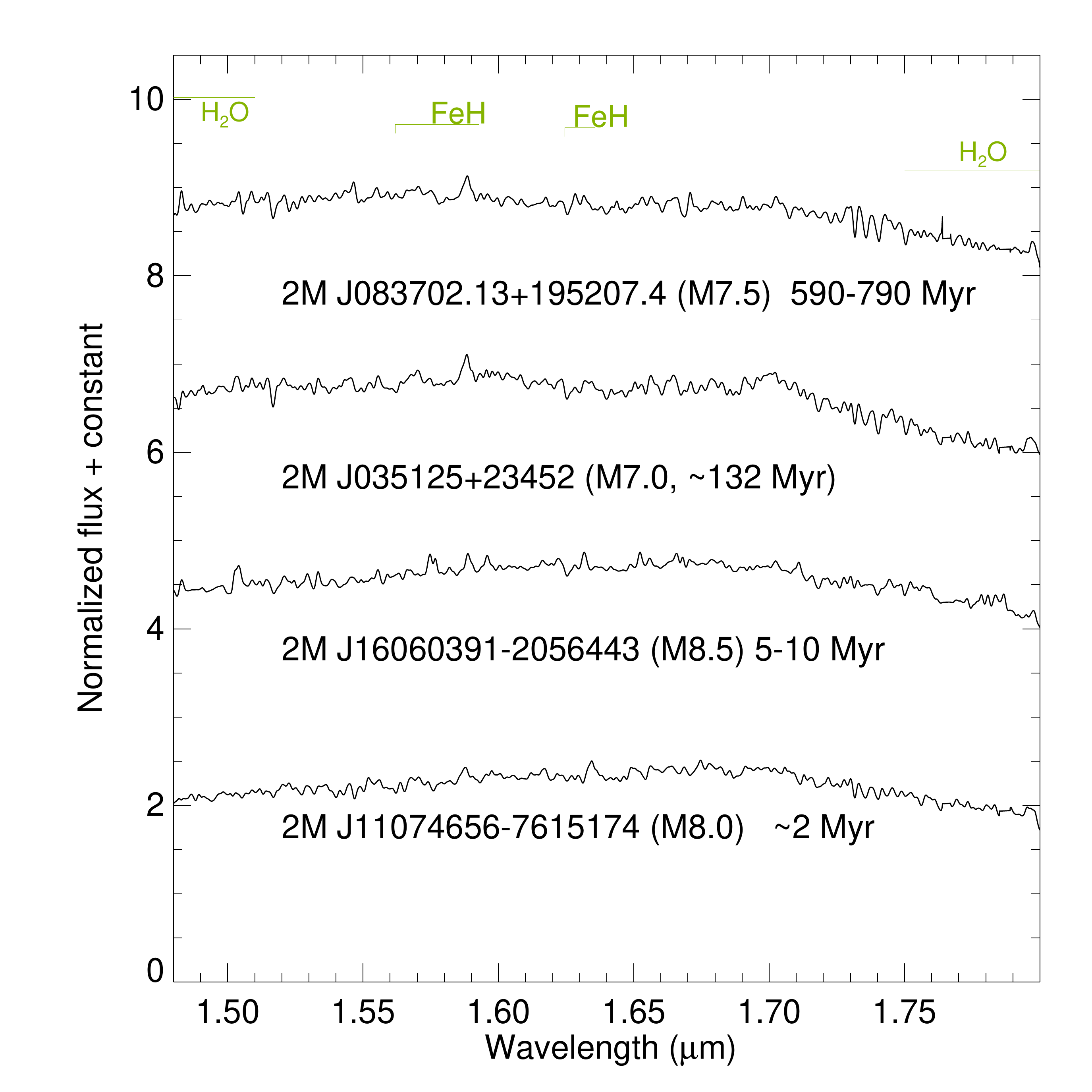}
\includegraphics[width=0.45\textwidth]{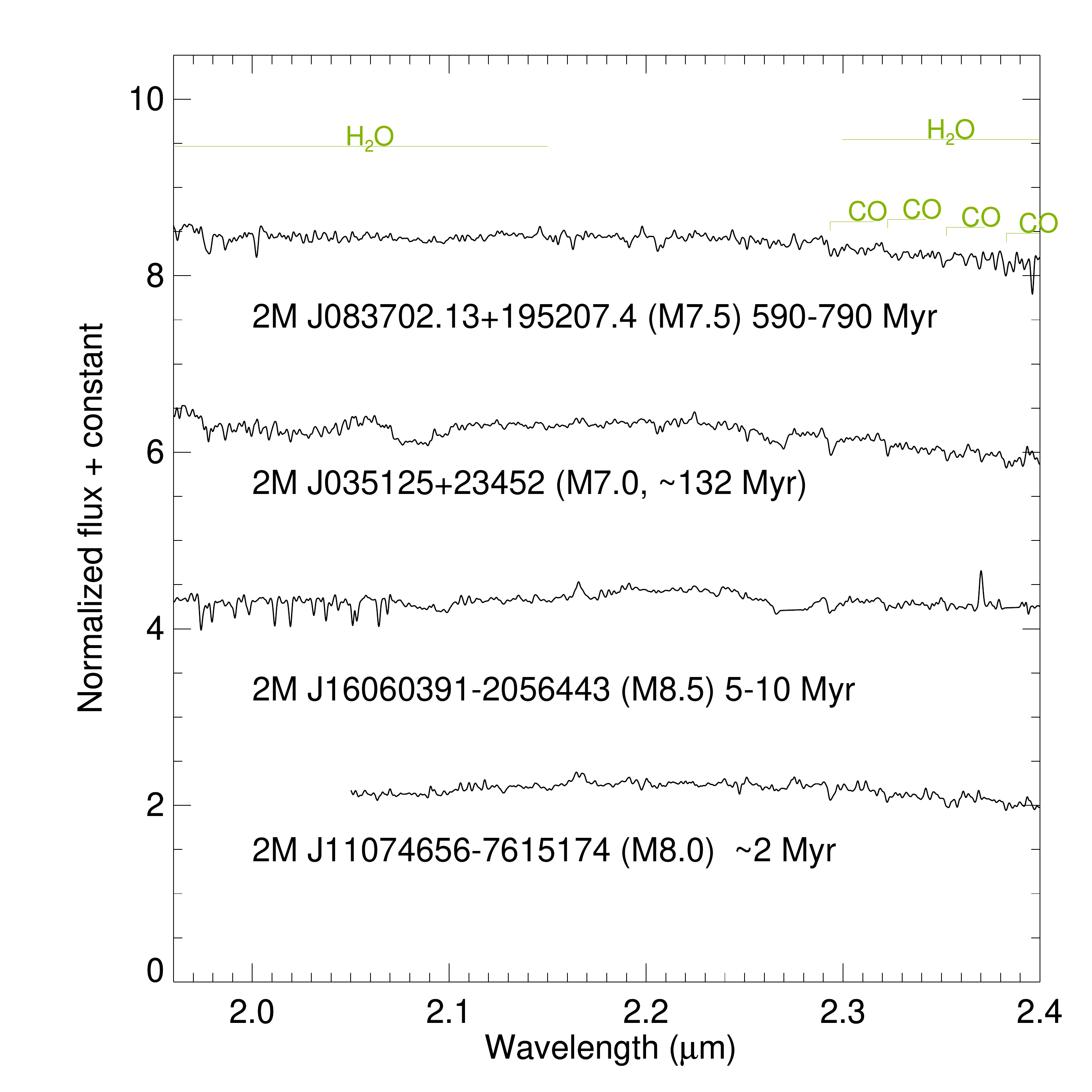}
\caption{\label{age_seqM8_VIS} Age sequence of spectra in the optical {and near-infrared} of M7.0--M8.0 dwarfs with ages from 2 Myr to 790 Myr. The spectral characteristics of of M7.0-M8.0 dwarfs are very similar for very young and intermediate age objects. }
\end{figure*}

We present the age sequences of M7.0--M8.5 brown dwarfs and low mass stars from our X-shooter sample with wavelengths from 0.6 to 2.5 $\mu$m from 2~Myr (corresponding to the VL-G gravity class) to 590$-$790~Myr (corresponding to the FLD gravity class). We degraded the resolution of our spectra to R$\sim$700 to reduce their noise. Their flux is normalised at 1.226~$\mu$m. We applied a positive shift to the spectra to be able to overplot all the age sequence in the same plot. In Figure \ref{age_seqM8_VIS},  we compare the optical and the $J$, $H$ and $K$ bands spectra of 2MASS~J11074656--7615174 (M8.0 Cha\,I member, age$\sim$2~Myr), 2MASS~J16060391--2056443 (M8.5 UppSco member, 5-11~Myr), 2MASS~J035125+23452 (M7, Pleiades  member, age$\sim$137~Myr), and  2MASS~J08372.13+195207.4 (M7.5, member to the Praesepe, 590$-$790~Myr). 

In Figure \ref{age_seqM8_VIS}, we observe the following evolution of spectral characteristics {of M7-M8 dwarfs} from younger to older age: in the optical, we {do not observe a clear tendency for the TiO (0.71, 0.77 and 0.79~$\mu$m), and VO (0.74, 0.76 and 0.84~$\mu$m) molecular bands with age}, as well as the alkali lines (Rb\,I at 0.794~$\mu4$m, Na\,I at 0.818 and 0.819~$\mu$m, and Cs\,I at 0.852~$\mu$m). {\cite{Allers2013} claimed that VO varies slightly with surface gravity for M-dwarfs at 1.06~$\mu$m, but much less than for L-dwarfs}. In the near-infrared, we observe that in the $J$-band the K\,I alkali lines doublets (1.169 and 1.177~$\mu$m, 1.243 and 1.253~$\mu$m) increase their pEW with age (i.e. increase {of} surface gravity). In the $H$-band, the FeH (1.59~$\mu$m) absorption appears deeper with age. %In the $K$-band, the most notorious spectral characteristics that appear with age are the CO molecular bands at 2.29, 2.32, 2.34, 2.35, 2.38, 2.41~$\mu$m.

%We observe that the overall shape of the spectra from 2 Myr to 112 Myr is similar: we only observe deep alkali lines in  2MASS~J08370215$+$1952074, as well as hints of the FeH band in the $H$-band. In the optical, the CaH (0.69~$\mu$m), TiO (0.71, 0.77 and 0.79~$\mu$m) and VO (0.74, 0.76 and 0.84~$\mu$m) molecular absorptions appear significantly stronger for the Cha I object and they disappear progressively as the age increments.   

\subsection{Spectral type-colour diagram}\label{SpT_colours}

\begin{figure*}
\includegraphics[width=0.95\textwidth]{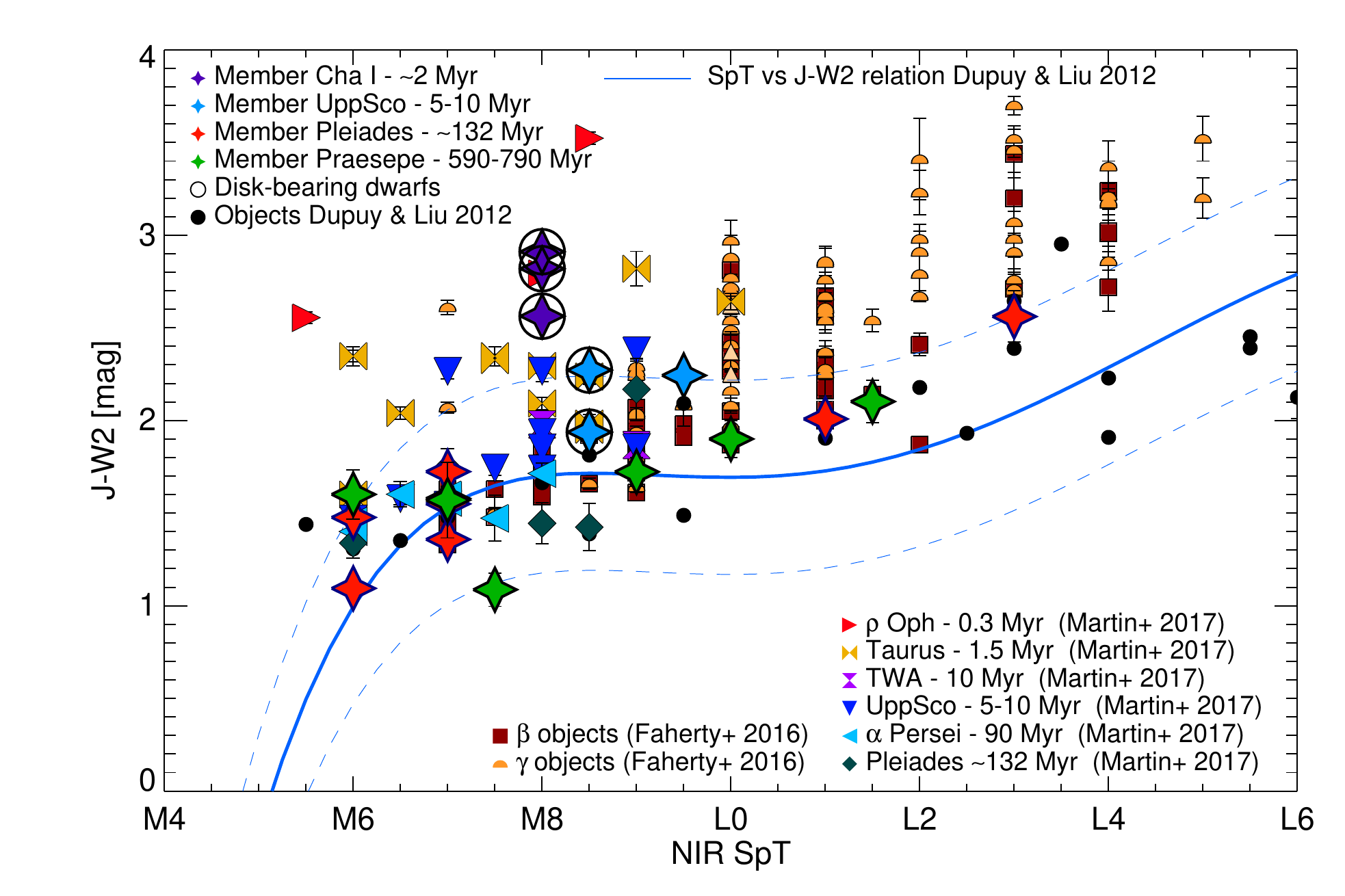}
\caption{\label{SpT_J_W2_m} Spectral type vs \textit{J--K} colour for  our sample, dwarfs from \citet{Dupuy_Liu2012}, objects with well-determined ages from \citet{Martin2017}, and objects members of young moving groups from \citet{Faherty2016}. We overplot circles over the targets that bear disks. We overplot with a blue line the spectral-photometric relationship from \citet{Dupuy_Liu2012} for field dwarfs with its typical rms  marked with a dashed blue line. We observe that only dwarfs with ages smaller than 10 Myr show \textit{J--K} red colour, lying above the upper rms line.}
\end{figure*}

We plot the near-infrared spectral types of our targets, objects with well-constrained ages from \cite{Martin2017}, and {members to young moving groups from \cite{Faherty2016} versus their $J-W2$ colour in Figure \ref{SpT_J_W2_m}. \cite{Faherty2016} targets are classified with gravity classes  corresponding approximately to intermediate gravity ($\beta$) and low gravity objects ($\gamma$ and $\delta$), respectively}. We overplot field dwarfs from \cite{Dupuy_Liu2012} as a comparison. In addition, {we overplot as a blue solid line the spectrophotometric relation of \cite{Dupuy_Liu2012} valid for field dwarfs with its associated rms (dashed lines)}. 

 We observe that Cha\,I,  $\rho$ Ophiuchi, Taurus and some UppSco members, all below $\sim$10~Myr, lie above the mean  $J-W2$ colours for field M and L dwarfs. We observe that objects from \cite{Faherty2016} with spectral types earlier than L0 do not show in general extremely red $J-W2$ colours. {In addition}, most of their objects with later spectral types show red colours independently of their gravity classification. This fact suggests that, whatever is the cause for red colours in members of young moving groups, it is {likely} not {a direct indication} of  the age of the source. In addition, it is important to note that the ages of the sample in \cite{Faherty2016} are estimated using the BANYAN tool \citep{Malo2013, Gagne2014} that provides the probability of one object to belong to a young moving group, and thus it has a determined age. The results given by the BANYAN tool should be interpreted with caution, and the resulting ages from that tool should be further supported by other indications of young age\footnote{Further details can be found in: \url{http://www.astro.umontreal.ca/~gagne/banyanII.php}}.

%This fact suggests that extremely red colours in brown dwarfs of our sample with M5.5 to L3.5 spectral types might be due to the existence of mid-infrared excesses due probably to the existence of disks that last typically for around 10 Myr \citep{Carpenter2006}.

\begin{figure}
\centering
\includegraphics[width=0.5\textwidth]{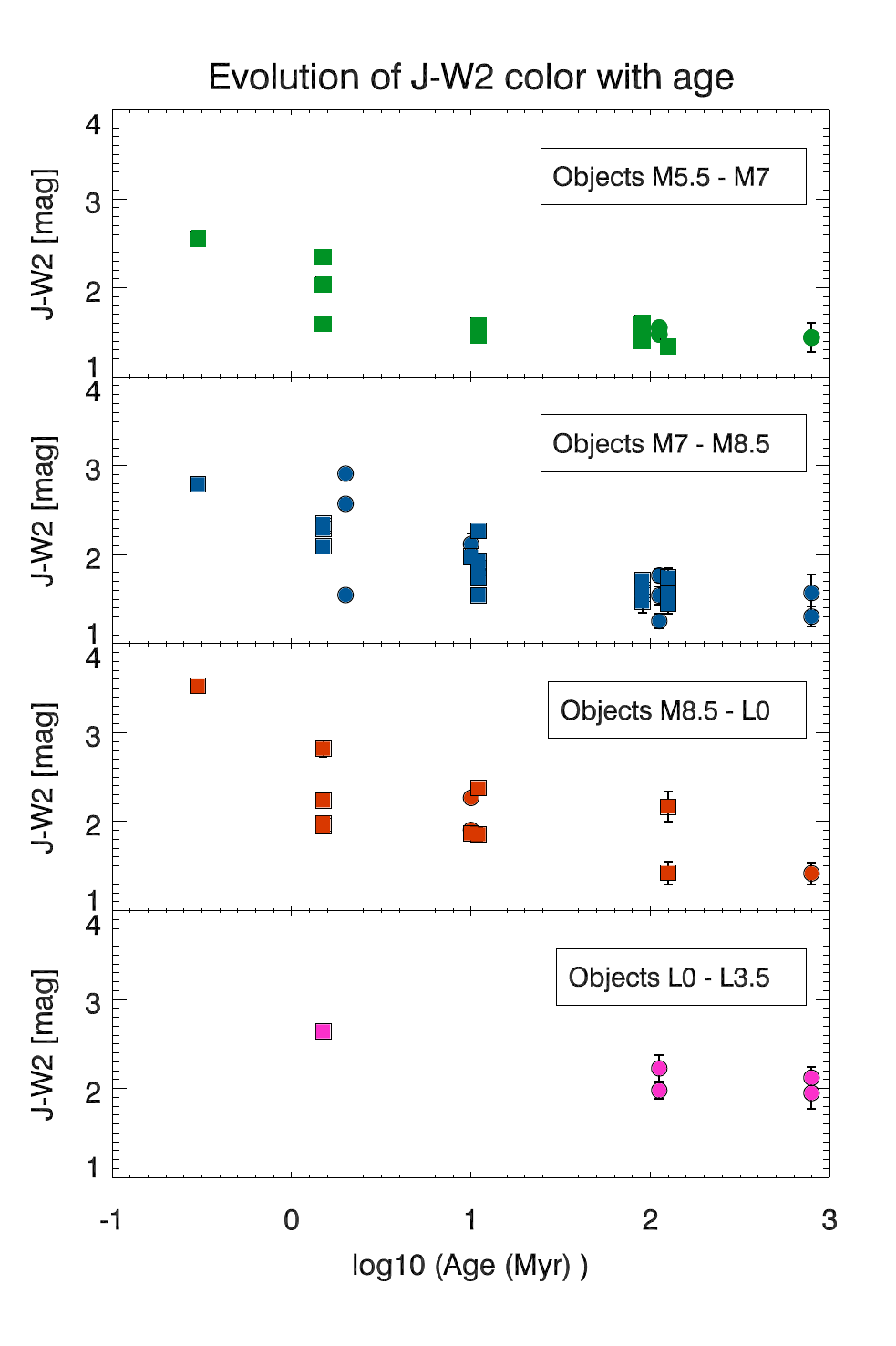}
\caption{\label{colour_age} Log(Age) versus $J-W2$ colour for  our sample (circles), and objects with well-determined ages from \citet{Martin2017}, i.e. members of Alpha Persei, of UppSco, of Taurus, and from Rho Ophiuchi, separated by ranges of spectral types (squares). We observe that after the 10 Myr the $J-W2$ colour does not change significantly from the $J-W2$ colour of a field object.}
\end{figure}

In Figure \ref{colour_age}, we plot ages from 0.3 to 590$-$790 Myr in logarithmic scale, versus $J-W2$ colour for four ranges of spectral types: M5.5--M7, M7--M8.5, M8.5--L0 and L0--L3.5. We observe that in general colours of brown dwarfs and low-mass stars are {bluer} when increasing age. Furthermore, in average, after 10 Myr, the average $J-W2$ colour is similar to the $J-W2$ for objects that are 590$-$790 Myr old for objects with a similar spectral type.  In Table \ref{colour_age_table}, we summarise the Kendall $\tau$ coefficient that show the moderate anticorrelation between $J-W2$ red colour and age, with $\tau$ values between $-$0.72 for M5.5--M7.0 spectral types, to $-$0.54 for L0.0--L3.5 spectral types. The fact that the red colours do not evolve significantly after 10~Myr suggests that the reddening we observe {might be} due to circumstellar disks, that can survive up to 10~Myr \citep{Carpenter2006}, {at least for objects with the spectral types we consider in this work {(M6.0-L3.0)}}. 
{Other alternative explanations for the red colours of some brown dwarfs include the existence of ring structures around these objects \citep{Zakhozhay2017}, {extinction through the star forming region}, a viewing angle with inclinations $>$20$^\circ$  \citep{Vos2017}, or the presence of sub-micron particle grains in the atmosphere of L dwarfs, {which are probably not included in brown dwarf cloud models with the correct number of particles and opacities} \citep{2014MNRAS.439..372M, Hiranaka2016, Bonnefoy2016}.}
 
%This fact, further supports the theory that red colours in objects younger than 10 Myrs might be caused by the existence of disks, at least for objects with spectral types between M5.5 and L3.5.

\begin{table}
\begin{center}
	\caption{Kendall coefficients indicating the correlation between  to $J-W2$ and log10(Age (Myr))  plots as shown in Figure \ref{colour_age}.}  
	\label{colour_age_table}
	\renewcommand{\footnoterule}{}  % to avoid a line before footnotes
		\begin{tabular}{lcc}
			\hline 
			
			SpT range   &  Kendall-$\tau$ coefficients & significance\\		
			\hline              		
			        M5.5--M7.0 & $-$0.72 & 1.8e-4\\
			        M7.0--M8.5 &   $-$0.69 & 1.8e-7 \\
			        M8.5--L0.0 &   $-$0.54 & 1.5e-2\\
                    L0.0--L3.5 &   $-$0.54 & 1.5e-2\\  
			\hline		
			
		\end{tabular}

	\end{center}
		
\end{table}

\section{Physical parameters of our sample}\label{physical_param}

Due to the age-mass degeneration for brown dwarfs, it is challenging  to estimate the ages, masses, radii, and  gravities of  brown dwarfs. Nevertheless, if we know the  ages, we are able to break the age-mass degeneracy and provide a complete physical characterisation of our sample using evolutionary models for low-mass stars and brown dwarfs. This is the aim of this Section.

\subsection{Bolometric luminosity and bolometric correction}\label{luminosity}

	We have optical and near-infrared spectroscopy from X-shooter, or flux-calibrated optical spectroscopy from OSIRIS, and mid-infrared photometry from the WISE catalog ($W1$ and $W2$) for the all  the objects in our sample, which allow us to calculate the bolometric luminosity ($\mathrm{L_{bol}}$),  and the bolometric corrections in $J$- and $K$-band ($\mathrm{BC}_{J}$ and $\mathrm{BC}_{K}$). We do not calculate $\mathrm{L_{bol}}$ for objects with detected disks: {three Cha\,I members (2MASS~J11062554--7633418, 2MASS~J11074656--7615174, 2MASS~J11085497--7632410), and two UppSco members (2MASS~15591135--23380002, and 2MASS~J16060391--2056443)}.

\begin{figure}
\centering
\includegraphics[width=0.5\textwidth]{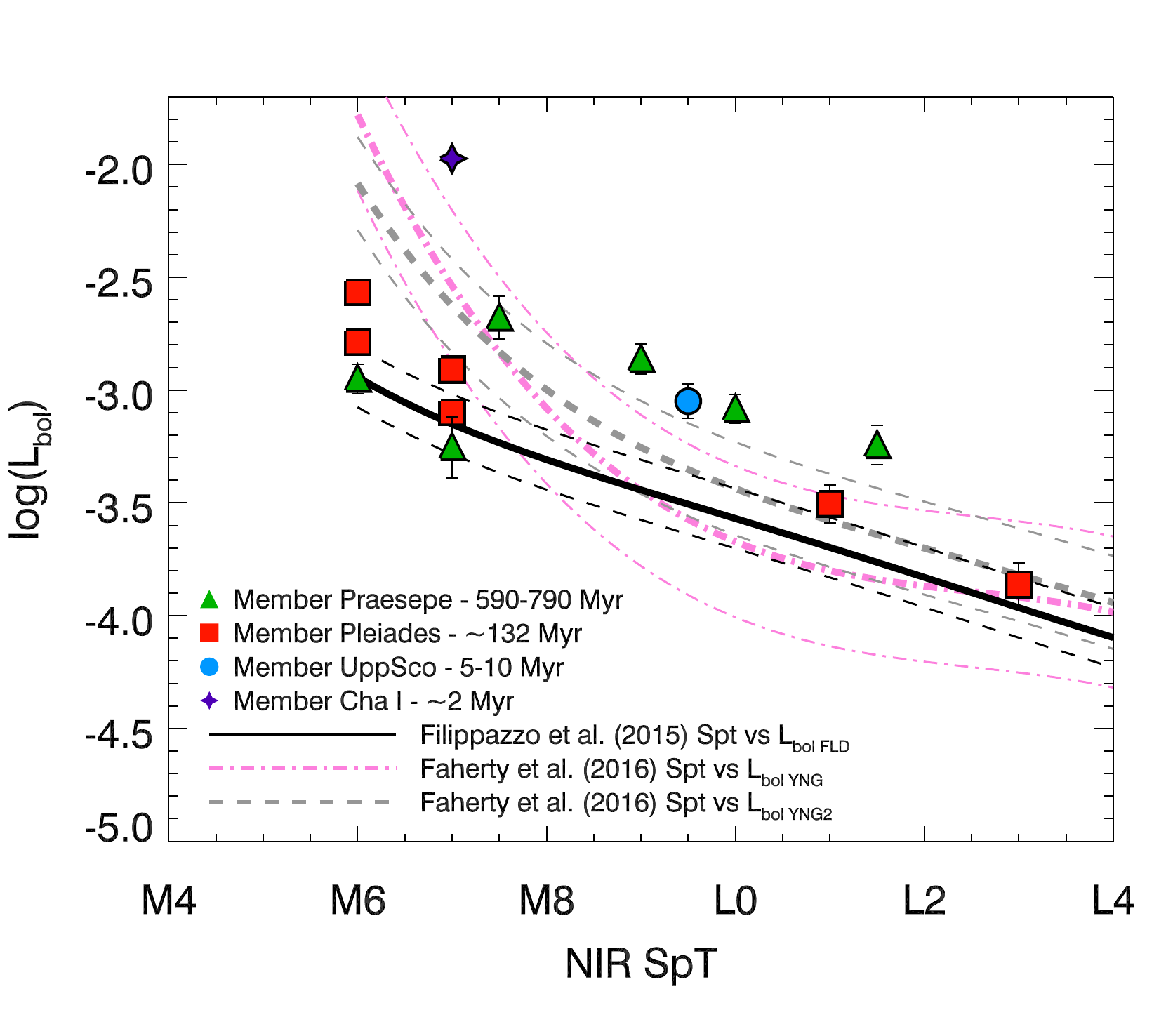}
\caption{ \label{Lbol} $\mathrm{L_{bol}}$  calculated for our sample following the procedure presented in \citet{Filippazzo2015}. The grey solid line represents the empirical relation derived in  the same work   for spectral types versus $\mathrm{L_{bol}}$, and the grey dashed lines delimit the rms of the empirical relation.  }
\end{figure}

  To calculate the $\mathrm{L_{bol}}$, we employed the method presented in \cite{Filippazzo2015}, doing a linear interpolation to fill in the gaps left between {0 flux at 0~$\mu$m}, and the X-shooter optical flux calibrated spectra. In case we did not have optical spectrum for the source, we fill in the optical wavelengths using its Pan-STARRS optical photometry. {Similarly}, we interpolate the water bands at 1.4~$\mu$m and at 1.8~$\mu$m, and the gap between the X-shooter spectra in the near-infrared and the WISE photometry, W1 and W2. Finally, we interpolate linearly between the reddest WISE photometric point available and the 1000~$\mu$m, following the procedure of \cite{Filippazzo2015}. To calculate the bolometric luminosity of each source we applied the equation:

\begin{equation}  
 L_{bol} = 4 \pi d^{2} \int_{0 \mu m}^{1000 \mu m} F_{\lambda} d\lambda
\end{equation}
 
  where $F_{\lambda}$ is the calibrated flux density in units of  erg $\mathrm{s^{-1} cm^{-2} \mu m^{-1}}$, and $d$ is the distance to the source in cm. {We present the $\mathrm{L_{bol}}$ in Table \ref{Lbol_table}, and we plot them vs their near-infrared spectral type in Figure \ref{Lbol}}. {We overplot the polynomial fit to the spectral types versus $\mathrm{L_{bol}}$ plot from \cite{Filippazzo2015}, obtaining similar results for the values of the $\mathrm{L_{bol}}$, with exception of 2MASS~J11123099--765334 from Cha\,I, and 2MASS~J16060629--2335133 from UppSco, that are young objects. In addition, objects UGCS J083748.00+201448.5, 2MASS J08410852+1954018, 2MASS J08370450+2016033, and UGCS J084510.65+214817.0, from the Praesepe, and objects 2MASS~J03484469+2437236, and 2MASS J03443516+2513429 from the Pleiades open cluster are also overluminous, as expected, given that they are  binary candidates (see Table~\ref{literature}).}

  In addition, we calculated the bolometric correction  for our sample using the equation:
  
  \begin{equation}  
BC_{band} = M_{bol} - M_{band} 
\end{equation}

\begin{figure}
\centering
\includegraphics[width=0.5\textwidth]{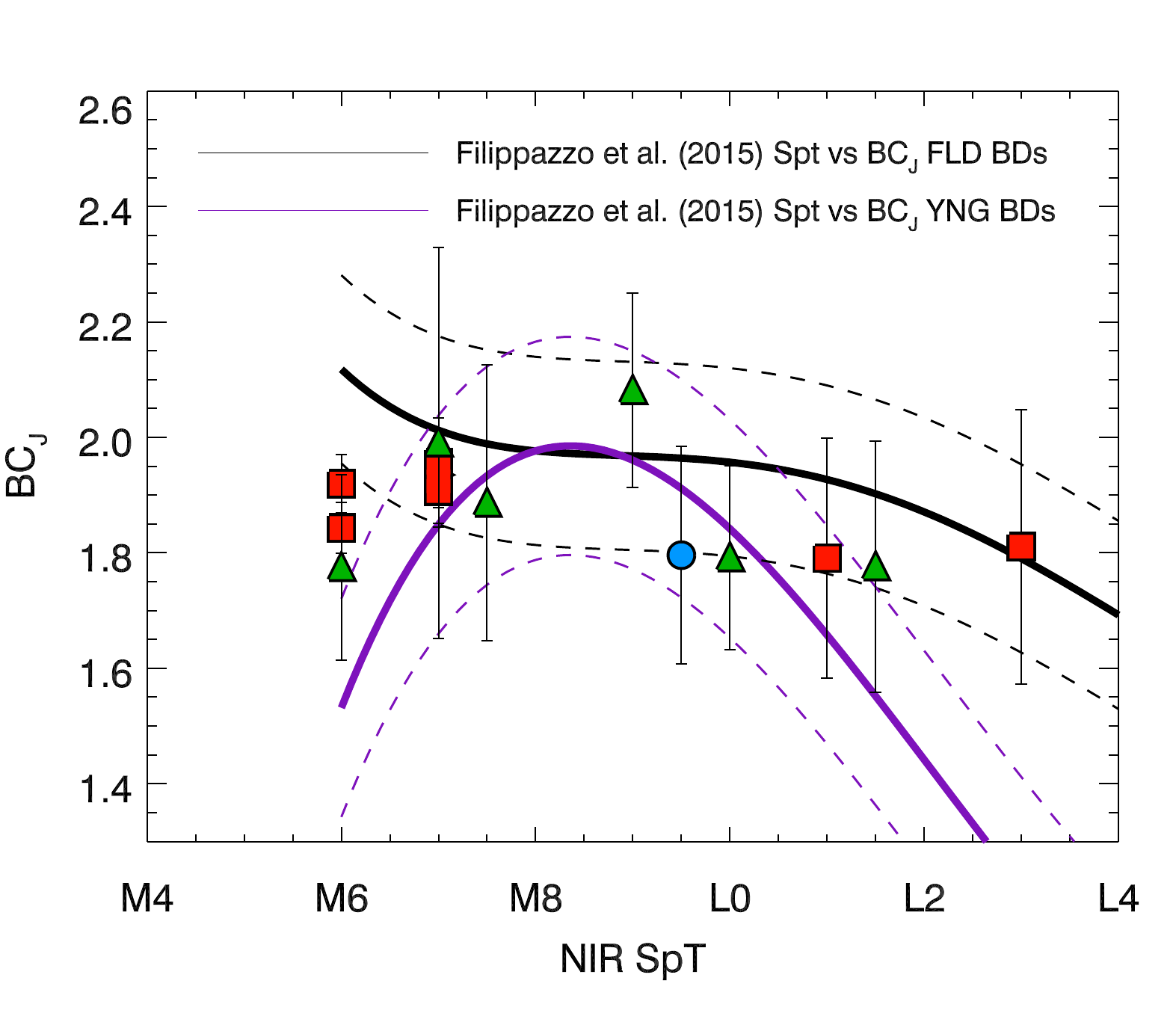}
\caption{ \label{BCJ} $\mathrm{BC_{J}}$  calculated for our sample following the procedure presented in \citet{Filippazzo2015}. The black solid line represents the empirical relation derived in  the same work   for spectral types versus $\mathrm{BC_{J}}$, and the black dashed lines delimit the rms of the empirical relation. The purple solid line with its respective purple dashed lines represent the same relation for young brown dwarfs.}
\end{figure}

\begin{figure}
\centering
\includegraphics[width=0.5\textwidth]{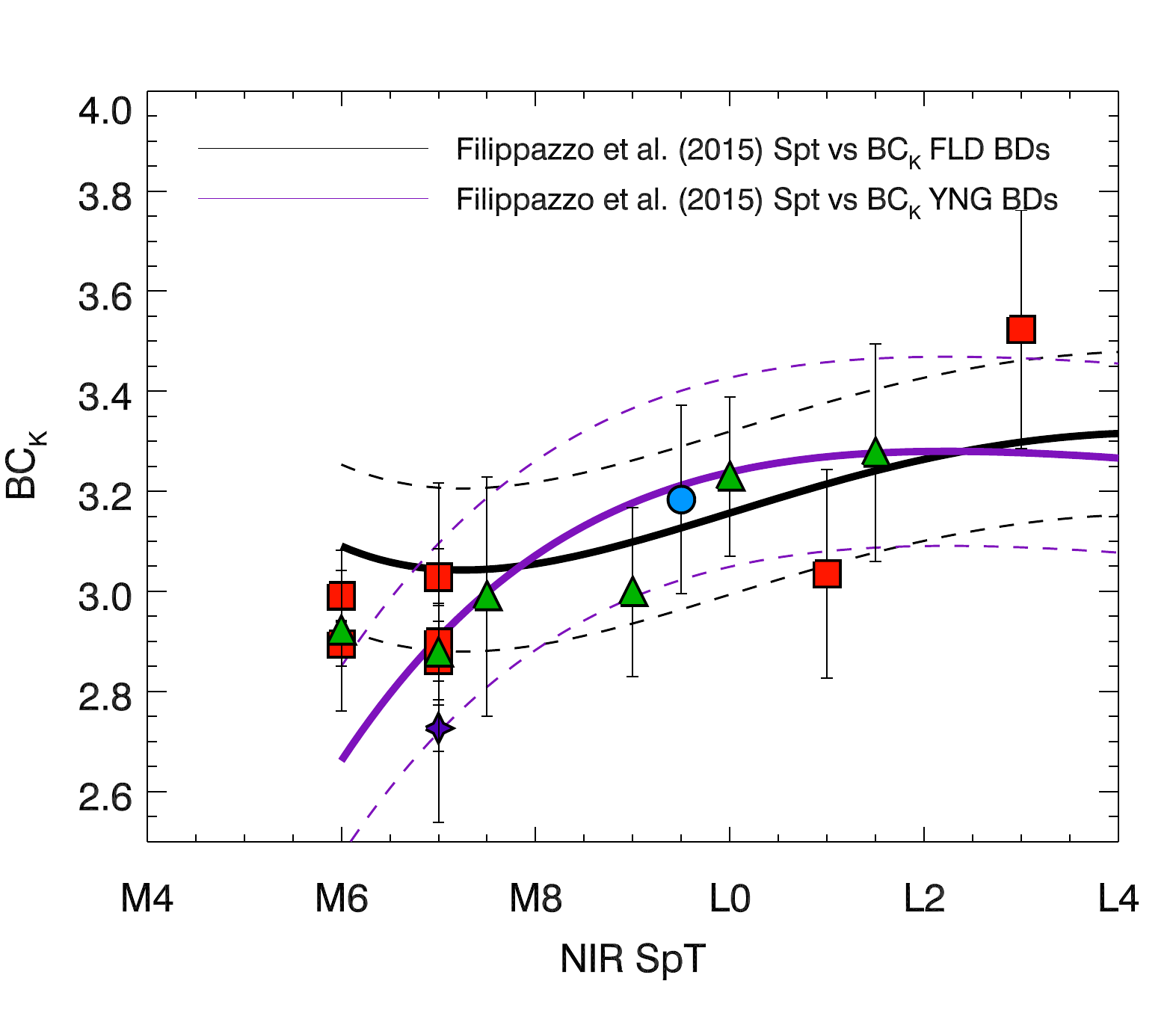}
\caption{ \label{BCK} $\mathrm{BC_{K}}$  calculated for our sample following the procedure presented in \citet{Filippazzo2015}. The black solid line represents the empirical relation derived in  the same work   for spectral types versus $\mathrm{BC_{K}}$, and the black dashed lines delimit the rms of the empirical relation. The purple solid line with its respective purple dashed lines represent the same relation for young brown dwarfs. }
\end{figure}

In Figures \ref{BCJ} and \ref{BCK} we {compare} our results with the BCs obtained using the polynomial fit derived by \cite{Filippazzo2015}, obtaining in {generally} {consistent between our BCs and the BCs obtained using the polynomial published in \cite{Filippazzo2015}. The values obtained for the BCs for our sample are shown in Table \ref{Lbol_table}.  This suggests that the binary candidates, if confirmed, might have two components with similar spectral types to have BCs consistent with a single component.} 

	\begin{table*}
		\small
		\caption{Bolometric luminosity and bolometric correction for our targets in $J$ and $K$-band.}  
		\label{Lbol_table}
		\centering
		\begin{center}
			\begin{tabular}{llllll}
				\hline
				\hline 
				
				Name & SpT & $Gaia$ $\pi^{a}$ (mas) & $\log(\mathrm{L_{bol}/L_{\odot}})$ & $ \mathrm{BC}_{K}$ & $ \mathrm{BC}_{J}$   \\		
				\hline              
				UGCS J083748.00$$+$$201448.5 & M6.0 & 3.53$\pm$2.01  & -2.95$\pm$0.06  & 2.92$\pm$0.16 &     1.77$\pm$0.16 \\
                2MASS J08370215$+$1952074    & M7.5 & 3.78$\pm$0.80  & -2.68$\pm$0.09  & 2.99$\pm$0.24 &     1.92$\pm$0.24\\
                UGCS J083654.60$+$195415.7   & M7.0 &                &-3.25$\pm$0.13  &	2.87$\pm$0.33 &     1.99$\pm$0.33 \\
                2MASS J08410852$+$1954018    & M9.0 &                &-2.86$\pm$0.06  & 2.99$\pm$0.17 &     2.08$\pm$0.17 \\
                2MASS J08370450$+$2016033    & L0.0 &                &-3.08$\pm$0.06 & 3.22$\pm$0.16 &     1.79$\pm$0.16 \\
                UGCS J084510.65$+$214817.0   & L1.5 &                & -3.24$\pm$0.09  & 3.28$\pm$0.22 &     1.77$\pm$0.22\\
                
                2MASS J03484469$+$2437236    & M6.0	& 6.81$\pm$0.34 & -2.57$\pm$0.02  &	2.89$\pm$0.04 &    1.84$\pm$0.04 \\
                2MASS J03491512$+$2436225    & M6.0 & 7.43$\pm$0.45 & -2.79$\pm$0.02  & 2.99$\pm$0.05 &    1.92$\pm$0.05 \\
                2MASS J03512557$+$2345214    & M7.0 & $\mathrm{4.38\pm1.35^{b}}$ & -3.09$\pm$0.03  &	2.89$\pm$0.08 &    1.96$\pm$0.08 \\
                2MASS J03443516$+$2513429    & M7.0 & 7.95$\pm$0.98 & -2.91$\pm$0.02  &	3.03$\pm$0.06 &    1.91$\pm$0.06 \\
                2MASS J03463425$+$2350036    & L1.0 &               & -3.50$\pm$0.08  &	3.04$\pm$0.21 &    1.79$\pm$0.21\\
                2MASS J03461406$+$2321565    & M7.0 & 7.22$\pm$0.58 & -2.91$\pm$0.03  & 2.86$\pm$0.08 &    1.92$\pm$0.08\\
                2MASS J03541027$+$2341402    & L3.0 &               & -3.86$\pm$0.10  & 3.52$\pm$0.23 &    1.81$\pm$0.23 \\
                
                2MASS  J16060629--2335133    &	M9.5&               & -3.05$\pm$0.08  &	3.18$\pm$0.19 &    1.80$\pm$0.19 \\
                
                2MASS J11123099--7653342     &	M7.0&               &-1.97$\pm$0.02  &	2.72$\pm$0.05 &    1.93$\pm$0.05 \\
              
				\hline

			\end{tabular}
		\end{center}
		\begin{tablenotes}
		\small
		\item {a: When the parallax for the individual object is not available, we use the average parallax to the open cluster/association to calculate the bolometric luminosity as provided in Section \ref{sample}}.\\
		\item {b: 2MASS J03512557$+$2345214 has a significantly different $Gaia$ parallax from the other Pleiades members, thus, it might not be a member of the Pleiades open cluster.}\\
		
	\end{tablenotes}
	\end{table*}

\subsection{Mass, Radii, effective temperature and surface gravity}\label{phys_parameters}

We use the {extension} of the evolutionary models from \cite{Baraffe2015} for brown dwarfs and extrasolar planets to estimate the values of the masses, effective temperatures, radii and surface gravities for our objects with well-determined ages. In addition, we will compare the values of the surface gravities with the gravity classes {defined by} \cite{Allers2013}. Because not all grids for all ages are available, we use the closest one in age for the estimation of the physical parameters of our objects: for members of Cha I ($\sim$2 Myr), we use the grid for 2~Myr, for the members of UppSco (5$-$10 Myr), we use the 10~Myr grid, for the Pleiades (132$\pm$27 Myr), we use the 120~Myr grid, and for members of the Praesepe (590$-$790 Myr), we use the 625~Myr grid. In Table \ref{physical}, we show  the values of the masses, effective temperatures, radii, surface gravities, gravity scores and gravity classes calculated as described in \cite{Allers2013}.  

After removing binary candidates, we use evolutionary models to predict the  masses for  our sample (see Table \ref{physical}). We estimate masses between 15.2$\pm$1.1~$\mathrm{M_{Jup}}$ and  113.6$\pm$21.4~$\mathrm{M_{Jup}}$. {Most of our objects, apparently single, have masses below the hydrogen burning limit}. We estimate radii between 1.11$\pm$0.14~$\mathrm{R_{Jup}}$ and 3.75$\pm$0.01~$\mathrm{R_{Jup}}$, and surface gravities between 3.51$\pm$0.01~dex and 5.22$\pm$0.01~dex.
The Praesepe members have surface gravities higher than log~g$\sim$5.20~dex, consistent with FLD-G gravity classification. The Pleiades members have surface gravities of log~g$\sim$4.79--4.92~dex, consistent with INT-G to FLD-G classification. The UppSco members have log~g$\sim$4.07~dex, consistent with VL-G to INT-G surface gravity classification. Finally, the Cha\,I member has a log~g$\sim$3.51~dex, consistent with VL-G gravity classification. 

%Comparing the ages of the members to the different open clusters to the gravity classification obtained using \cite{Allers2013} gravity indices, we can conclude that they are succeeded at spotting the very-low gravity members of our sample (Chamaeleon\,I and Upper Scorpius members). In the other hand, the indices were not always successful at providing an accurate gravity classification for the Pleiades or the Praesepe members, as concluded from the different  gravity classifications for objects of the same open cluster, that varied between the INT-G and FLD-G gravity classification.

	\begin{table*}

		\caption{Masses, effective temperatures, surface gravities and radii given by the evolutionary models of \citet{Baraffe2015}. {We exclude binary candidates and targets with reported circumstellar disks from this list}.}  
		\label{physical}
		\centering
		\begin{center}
			\begin{tabular}{llccllll}
				\hline
				\hline 

	Name &  SpT &       M ($\mathrm{M_{Jup}}$)	& $\mathrm{T_{eff}}$ (K) &		R ($\mathrm{R_{Jup}}$) & Log~g  & $\mathrm{GC^{c}}$ & OC/A  \\
    \hline 

	UGCS J083748.00$$+$$201448.5 & M7.0 &	113.6$\pm$21.4 &	2898$\pm$210 &		1.33$\pm$0.20 & 5.22$\pm$0.06 & FLD-G & Praesepe\\
	%2MASS J08370215$+$1952074 & M8.0   &	142.9$\pm$17.4 & 3072$\pm$70 &		1.61$\pm$0.18 & 5.15$\pm$0.03 \\
	UGCS J083654.60$+$195415.7& M8.0   &	91.2$\pm$15.3 &	2590$\pm$320 &		1.11$\pm$0.14 &  5.29$\pm$0.04 & INT-G & Praesepe\\
	%2MASS J08410852$+$1954018& M9.0    &	111.2$\pm$0.11 &  2875$\pm$1   &		1.31$\pm$0.01 & 5.23$\pm$0.01  \\
	%2MASS J08370450$+$2016033&  L0.0   &	104.0$\pm$11.3 &	2804$\pm$110	 &   	1.24$\pm$0.11 & 5.25$\pm$0.03   \\
	%UGCS J084510.65$+$214817.0& L1.5   &	94.4$\pm$15.7 & 2657$\pm$233 &		1.14$\pm$0.16 & 5.28$\pm$0.04 \\

	%2MASS J03484469$+$2437236 & M6.0&	89.0$\pm$7.2	& 2888$\pm$5 &	1.62$\pm$0.06 & 4.94$\pm$0.01 \\
	2MASS J03491512$+$2436225& M6.0 &	70.3$\pm$3.0	& 2723$\pm$33 &	1.46$\pm$0.03 & 4.92$\pm$0.01 & INT-G & Pleiades\\
	%2MASS J03512557$+$2345214 & M7.0&	47.8$\pm$5.9 &  2364$\pm$84 &	1.29$\pm$0.03 & 4.87$\pm$0.02\\
	2MASS J03443516$+$2513429& M8.5 & 61.7$\pm$6.5 &	 2621$\pm$76 &	1.40$\pm$0.03 & 4.91$\pm$0.01 & FLD-G & Pleiades\\
	2MASS J03463425$+$2350036 & L1.0& 37.0$\pm$4.5 &	 2017$\pm$192 &  1.23$\pm$0.02 & 4.80$\pm$0.03 & INT-G & Pleiades\\
	2MASS J03461406$+$2321565 & M6.0&	56.6$\pm$4.7 & 2537$\pm$72 &	1.36$\pm$0.04 & 4.90$\pm$0.01 & INT-G& Pleiades \\
	2MASS J03541027$+$2341402 & L3.0& 36.2$\pm$2.3 &	 1983$\pm$100 &	1.22$\pm$0.01 & 4.79$\pm$0.02 & VL-G & Pleiades\\

	2MASS  J16060629--2335133 &	M9.5 &	15.2$\pm$1.1 &	2238$\pm$55 &	1.84$\pm$0.12  & 4.07$\pm$0.02 & VL-G & UpSco\\

	2MASS J11123099--7653342&	M7.0& 27.3$\pm$1.6  & 2653$\pm$21     & 3.75$\pm$0.01     &  3.51$\pm$0.14 & VL-G & Cha\,I \\

\hline

			\end{tabular}
		\end{center}
        
        		\begin{tablenotes}
		\small
		\item c:  GS: Gravity scores as described in \citet{Allers2013}. The first three gravity scores belong to the result of the pEWs of the K\,I lines at 1169, 1177 and 1253~nm. The rest belong to the gravity scores obtained from the spectral indices described in Section \ref{spectral_indices}.\\
		
	\end{tablenotes}
			
	\end{table*}

\section{Conclusions and Final Remarks}\label{conclusions_remarks}

We obtained 0.6$-$2.5~$\mu$m VLT/X-shooter spectra of 20 low-mass stars and brown dwarfs members to the Cha\,I, (2~Myr), UppSco (5$-$10~Myr), Pleiades {(132$\pm$27~Myr)} and Praesepe (590$-$790~Myr) open cluster. Our {targets} have spectral types between M6.0 and L3.0.

We performed a consistent spectral classification of our targets by {comparing} their optical and near-infrared spectra {to both young and field M and L templates in both wavelength ranges independently}. We obtained a consistent spectral type for most of our objects in all the comparisons with a maximum {discrepancy} of $\pm$2.5 spectral types. The maximum dispersion in spectral type classification was reached for very young or field objects, for which the match to field or young objects respectively was more challenging to find. Finally, we adopted the spectral type obtained when comparing to near-infrared field objects as their final spectral classification.

We measured the pEWs of the alkali lines in the optical and in the near-infrared. For completion, we added other measurements for other objects with different ages (see Section \ref{pEWs}). We {found} that the pEWs of the alkali lines in the optical and near-infrared increases with age, as found by previous works \citep[and references therein]{Steele_Jameson1995,Martin1996,Gorlova2003,Cushing,Allers&Liu}.  We found that members of the Pleiades have similar pEW for most of the alkali lines than field objects. The same is observed for objects of the $\alpha$-Persei group (90 Myr). This fact suggests that at approximately 100 Myr, low-mass stars and brown dwarfs have {nearly} reached their final surface gravities, as concluded by \cite{Martin2017}, and as predicted by evolutionary models for brown dwarfs \citep{Baraffe2015}.

In addition, we further investigated the relationship between pEWs of the alkali lines in the near-infrared and the age. We found a moderate correlation between the pEWs of the alkali lines in the $J$-band and age, for all ranges of spectral types. We observed that the increase of the pEWs with age is more remarkable for {objects with spectral type later than M8}. We provided relationships between pEWs and age that can be used in the future to {roughly} estimate ages given the pEWs of the alkali lines in the $J$-band.

We calculated the gravity scores defined by \cite{Allers&Liu} using the alkali lines in the $J$-band, and the $\mathrm{FeH_{J}}$, the $\mathrm{KI_{J}}$, the $H$-cont, the $\mathrm{FeH_{z}}$, and the $\mathrm{VO_{z}}$ band. We obtained a final surface gravity classification using {all the indices, that reproduce reasonably well the youngest objects (Cha\,I and UppSco) but do not predict accurately intermediate and field M and L dwarfs}. Thus, surface gravity indices must be used with caution, specially when a non low-surface gravity classification is obtained.
{Red colours} have been believed to be a potential indicator of young age and low surface gravity \citep[and references therein]{Allers2013,Bonnefoy2014b,Manjavacas2014}. Thus, we investigated the correlation between the $J-W2$ colour and the age. We found a moderate correlation between those two quantities {for the spectral types we consider in our work}. In Figures \ref{SpT_J_W2_m} and \ref{colour_age}, we found that objects that were mostly red, have ages below $\sim$10~Myr, that might be linked to the existence of a protoplanetary disk, {ring structures, extinction through the star forming region, or viewing angle, but} not necessarily to the influence of low surface gravity in their atmospheres. 
{In addition, some of the members of moving groups older than 100~Myr presented in \cite{Faherty2016} show very red colours. This fact suggests that colours do not scale linearly with ages \citep{2014MNRAS.439..372M,Liu2016,Bowler2017,Zapatero-Osorio2017}}.

%We calculated the  $\mathrm{FeH_{J}}$, $\mathrm{KI_{J}}$, $\mathrm{FeH_{z}}$, $\mathrm{VO_{z}}$ and $H$-cont spectral indices as defined in \cite{Allers2013}, and their respective gravity scores. We found that spectral indices sometimes fail to provide an accurate estimation of surface gravity. \cite{Lodieu2017} found that the $H$-cont index is the most sensitive to surface gravity, nonetheless, we find that this index was not significantly different for the VL-G and the FLD-G objects in our sample.  Instead, we find that  $\mathrm{FeH_{J}}$ is more correlated with age.

We calculated the bolometric luminosity of our sample using the optical, or optical Pan-STARRS photometry,  near-infrared X-shooter flux calibrated spectra, and {WISE photometry}. We calculated the bolometric corrections in the $J$- and $K$-band, {and compared them to the polynomial fits from \citet{Filippazzo2015}, and \cite{Faherty2016}}. With exception of binary candidates, most of our targets followed the spectral type vs $\mathrm{L_{bol}}$ relation from \citet{Filippazzo2015}. Praesepe and Pleiades non-binary candidate members followed the relation derived by \cite{Filippazzo2015} for field objects. {Members of the Cha\,I or the UppSco association are overluminous with respect to the field and the young sequence, which is in agreement with the expectations for such young sources}.

Similarly, we plot the relation between near-infrared spectral types and the $\mathrm{BC_{J}}$, and $\mathrm{BC_{K}}$. We overplot the spectral type vs $\mathrm{BC_{J}}$ and $\mathrm{BC_{K}}$ relation for field and young objects. All objects of our sample follow these relations, even binary candidates. This suggests that, in case our binary candidates are actually binaries, their components should have similar spectral type, so that the difference between the absolute flux in $J$ and $K$ bands remains similar to the difference for a single object of the same spectral type.

{The surface gravities obtained for our targets using \cite{Baraffe2015} evolutionary models did not always agree with the gravity classification given by \cite{Allers2013} indices. The gravity classification was consistent only for all objects from the Cha\,I and from the UppSco association. Thus, as suggested before, gravity indices should be used with caution, considering that they might not always provide an accurate spectral classification.}

	\section*{Acknowledgements}

Based on observations collected at the European Organisation for Astronomical
Research in the Southern Hemisphere under ESO programmes: 098.C-0277(A), 093.C-0109(A), and 095.C-0812(A), 093.C-0769(A), and 095.C-0378(A).

This publication makes use of data
products from the Two Micron All Sky Survey, which is a joint
project of the University of Massachusetts and the Infrared
Processing and Analysis Center/California Institute of Technology,
funded by the National Aeronautics and Space
Administration and the National Science Foundation

This work is based on observations (programme GTC66-12B) 
made with the Gran Telescopio Canarias (GTC), operated on the island of La Palma
in the Spanish Observatorio del Roque de los Muchachos of the Instituto de
Astrof\'isica de Canarias. 

This work has made use of data from the European Space Agency (ESA) mission
{\it Gaia} (\url{https://www.cosmos.esa.int/gaia}), processed by the {\it Gaia}
Data Processing and Analysis Consortium (DPAC,
\url{https://www.cosmos.esa.int/web/gaia/dpac/consortium}). Funding for the DPAC
has been provided by national institutions, in particular the institutions
participating in the {\it Gaia} Multilateral Agreement.

N. L. and V. J. S.-B. are supported by programme AYA2015-69350-C3-2-P and MRZO by programme
AYA2016-79425-C3-2-P from Spanish Ministry of Economy and Competitiveness (MINECO).

%We gratefully acknowledge CAHA allocation time committee and CAHA Observatory staff for assiting the PI during the observations.  This work was supported by Sonderforschungsbereich SFB 881 "The Milky Way System" (subproject B6) of the German Research Foundation (DFG). This research has made use of the SIMBAD database, operated at CDS, Strasbourg, France. 

%\newpage%%%%%%%%%%%%%%%%%%%%%%%%%%%%%%%%%%%%%%%%%%%%%%%%%%%%%%

%\begin{thebibliography}{}
%  \bibitem{Cushing2005} Author1, A.B., Author2, C.D.: 2001, AN 322, 1
%  \bibitem{} Author3, E.F., Author4, G.H.: 2001, AN 322, 10
% \bibitem{} Author5, I.: 2001, AN 322, 20
% \bibitem{} Author6, J.: 2001, AN 322, 30

\bibliographystyle{mn2e_fix}
\bibliography{young_xshooter}

%\end{thebibliography}
\newpage
\appendix

\begin{figure*}
\centering
\includegraphics[width=0.95\textwidth]{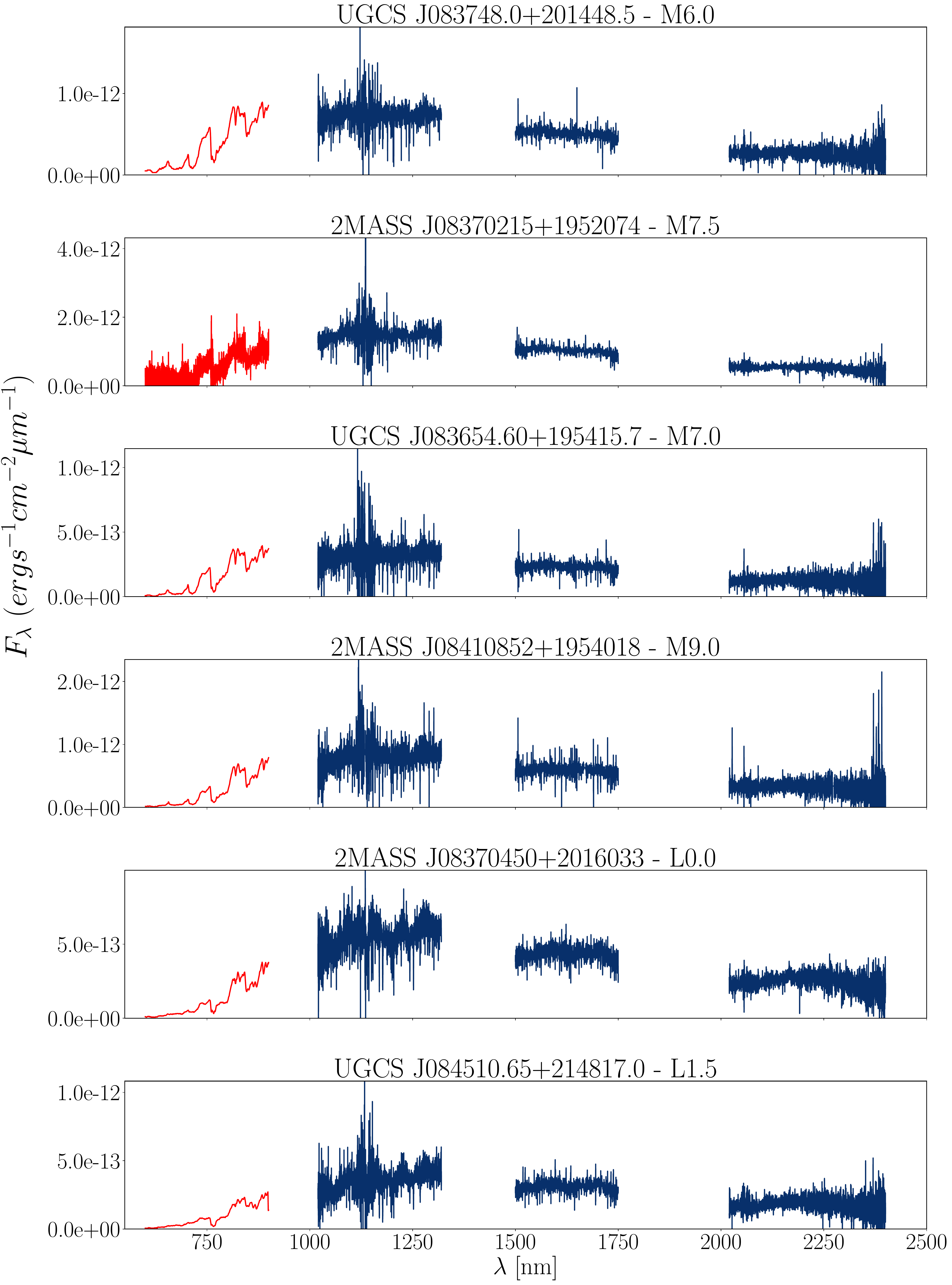}
\caption{\label{all_spectra_pra} X-shooter/VLT and OSIRIS/GTC spectra for objects members from the Praesepe open cluster listed in Table \ref{literature}. The OSIRIS or X-shooter optical spectra is shown in red for clarity. The flux density is given in $F_{\lambda}$ in $\mathrm{erg\;  s^{-1} cm^{-2} \mu m^{-1}}$.}
\end{figure*}

\begin{figure*}
\centering
\includegraphics[width=0.99\textwidth]{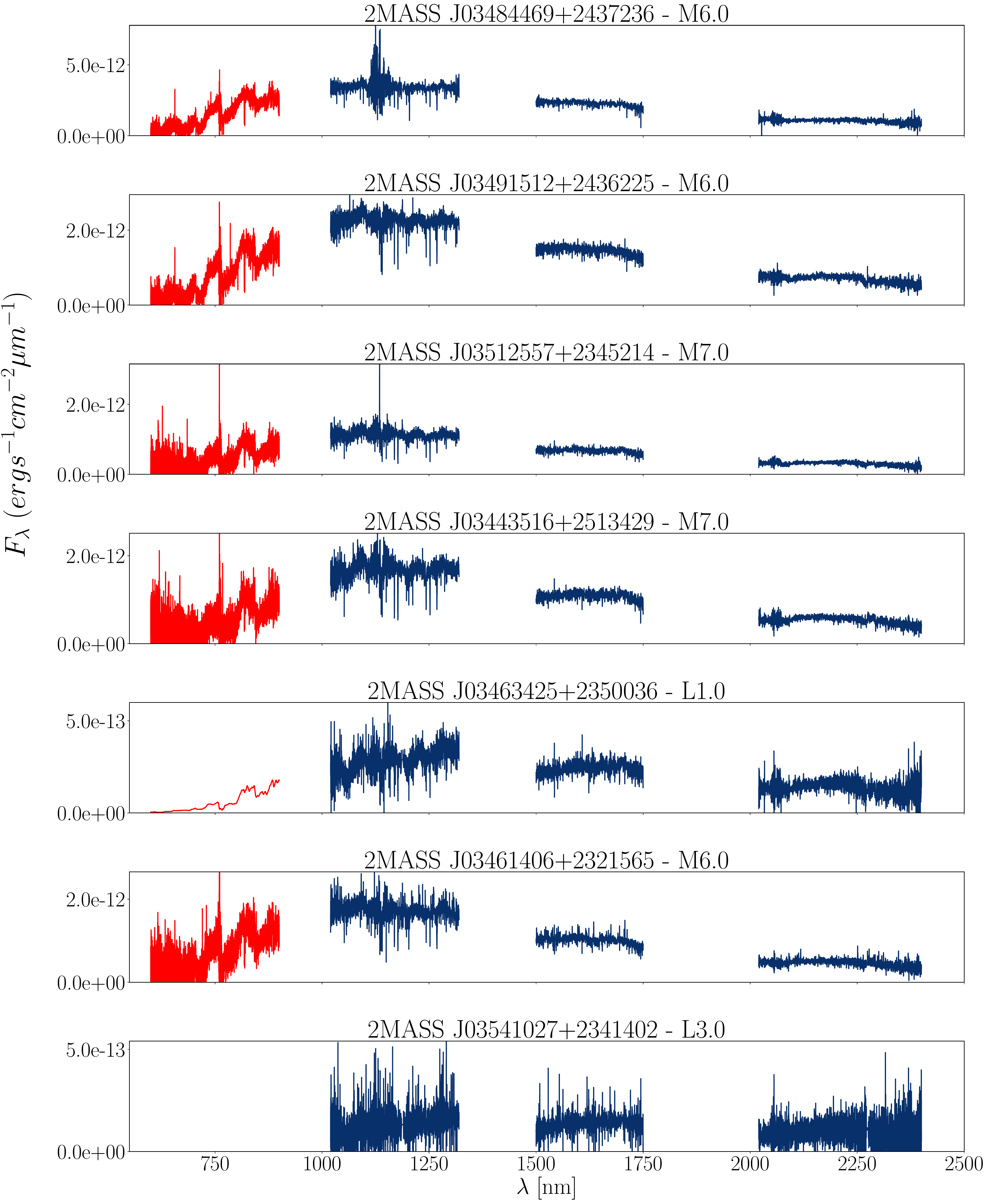}
\caption{\label{all_spectra_ple} X-shooter/VLT and OSIRIS/GTC spectra for objects members from the Pleiades open cluster listed in Table \ref{literature}. The  X-shooter optical spectra is shown in red for clarity. The flux density is given in $F_{\lambda}$ in $\mathrm{erg\;  s^{-1} cm^{-2} \mu m^{-1}}$.}
\end{figure*}

\begin{figure*}
\centering
\includegraphics[width=0.95\textwidth]{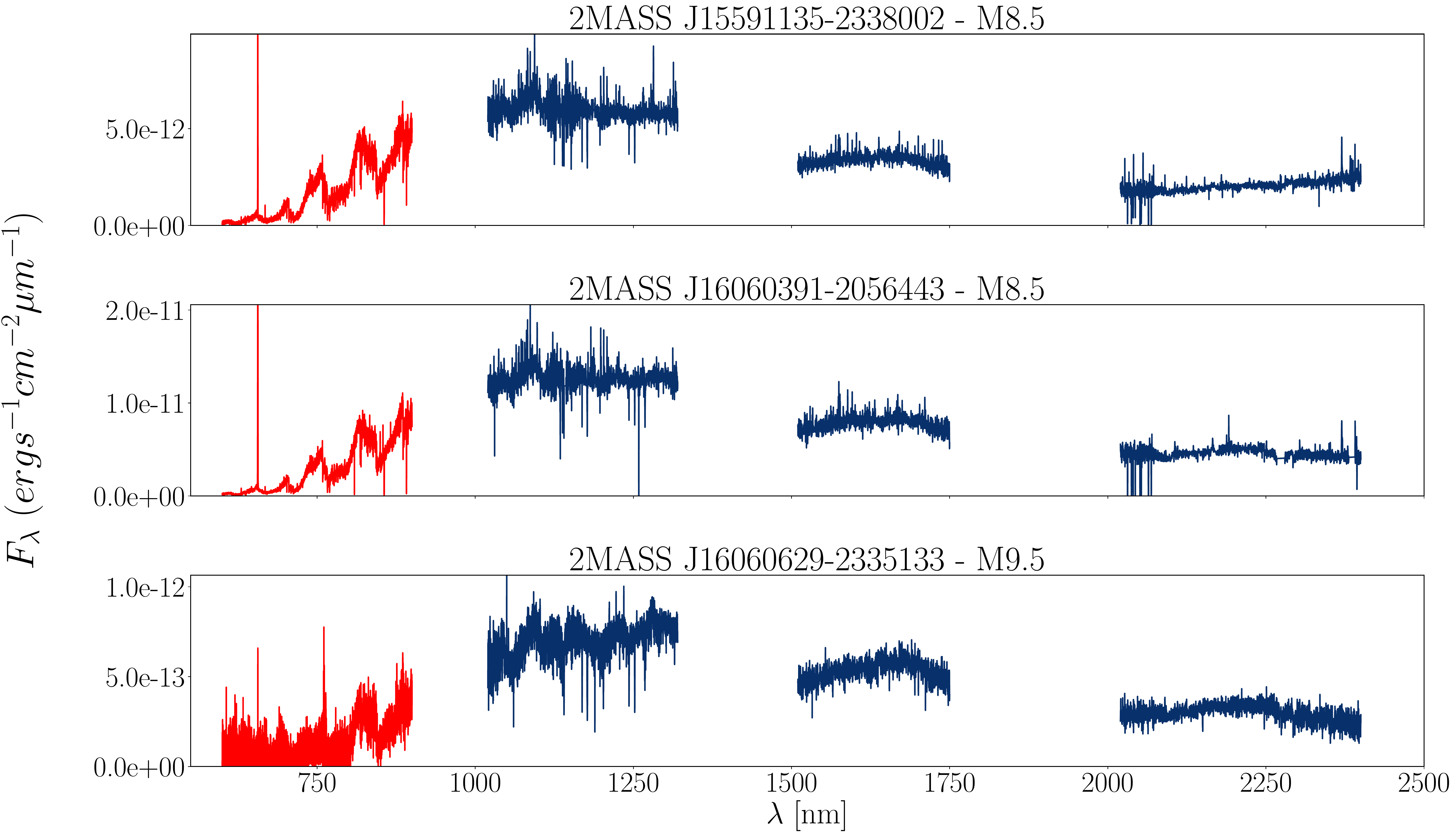}
\caption{\label{all_spectra_uppsco} X-shooter/VLT and OSIRIS/GTC spectra for objects members from the UppSco association listed in Table \ref{literature}. The  X-shooter optical spectra is shown in red for clarity. The flux density is given in $F_{\lambda}$ in $\mathrm{erg\;  s^{-1} cm^{-2} \mu m^{-1}}$.}
\end{figure*}

\begin{figure*}
\centering
\includegraphics[width=0.95\textwidth]{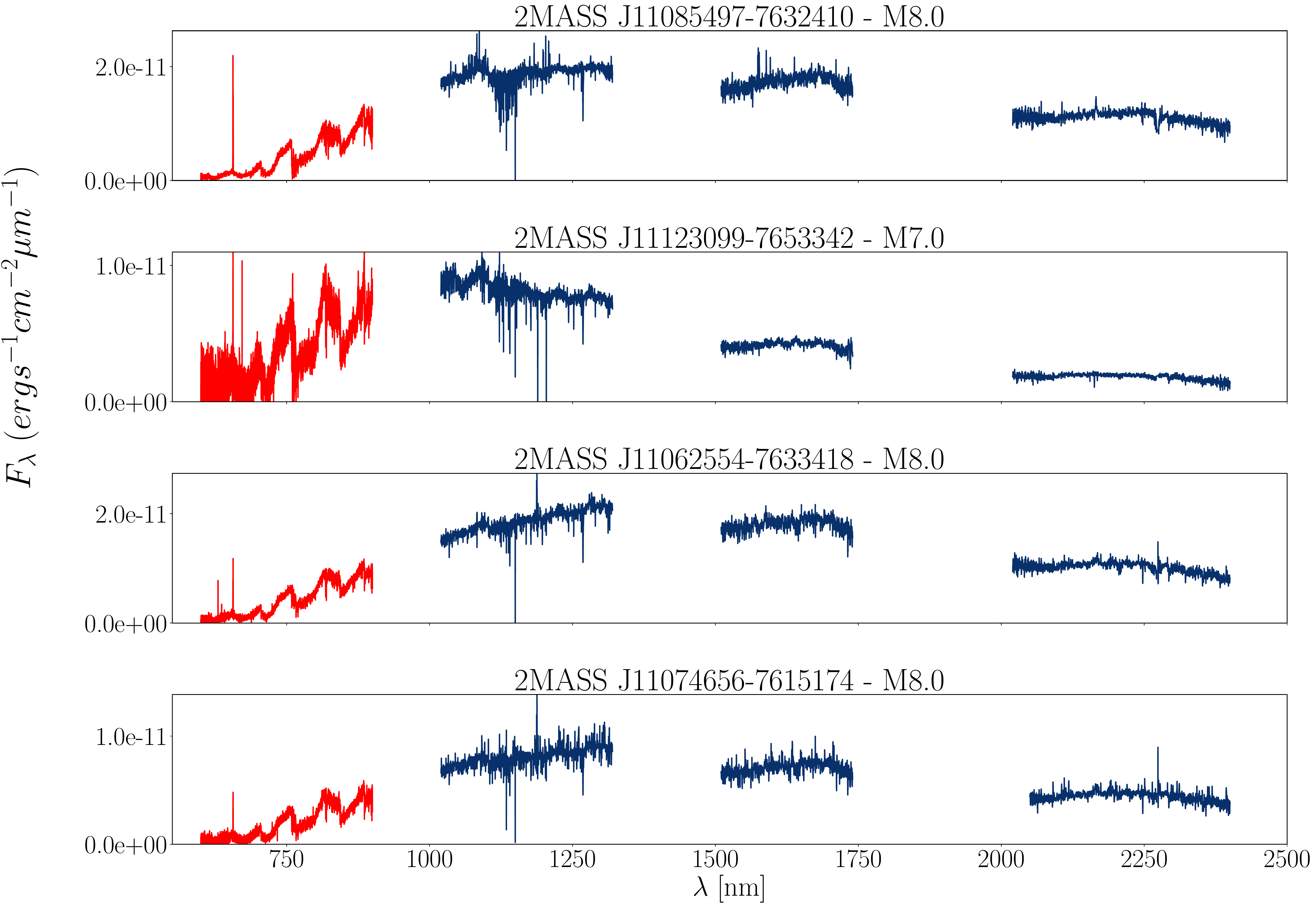}
\caption{\label{all_spectra_cha} X-shooter/VLT and OSIRIS/GTC spectra for objects members from the Cha\,I association listed in Table \ref{literature}. The  X-shooter optical spectra is shown in red for clarity. The flux density is given in $F_{\lambda}$ in $\mathrm{erg\;  s^{-1} cm^{-2} \mu m^{-1}}$.}
\end{figure*}

 \begin{landscape}
\begin{table}
	\small
	\caption{Log of the observed targets with VLT/Xshooter.}  
	\label{log0}
	\centering
	\renewcommand{\footnoterule}{}  % to avoid a line before footnotes
	\begin{center}
		\begin{tabular}{lllllllllcl}
			\hline 
				\hline
				Name      & Date             & Arm     & DIT (s) & NINT & Seeing ('') & SNR VIS/NIR  &  Airmass & Notes \\
				\hline              
				UGCS J083748.00$+$201448.5  & Jan 29, 2017 & VIS/NIR & 305/300 & 4/4 & 0.73 & 1.9/7.2 & 1.49 &  \\
				Hip040881 & Jan 29, 2017    & VIS/NIR.     & 6.25/30 & 2/2 & 0.94& &1.55  &B9.5V Telluric Standard\\
				\hline	
				2MASS J08370215$+$1952074 & Jan 28, 2017 & VIS/NIR & 305/300 & 10/7 & 0.9 & 3.5/12.1 & 1.58 & \\
				Hip040881 & Jan 28, 2017 & VIS/NIR & 5/10 & 2/2 & 0.66 & & 1.62  &B9.5V Telluric Standard\\		
                
               \hline	
				UGCS J083654.60$+$195415.7 & Jan 23, 2017 & VIS/NIR & 305/300 & 10/7 & 0.83 & 0.9/4.5 & 1.51 & \\
				Hip043564 & Jan 23, 2017  & VIS/NIR & 40/90 & 2/2 & 0.8 & &1.50  &B5V Telluric Standard\\		
				\hline
				J08410852$+$1954018 & Jan 23, 2017 & VIS/NIR & 305/300 & 8/8 & 1.02 & 1.2/7.9 & 1.41& \\
				Hip043564     & Jan 23, 2017 & VIS/NIR & 40/90 & 2/2 & 0.92 & &1.5  &B5V Telluric Standard\\
				
				\hline
				2MASS J08370450$+$2016033 & Jan 1, 2017 & VIS/NIR & 305/300 & 8/8 & 0.51 & 1.0/8.2 & 1.50 & \\
				Hip022597       & Jan 1, 2017 & VIS/NIR & 12.5/30 & 1/1 & 1.1 & &1.05  &B5V Telluric Standard\\
				
				\hline
				UGCS J084510.65$+$214817.0  & Jan 2, 2017 & VIS/NIR & 305/300 & 10/10 & 0.6  & 0.6/5.0 & 1.46 &\\
				Hip026545      & Jan 2, 2017 & VIS/NIR & 12.5/30   & 2/2  & 0.74 & &1.49  &B6V Telluric Standard \\
				
				\hline
				2MASS J03484469$+$2437236 & Nov 30, 2016 & VIS/NIR & 305/300 & 6/6 & 0.67 & 9.0/24.6 & 1.53 & \\
				Hip009534     & Jan 4, 2010 & VIS/NIR & 6/5 & 1/1 & 1.1 & & 1.1  &B6V Telluric Standard\\
				
				\hline
				2MASS J03491512$+$2436225 & Nov 27, 2016 & VIS/NIR & 305/300 & 6/6 & 0.34 & 6.7/23.2 & 1.61 & \\
				Hip017900     & Nov 27, 2016 & VIS/NIR & 6/10   & 2/2 & 0.6 &  & 1.51  &B8V Telluric Standard \\
				
				\hline
				2MASS J03512557$+$2345214 & Nov 26, 2016 & VIS/NIR & 305/300 & 8/8 & 0.4 & 2.8/13.4 & 1.51 & \\
				Hip021013     & Nov 26, 2016 & VIS/NIR & 8/10 & 2/2 & 0.45 & & 1.51  &B8III Telluric Standard\\
				
				\hline
				2MASS J03443516$+$2513429 & Nov 20, 2016 & VIS/NIR & 305/300 & 6/6 & 1.22 & 2.8/18.9 & 1.82 & \\
				Hip022527     & Nov 20, 2016 & VIS/NIR & 12.5/30 & 2/2 & 1.0 & & 1.72  & B3V Telluric Standard\\
				\hline
				2MASS J03463425$+$2350036 & Nov 11, 2016 & VIS/NIR & 305/300 & 10/10 & 0.28 & 0.5/5.2 & 1.58 & \\
				Hip020789     & Nov 11, 2016 & VIS/NIR & 6/7 & 2/2 & 0.27 & & 1.65  &B7V Telluric Standard\\
				\hline
				2MASS J03461406$+$2321565 & Nov 3, 2016 & VIS/NIR & 305/300 & 6/14 & 0.84 & 4.1/15.4 & 2.09 & \\
				Hip054006     & Nov 3, 2016 & VIS/NIR & 15/23   & 2/2 & 0.9 & & 1.91  &B5V Telluric Standard\\
				\hline
				2MASS J03541027$+$2341402 & Nov 3, 2016 & VIS/NIR & 305/300 & 14/14 & 1.04 & 0.1/3.6 & 1.68 & \\
				Hip045125     & Nov 3, 2016 & VIS/NIR & 15/12   & 2/2 & 0.77 & & 1.52  &B9V Telluric Standard\\
				\hline
				2MASS  J15591135--2338002& Apr 25, 2014 & VIS/NIR & 866/227.5 & 3/3 & 1.64 & 25.2/21.6 & 1.09 & \\
				Hip079073     & Apr 25, 2014 & VIS/NIR & 20/20   & 2/2 & 1.50 & &1.09  &G2V Telluric Standard\\
				\hline
				
				2MASS J16060391--2056443   & Apr 08, 2014 & VIS/NIR & 866/27.5 & 3/3 & 1.47 & 38.8/45.7 & 1.00 & \\
				Hip074389     & Apr 08, 2014 & VIS/NIR & 6.5/8   & 2/2 & 1.01 & &1.01 &G1.5V Telluric Standard\\
				\hline
				2MASS  J16060629--2335133   & Jun 25, 2014 & VIS/NIR & 197/190 & 14/14 & 0.99 & 2.2/13.8 & 1.21 & \\
				Hip087314     & Jun 25, 2014 & VIS/NIR & 6.25/30   & 2/2 & 1.00 & & 1.62  &B2/3Vnn Telluric Standard\\
				\hline
			%	{J16082847-23151032} & June 20, 2014 & VIS/NIR & 290/300 & 5/5 & 0.8 & 1.4 & \\
			%	Hip030175     & December 16, 2009 & VIS/NIR & 6/5 & 5/5 & 0.8 & 1.4 & B9.5V Telluric Standard\\
                
%                \hline
	%			{J16073799-2242468} & June 20, 2014 & VIS/NIR & 290/300 & 5/5 & 0.8 & 1.4 & \\
	%			Hip030175     & December 16, 2009 & VIS/NIR & 6/5 & 5/5 & 0.8 & 1.4 & B9.5V Telluric Standard\\
    				2MASS J11085497--7632410 & May 5, 2015 & VIS/NIR & 645/245 & 4/4 & 1.03 & 13.0/21.1 & 1.69 & \\
				Hip061066     & May 5, 2015 & VIS/NIR & 25/20 & 2/2 & 1.17 & & 1.58  &B9V Telluric Standard\\
				
				\hline
				2MASS J11123099--7653342& Jan 14, 2017 & VIS/NIR & 305/300 & 6/6 & 0.8 & 11.5/53.6 & 1.63& \\
				Hip053024     & Jan 14, 2017 & VIS/NIR & 9/20   & 2/2 & 0.75 & &1.65  &B4V Telluric Standard\\
				\hline

				2MASS J11074656-7615174 & Apr 3, 2015   & VIS/NIR & 585/225 &4/4 & 1.4 &  30.2/16.4&1.67 & \\
				Hip061066     & Apr 3, 2015    & VIS/NIR &  12.5/15  &2/2 & 0.83 & &1.59 &B9V Telluric Standard\\
				\hline
				Cha J11070768--7626326& Apr 5, 2015   & VIS/NIR & 585/225 &4/4 & 1.35 & 17.7/39.4 & 1.65& \\
				Hip061066     & Apr 5, 2015  & VIS/NIR & 15/20   &2/2 & 1.63 & &1.58  &B9V Telluric Standard\\

			\hline

		\end{tabular}
	\end{center}
	
\end{table}
\end{landscape}

	\begin{table*}

			\caption{Observing log for objects observed with GTC/OSIRIS{:} . }  
			\label{log1}
			\centering
			\begin{tabular}{llcl}
				\hline
				\hline
Name &  Resolution & Num. exp. $\times$ exp. time (s) & Observing date \\
\hline
UGCS J083748.00$+$201448.5 &  R300R & 1$\times$900 s & 17 Dec 2012 \\
2MASS J08410852$+$1954018 & R300R & 1$\times$900 s  &15 Dec 2012\\
UGCS J083654.60$+$195415.7 & R300R &3$\times$600 s& 03 Feb 2013 \\
2MASS J08370215$+$1952074 & R300R &1$\times$900 s& 20 Jan 2013  \\
UGCS J084510.65$+$214817.0& R300R &6$\times$700 s& 16 Jan 2013  \\
2MASS J08370450$+$2016033 & R300R &3$\times$700 s& 16 Jan 2013 \\
2MASS J03463425$+$2350036 &  R300R & 3$\times$700 s & 17 Jan 2013\\
\hline
			\end{tabular}

	\end{table*}

\begin{figure*}
\centering
\includegraphics[width=0.99\textwidth]{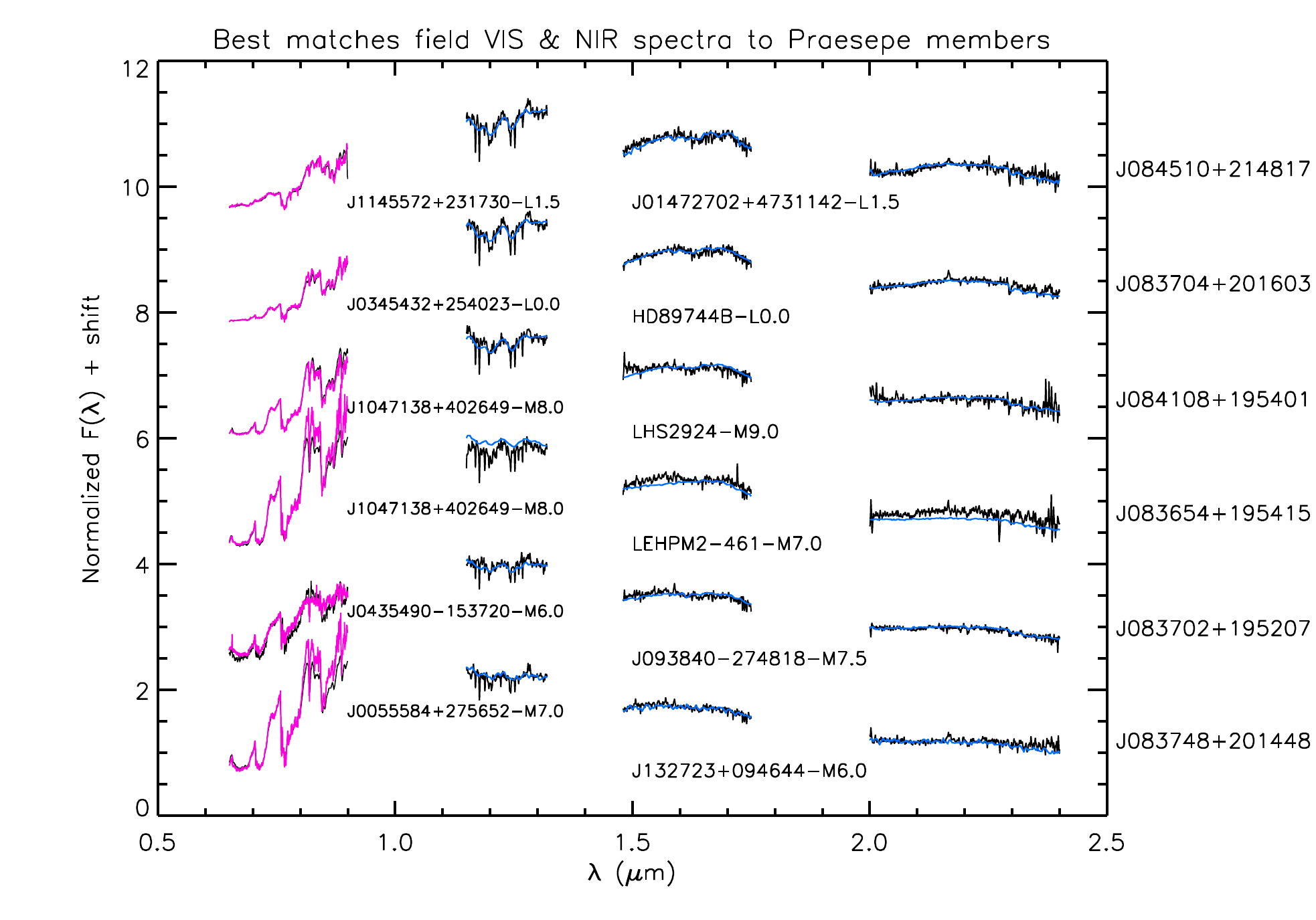}
\caption{{Best field dwarf matches (coloured lines) in the optical and near-infrared for Praesepe members (black).}}
\label{field_prae}
\end{figure*}

\begin{figure*}
\centering
\includegraphics[width=0.99\textwidth]{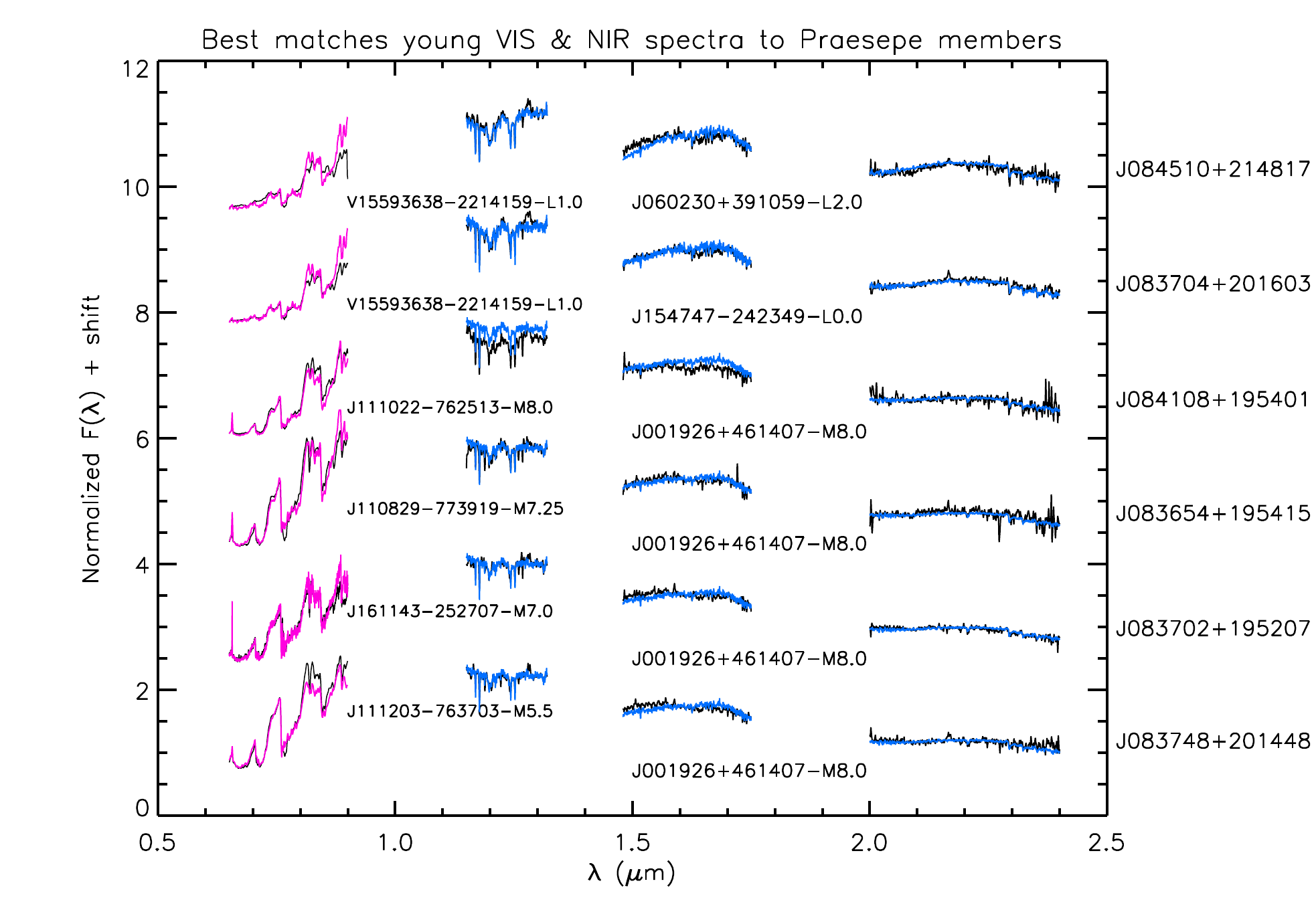}
\caption{{Same as Figure \ref{field_prae} but comparing with the library of young M and L dwarfs.}}
\label{young_prae}
\end{figure*}

\begin{figure*}
\centering
\includegraphics[width=0.99\textwidth]{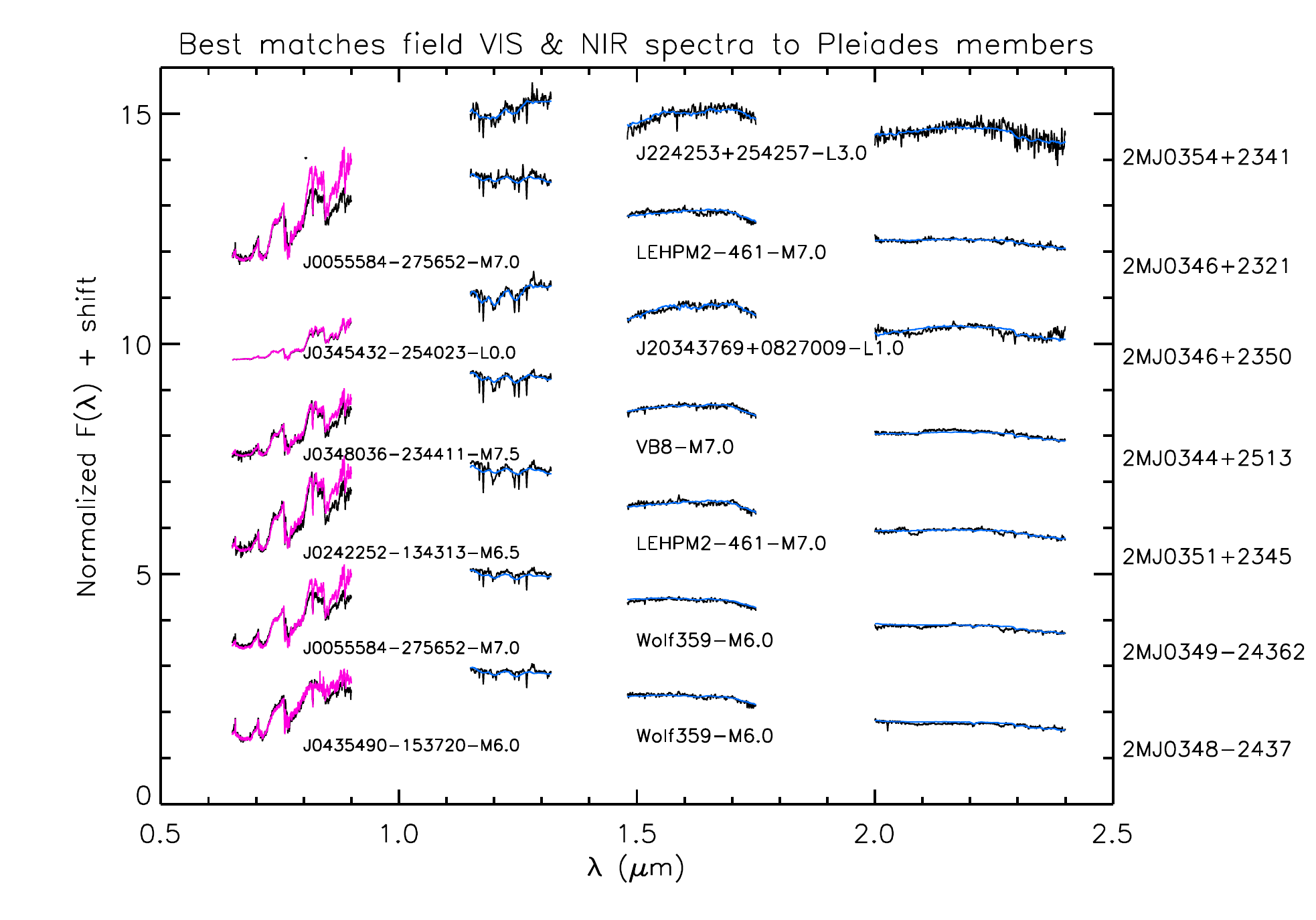}
\caption{Best field dwarf matches (coloured lines) in the optical and near-infrared for Pleiades members (black).}
\label{field_plei}
\end{figure*}

\begin{figure*}
\centering
\includegraphics[width=0.99\textwidth]{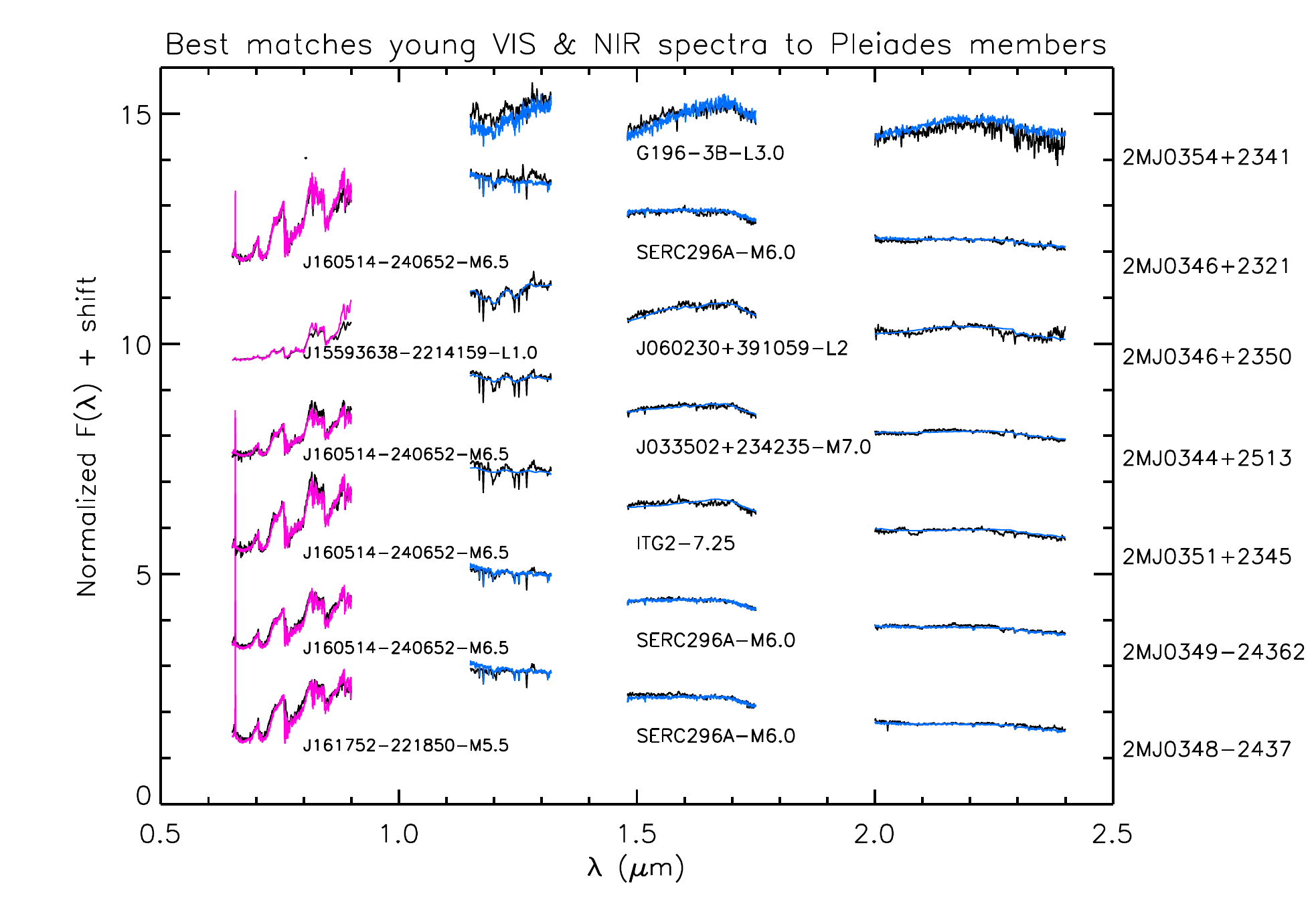}
\caption{Same as Figure \ref{field_plei} but comparing with the library of young M and L dwarfs.}
\label{young_plei}
\end{figure*}

\begin{figure*}
\centering
\includegraphics[width=0.99\textwidth]{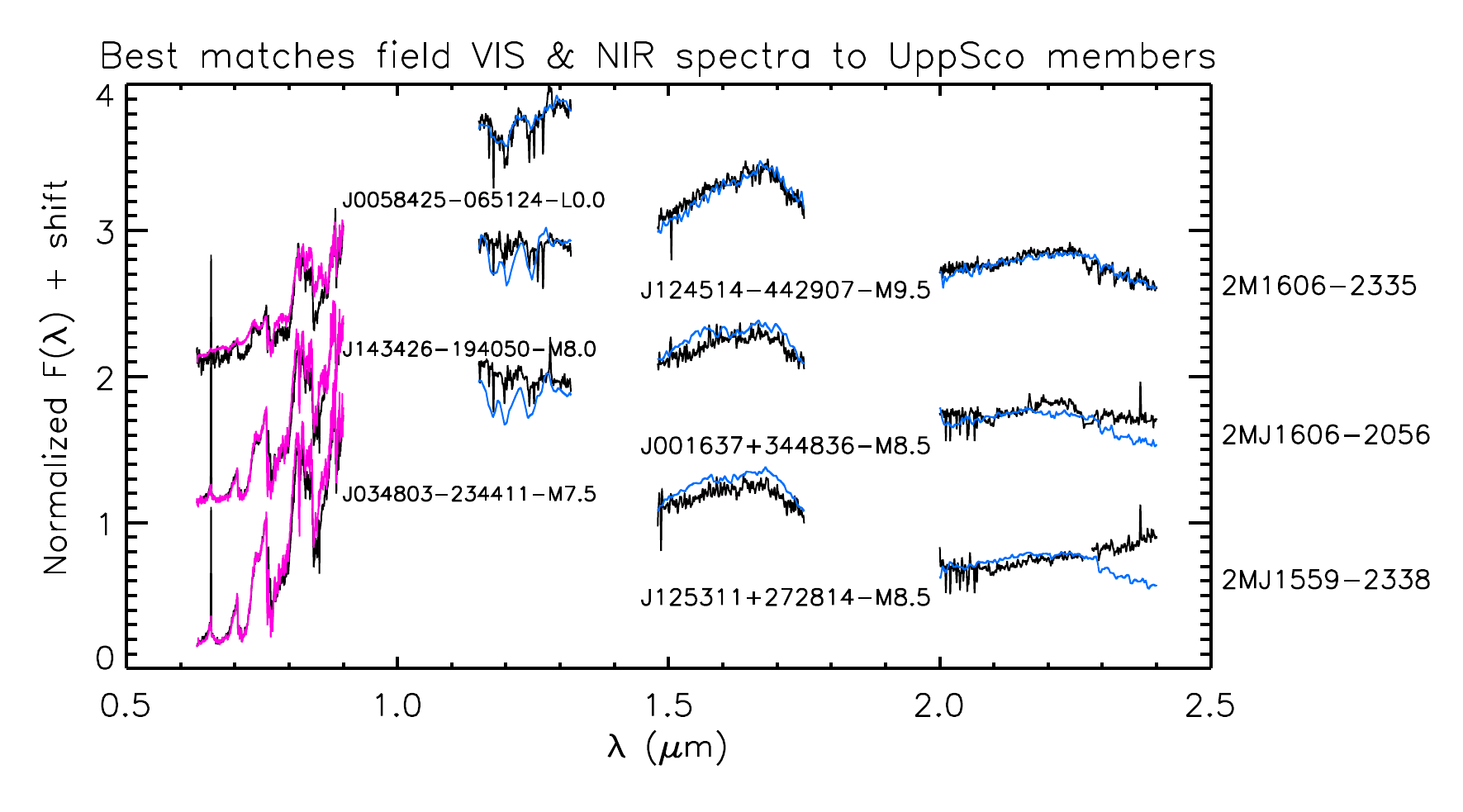}
\caption{Best field dwarf matches (coloured lines) in the optical and near-infrared for UppSco members (black).}
\label{field_upps}
\end{figure*}

\begin{figure*}
\centering
\includegraphics[width=0.99\textwidth]{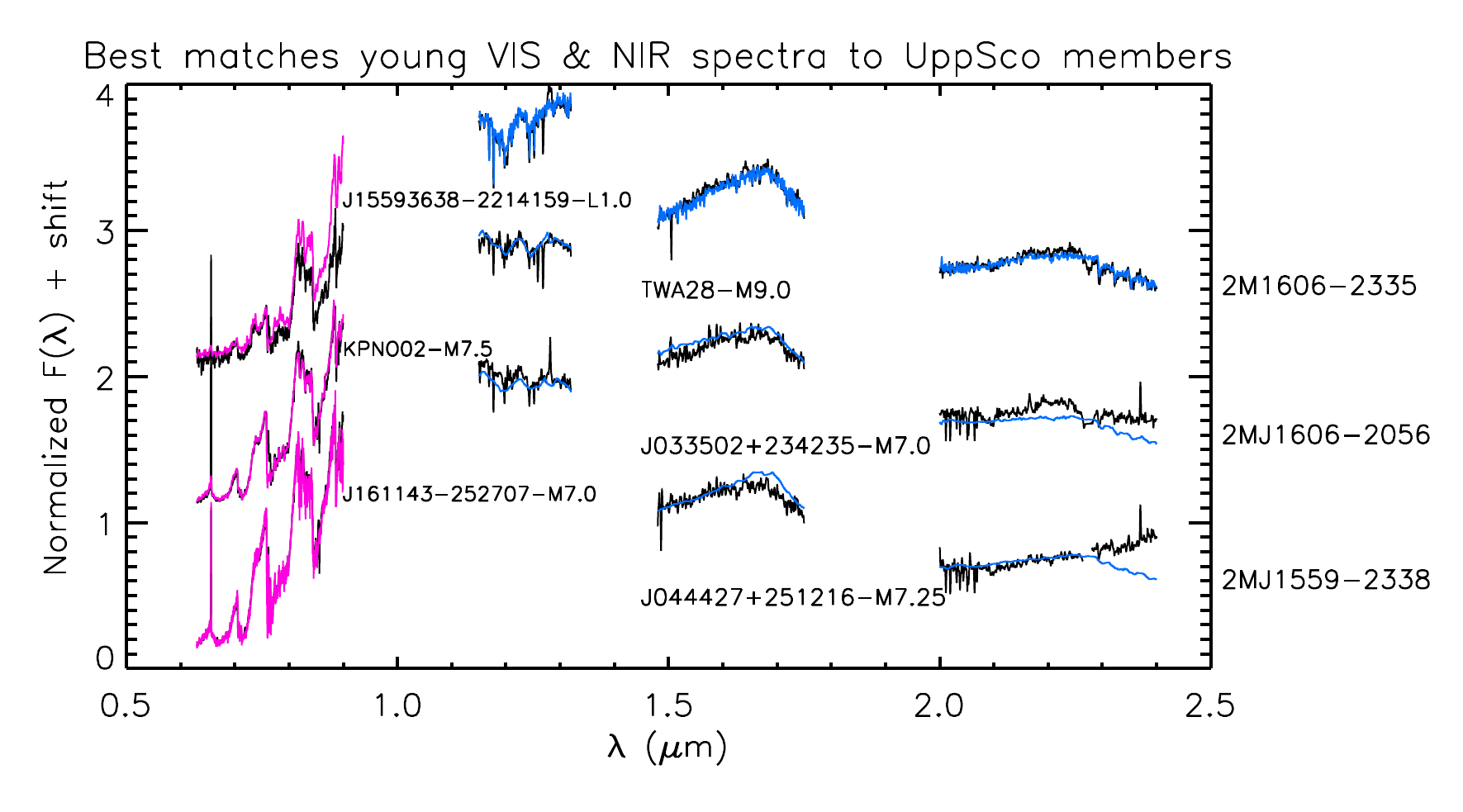}
\caption{Same as Figure \ref{field_upps} but comparing with the library of young M and L dwarfs.}
\label{young_upps}
\end{figure*}

\begin{figure*}
\centering
\includegraphics[width=0.99\textwidth]{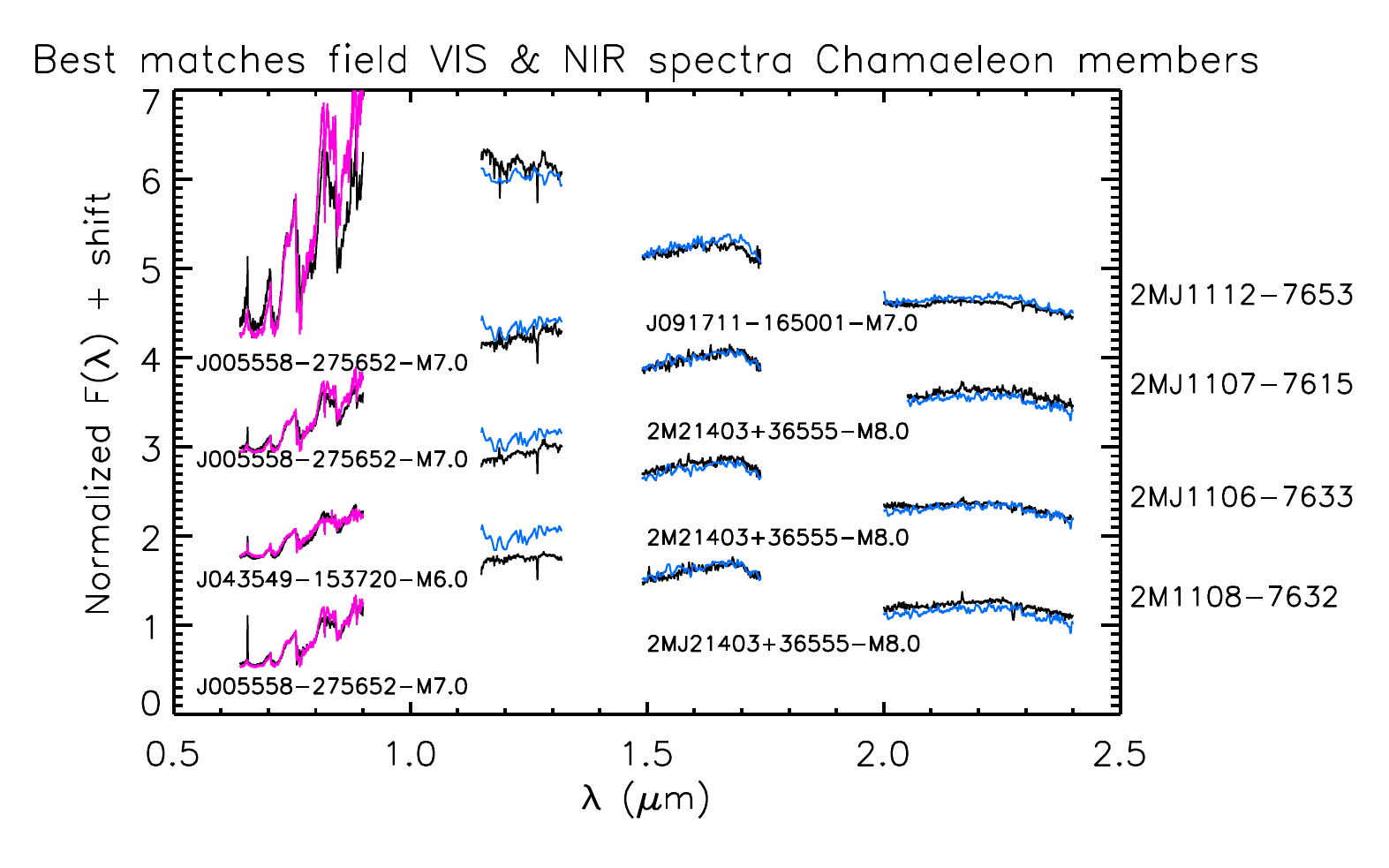}
\caption{Best field dwarf matches (coloured lines) in the optical and near-infrared for Cha\,I members (black).}
\label{field_cha}
\end{figure*}

\begin{figure*}
\centering
\includegraphics[width=0.99\textwidth]{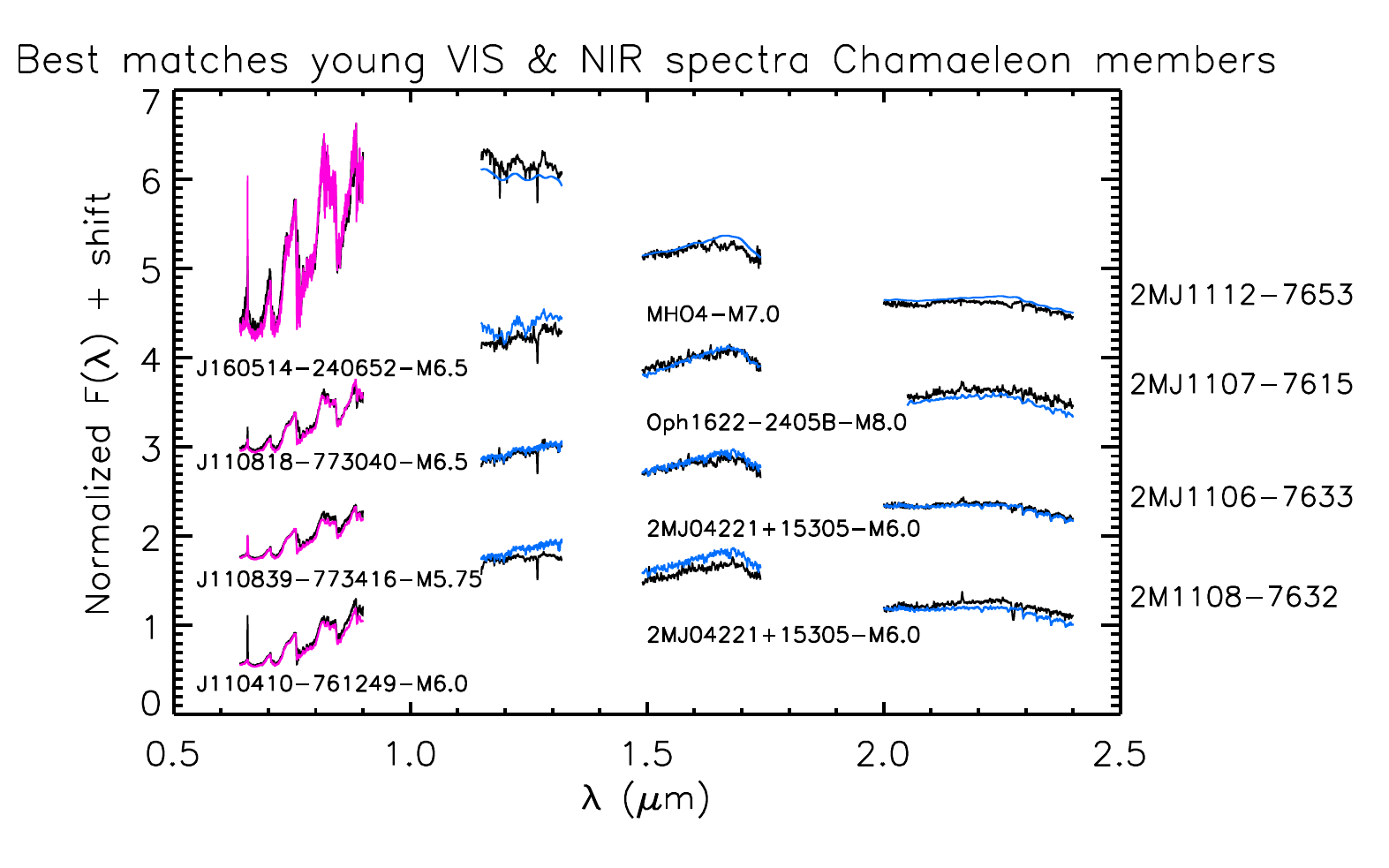}
\caption{Same as Figure \ref{field_cha} but comparing with the library of young M and L dwarfs.}
\label{young_cha}
\end{figure*}

%\begin{landscape}

%\end{landscape}

%\begin{landscape}

%\end{landscape}

%\section{This is the title of the first appendix}
%Larger tables, collections of images, spectra or similar kind of data shall be 
%presented in the appendix section rather than in the main body of the 
%text. Several appendices can be separated by the \verb$+$\section{$+${\it title
%of appendix}\verb$+$}$+$ command. They are enclosed in the 
%\verb$+$appendix$+$ environment.

	\bsp

\label{lastpage}

\end{document}